\documentclass[twocolumn, natbib]{svjour3}
\smartqed

\usepackage{graphicx}
\usepackage{placeins}
\usepackage{color}

\usepackage{amsmath}
\usepackage{amsfonts}
\usepackage{amssymb}

\usepackage{hyperref}
\usepackage{subcaption}

\begin{document}

\title{$\text{HVC}_\text{I}$ Neuron Properties from Statistical Data Assimilation}

\author{Daniel Breen \and Sasha Shirman \and Eve Armstrong \and Nirag Kadakia \and Henry Abarbanel}

\institute{Daniel Breen \and Sasha Shirman \and Eve Armstrong \and Nirag Kadakia \and Henry Abarbanel \at Department of Physics, University of California, San Diego, 9500 Gilman Drive, La Jolla, CA 92093-0402, USA}


\maketitle
\begin{abstract}
Data assimilation (DA) solves the inverse problem of inferring initial conditions given data and a model. Here we use biophysically motivated Hodgkin-Huxley (HH) models of avian $\text{HVC}_\text{I}$ neurons, experimentally obtained recordings of these neurons, and our data assimilation algorithm to infer the full set of parameters and a minimal set of ionic currents precisely reproducing the observed waveform information. We find many distinct validated sets of parameters selected by our DA method and choice of model. We conclude exploring variations on the inverse problem applied to neurons producing accurate or inaccurate results; by manipulating data presented to the algorithm, varying sample rate and waveform; and by manipulating the model by adding and subtracting ionic currents.
\end{abstract}

\section{Introduction}

Male zebra finches sing a short, stereotyped motif for the span of their adult life, a behavior that is learned over several months as a juvenile. One goal of the present research is to explore the single cell and network mechanisms underlying learned sequential behaviors with a biophysically grounded simulation of the song motor pathway. Conductance based models represent membrane dynamics in terms of ion currents through passive, voltage-gated, and ligand-gated conductances. The dynamics of these conductances can be expressed mathematically in terms of biophysical descriptions of specific ion channels of the cell. This provides a mechanistic link between molecular cell properties and behavior \citep{bean2007action, marder2007understanding}. Previous research strongly suggests that the membrane dynamics of single neurons within nucleus HVC are crucial mechanistic components encoding the zebra finch song motif~\citep{hahnloser2002ultra, jin2007intrinsic, long2010support}. Two classes of neuron constitute the building blocks of the motor pathway responsible for song production. The first is a class of neurons called $\text{HVC}_\text{RA}$ neurons which send excitatory projections to the nucleus RA from HVC. These neurons conspicuously feature at most one burst per song motif. The other kind of neuron is called an HVC Interneuron ($\text{HVC}_\text{I}$ neuron) and does not project to other nuclei, but instead inhibits neurons local to HVC.

In this paper we implement our methods of statistical data assimilation (DA) to construct a model of the $\text{HVC}_\text{I}$ neuron as an isolated building block. We demonstrate the power of DA to extract information about $\text{HVC}_\text{I}$ neurons given a biophysical model and data, such as a minimal set of necessary ionic currents and sets of parameters reproducing essential characteristic electrophysiological features. We also demonstrate the present limitations of our DA procedure that may be due to deficiencies in the model, an insufficient number of measurements, or the numerical difficulty of the DA computation. We do not attempt to study models of these cells within the context of a simulated network of neurons. The companion to this paper \citep{nirag2016hvcraneurons} studies the single cell mechanisms of an $\text{HVC}_\text{RA}$ neuron model. Future work will be aimed at placing these model $\text{HVC}_\text{RA}$ and $\text{HVC}_\text{I}$ cells as nodes within a network.

The mathematical form of our model is a Hodgkin Huxley (HH) conductance based model. The ion currents included are biologically grounded in pharmacological experiments and computer modeling .
Our data assimilation algorithm estimates all of the unobserved states and parameters of the model conditioned on measurements recorded from $\text{HVC}_\text{I}$ neurons in vitro. The present work is significant because it extends previous work \citep{toth2011dynamical, kostuk2012dynamical, knowlton2014dynamical, meliza2014estimating}, demonstrating that the time course of several unobservable variables and dozens of parameters that enter the dynamical equations nonlinearly, can be accurately estimated, using recently developed methods of data assimilation \citep{ye2015improved}. It also demonstrates the capability of the data assimilation methods to estimate \textit{all} the parameters of conductance-based Hodgkin Huxley models, including those which enter nonlinearly such as the parameters describing the gating kinetics, given voltage recordings of real $\text{HVC}_\text{I}$ neurons. Neuronal properties described by these parameters change over time \citep{cerda2010analysis}, have a spatial dependence over the different parts of the neuron\citep{bean2007action}, while also varying among cells of the same class \citep{schulz2006variable}. Because our method is capable of estimating all the parameters which enter into HH models, we can better characterize variability, such as the shape of I-V curves, across neurons. Data assimilation also allows one to characterize properties of a model that would not otherwise be readily apparent, such as the existence of multiple distinct parameter sets producing nearly indistinguishable voltage traces given a stimulating current protocol. We discuss such parameter set degeneracies for our $\text{HVC}_\text{I}$ neuron model.

Our data assimilation algorithm formulates the problem as one of nonlinear optimization over a high dimensional path integral and has been explored both in its exact and approximate form on various chaotic and neural models~\citep{toth2011dynamical, kostuk2012dynamical, abarbanel2013predicting, knowlton2014dynamical, meliza2014estimating, ye2014estimating, ye2015improved, nirag2016hvcraneurons}. Experiments done on simulated systems in which all of the parameters and states are known to the experimenter we call \textit{twin experiments}. The purpose of such experiments is to inform the use of the algorithm on real neural systems in order to estimate model parameters and unknown states which produce accurate predictions. In real physical systems, only sparse measurements are typically available and errors in the model and measurements are inevitable. If the data assimilation algorithm can recover unknown states and parameters in controlled conditions where the parameters and state of the system at all times are known, then we are more confident of its ability to do so with real systems.

First we motivate and validate our model with results of data assimilation on voltage recordings of $\text{HVC}_\text{I}$ neurons \textit{in vitro}, showing that it exhibits key qualitative features and biophysical mechanisms found in other work~\citep{kubota1998electrophysiological,daou2013electrophysiological}. We then show results of applying our data assimilation procedure on synthetic and real voltage recordings of single $\text{HVC}_\text{I}$ neurons.

\section{Methods}
\label{sec:Methods}
\subsection{$\text{HVC}_\text{I}$ Neuron Model}
\label{sec:Methods_model}

Our Hodgkin Huxley model for an isolated $\text{HVC}_\text{I}$ neuron derives from the model of \cite{daou2013electrophysiological}. The current balance equation for the membrane voltage is:

\begin{align*}
C_m\dfrac{dV(t)}{dt} =& I_K(V(t)) + I_{Na}(V(t)) + I_{CaL}(V(t),[Ca](t)) \\
&+ I_{CaT}(V(t),[Ca](t)) + I_A(V(t)) \\
&+ I_{SK}(V(t),[Ca](t)) + I_{KNa}(V(t)) + I_H(V(t)) \\
&+ I_{Nap} + I_L(V(t)) + I_{inj}(t)
\end{align*}

$C_m$ is the capacitance of the membrane and $I_{inj}$ is a custom built drive current. The ion currents are $I_K$, the delayed rectifier potassium current; $I_{Na}$, the inactivating sodium current; $I_{CaL}$, a high threshold L-type calcium current; $I_{CaT}$, a low threshold voltage gated calcium current; $I_A$, an A-type potassium current; $I_{SK}$, a small-conductance calcium activated potassium current; $I_{KNa}$, a sodium dependent potassium current; $I_H$, a hyperpolarization activated cation current; $I_{Nap}$, a persistent sodium current; and $I_L$, the leak current. Our analysis is restricted to isolated single $\text{HVC}_\text{I}$ neurons, so we do not include a synaptic current term $I_{syn}$.

We reduce this model to a simplified form with suitability for insertion into nodes within a simulation of HVC in mind. A simpler model is also easier to understand, reduces the computational difficulty of data assimilation and the number of parameter set degeneracies, or model symmetries (to be discussed below), present given the limitations of our present computational resources and methods to transfer the information in the data to our model. We eliminate $I_A$, $I_{SK}$, $I_{KNa}$, and $I_{Nap}$ because these currents have been shown to be small in $\text{HVC}_\text{I}$ neurons in other work \citep{daou2013electrophysiological}. $\text{HVC}_\text{I}$ neurons are highly excitable with high resting membrane potentials and very little afterhyperpolarization (AHP), so the absence of these currents can be partially understood as a consequence of $I_A$, $I_{SK}$, $I_{KNa}$ being currents which depress the resting membrane voltage of neurons and contribute to AHP. $I_{CaL}$ may contribute to the dynamics of $\text{HVC}_\text{I}$ neurons, but we did not find a significant difference in the quality of data assimilation results when $I_{CaL}$ was and was not present in addition to the remaining currents. We conclude that it can be justified for removal for our purposes, modeling the time evolution of the voltage dynamics for durations not exceeding $\approx$ 10 seconds, as not important in causing the most conspicuous features of $\text{HVC}_\text{I}$ neuron electrophysiology: a hyperpolarization induced sag upon hyperpolarizing current injection and rebound spiking upon release from hyperpolarizing current injection. $I_{CaT}$ and $I_H$ are both critical mechanisms in this behavior as discussed in other work \citep{daou2013electrophysiological}, so we retain them in our model.

Upon simplification our Hodgkin Huxley model for an isolated $\text{HVC}_\text{I}$ neuron becomes:
\begin{equation}
\begin{aligned}
C_m\dfrac{dV(t)}{dt} &= I_K(V(t)) + I_{Na}(V(t)) + I_H(V(t)) \\
&+ I_{CaT}(V(t), [Ca](t)) + I_L(V(t)) + I_{inj}(t) \\
\dfrac{d[Ca]}{dt} &= \phi I_{CaT}(V(t), [Ca](t)) + \frac{[Ca]_{eq} - [Ca](t)}{\tau_{[Ca]}} \\
\dfrac{dx}{dt} &= \frac{x_{\infty}(V) - x(t)}{\tau_x(V)} 
\end{aligned}
\end{equation}

  The variables ${n(t), m(t), h(t), H(t), a(t), b(t)} \in x$ are gating variables which regulate the conductance of ions through the membrane of a neuron, described by the following equations:

\begin{equation}
  \begin{aligned}
  x_{\infty}(V) &= 0.5(1 + \tanh(\frac{V - \theta_x}{\sigma_x})) \\
  \tau_x(V) &= t_1 + t_2(1 - \tanh^2(\frac{V - \theta_{x_{t}}}{\sigma_{x_{t}}}))
  \end{aligned}
  \end{equation}
  The $\theta$, $\sigma$, and $t_i$ are the kinetic parameters, properties of individual neurons.

The ion currents have the following form:
\begin{equation}
  \begin{aligned}
  \label{currents}
  I_K(V(t)) &= g_Kn(t)^4(E_{K} - V(t)) \\
  I_{Na}(V(t)) &= g_{Na}m(t)^3h(t)(E_{Na} - V(t)) \\
  I_H(V(t)) &= g_HH(t)^2(E_H - V(t)) \\
  I_{CaT}(V(t), [Ca](t)) &= g_{Ca_T}a(t)^3b(t)^3\Phi_{GHK}(V(t), [Ca](t)) \\
  \Phi_{GHK}(V(t), [Ca](t)) &= V(t)\frac{[Ca]_{ext}\exp(\frac{-V(t)}{V_T}) - [Ca](t)}{1-\exp(\frac{-V(t)}{V_T})}
  \end{aligned}
  \end{equation}
  
  The various $g_X$ and $E_X$ are the maximal conductances and reversal potentials of the ion currents. 

  $\Phi_{GHK}(V(t),[Ca](t))$ is the Goldman-Hodgkin-Katz equation for the ionic flux through the neuron membrane. Here it is used instead of the ohmic form to accurately model the calcium current.
  
  Calcium appears as a dynamical variable in the model. The equations describing the calcium dynamics are informed by calcium ion conservation. These equations balance the change caused by calcium ion current influx and decay to equilibium calcium concentration.

  \begin{align*}
  \dfrac{d[Ca](t)}{dt} &= \phi I_{CaT}(V(t), [Ca](t)) + \frac{[Ca]_{eq} - [Ca](t)}{\tau_{[Ca]}} 
  \end{align*}
  
  $[Ca](t)$ is the cytosolic, or internal, calcium concentration, $[Ca]_{eq}$ is the equilibrium cytosolic calcium concentration, $\tau_{[Ca]}$ is the time constant describing the rate at which the internal concentration of calcium tends towards its equilibrium concentration, $[Ca]_{ext}$ is the concentration of calcium outside the cell membrane, and $V_T$ is the thermal voltage. With our $\text{HVC}_\text{I}$ model, parameters governing only the time evolution of $[Ca](t)$ cannot be determined from data assimilation using a measurement of $V(t)$ alone, as $[Ca](t)$ is only weakly coupled to the dynamics of membrane voltage. Virtually no calcium is available inside the cell to flow outwards at physiological concentrations, so $\Phi_{GHK}(V(t), [Ca](t)) \approx \Phi_{GHK}(V(t))$.

  $\text{HVC}_\text{I}$ neurons recorded in vitro are highly excitable, have a slow increase in resting membrane potential in response to hyperpolarizing current injection, and fire rebound action potentials in response to release from hyperpolarizing current injection~\citep{kubota1998electrophysiological, daou2013electrophysiological}. With the support of pharmacological manipulation and computational modeling, ~\cite{daou2013electrophysiological} suggests the underlying mechanisms are two voltage gated ion currents, a hyperpolarization activated current $I_H$ and a low threshold T type calcium current $I_{CaT}$. With these currents, our model reproduces the above qualitative features as shown in \autoref{fig:threefigs}. The parameters of the model in \autoref{table:real_estimated_parameters} exhibiting these features are derived from data assimilation on a real $\text{HVC}_\text{I}$ neuron. Results of estimation and prediction on the voltage trace, with corresponding stimulating current, are plotted in \autoref{figure:real_hvci_assimilation}. As shown in \autoref{fig:threefigs}, when $I_H$ alone is blocked, the sag is eliminated, but a delayed rebound spike is preserved. When both $I_H$ and $I_{CaT}$ are blocked, the sag and rebound spiking are eliminated. Additionally, when $I_{CaT}$ is blocked, the membrane potential is depressed by about 10 mV, in agreement with experimental observations\citep{daou2013electrophysiological}. This demonstrates that an HH model with currents defined in \autoref{currents} is sufficient to describe experimentally obtained voltage recordings of $\text{HVC}_\text{I}$ neurons.

\subsection{Path Integral Methods of Data Assimilation}
\label{sec:path_integral_methods}

Data assimilation refers to analytical and numerical procedures in which information in measurements is transferred to model dynamical equations seleected to describe the processes thought to produce the data. In the absence of measurements or first principles (or even second principles!) to derive parameters and unknown states in a model, data assimilation provides a method to obtain them systematically.

The implementation of data assimilation algorithms runs into numerical difficulties when the system is nonlinear and of high dimension, and when the measurements are sparse and noisy. In neuronal systems, typically the time course of the membrane $V(t)$ at the soma can be measured, but not the activation of the gating variables or most ionic concentrations. This renders difficult the algorithmic implementation of the inference of unobserved parameters and state variables coupled to the measured variables through the model equations.

We formulate our problem as a path integral realization of a statistical data assimilation procedure \cite{abarbanel2013predicting}. Because measurements and models will always have errors, our data assimilation method formally represents the time evolution of the unknown quantities with probability distributions. State variables of the models evolve in time according to a dynamical rule specified by the values of model parameters.


The vector of states is a D dimensional vector $\textbf{x}(t)$ defined at times $t_0,t_1,...,t_M$. These times constitute the estimation window $[t_0,t_M]$. Usually, the number L of measured states is much smaller than D. These measurements we denote $\textbf{y}(t)$. The goal of data assimilation is to estimate unknown state variables at the end of the estimation window $\textbf{x}(t_M)$ and the unknown model parameters $\textbf{p}$.

Skipping the derivation \cite{abarbanel2013predicting}, the probability for the configuration of the state vector at the end of the window given the observations is of the form:

\begin{equation}
\label{eq:prob_dis}
P(x(t_M)|Y(t_M)) = \int dX \exp(-A_0(X,Y))
\end{equation}

Here X and $Y$ denote the collection of all state variables $\textbf{x}(t)$ and set of measurements $\textbf{y}(t)$ at every time point in the estimation window, respectively. $A_0$ is a cost function that we minimize, which we call the action by analogy with path integral formalisms used in statistical physics and quantum mechanics \cite{abarbanel2013predicting}.

There are two ways to evaluate the integral. One is to sample the probability distribution directly using a Metropolis-Hastings Monte Carlo (MHMC) algorithm \cite{kostuk2012dynamical}. The other is to use Laplace's method and expand around stationary paths. This is the method that is used in this paper. This shifts the numerical difficulty of the problem into one of optimization, in the present case of finding the lowest minima of the cost function $A_0$, a non-convex problem.

$A_0$ can be written as a sum of two terms; the deviation of estimated states from their measured value and the deviation in estimated states from the value obtained from evolving estimated states at the previous time step forward with the estimated dynamical map. When the measurement noise and model error is assumed additive and Gaussian, the action has a particular form:

\begin{multline}
A_0(X|Y(t_M)) =\frac{R_m}{2}\sum_{n=0,l=1}^{M,L} (h_l(\textbf{x}(t_n)) - y_l(t_n))^2 \\ + \frac{R_f}{2}\sum_{n=0,d=1}^{M-1,D} (x_d(t_{n+1}) - f_d(\textbf{x}(t_n)))^2
\end{multline}

The $h_l$ are measurement operators which operate on the state of the system at every time point $\textbf{x}(t_n)$. $R_m$ is the inverse variance of the measurement error, and $R_f$ is the inverse variance of the model error. The relative values of $R_m$ and $R_f$ are assigned before the beginning of the optimization procedure.

\subsection{Annealing}

The way in which $R_m$ and $R_f$ are weighted relative to one another in the cost function influences the result of minimizing the cost function. Manipulating the cost function by varying these values forms the basis of our annealing method, shown to be effective in state and parameter estimation in archetypal chaotic models such as the Lorenz '96 model \cite{ye2015improved}. When $R_m \gg R_f$, the measurement error is assumed small while the model error is large. Such an assumption causes the cost function to form minima in the high dimensional landscape defined over the state and parameter space where measured states in the model fit the data closely. The model error is assumed large, so the model is enforced weakly, effectively decoupling parameters and unmeasured states from the data. These are unlikely to be estimated correctly\cite{ye2015improved}.


When $R_m \approx R_f$, both terms contribute equally to the cost function, so minimizing the cost function will tend to satisfy the data while simultaneously enforcing the dynamical map. However, when the model error is forced, by large $R_f$, to be small, the nonlinearity of the vector field \textbf{f(x)} manifests itself at the smallest scales in the phase space of the paths $X$ over which we are searching. This results in complicated fine structure seen as multiple local minima \cite{abarbanel2013predicting} in the action, especially when the number of measurements L is too small. It is unlikely that directly minimizing the cost function under this condition will yield good estimates of the system's parameters and state variables.

An annealing method has been developed \cite{ye2015improved} that uses information available in educated initial guesses about where the minima are to attempt to fit the data and enforce the model simultaneously. In this method, $R_m \gg R_f$ initially. The cost function is then minimized, which is typically easy. Then $R_f$ is increased in magnitude by a factor $\alpha > 1$, and the cost function is minimized again, starting the search for minima at the previous solution. The process is repeated until $R_f \gg R_m$. In this way, the algorithm creeps gradually towards a minimum which fits the data and the model better than other options.

The implementation of the algorithm was accomplished through the use of the open source software package IPOPT (Interior Point OPTimizer) with the linear solver ma57 \cite{wachter2006implementation}.

\subsection{Twin Experiments}
\label{sec:twinexperiments}

To build confidence in the ability of our algorithm to return the correct values of unknown parameters and states on a given system with sparse measurements, we attempt experiments in which we have as much knowledge and control of the system and experimental data as possible. We generate synthetic data so that the system, including all unknown states and parameters, are known to the experimenter. The experimenter compares the output of the algorithm with the true underlying dynamics. Because the model is exact and mirrors the true system, we call such an experiment a "twin experiment" ~\citep{abarbanel2013predicting}.

In a real experiment, the experimenter will not know the value of most parameters and states. The experimenter will usually have data for measured states for times greater than the end of the estimation window. When this is the case, the experimenter can test the algorithm's estimates by integrating the model forward using the estimated dynamical map from the configuration of the system at the end of the estimation window. This is called the model prediction. The prediction is compared with the measurements to evaluate the quality of state and parameter estimates within the estimation window. Good 'fits' to the data are easy to achieve, and are not considered a good measure of the quality of the estimated model.

In neural systems, injected currents that are high enough in amplitude to drive spiking behavior, long enough in duration to sample all the degrees of freedom of a model neuron, and low enough in frequency to not be absorbed into the $RC$ time constant of the membrane 
are 
necessary for the algorithm to succeed in a twin or real experiment. This is a result of the fact that nothing can be inferred about a component of a process if it is not influencing the behavior of the system in a data set. In order for the dynamical map to be inferred from the data, all of its degrees of freedom must be activated for the algorithm to have a chance to infer values of parameters and unknown states that generalize outside of the training set~\citep{hobbs2008using}. A chaotic stimulus waveform of sufficiently low frequency, such as the trajectory of one of the states generated from the Lorenz `63 dynamical equations, will satisfy the above conditions. In our twin experiments on the $\text{HVC}_\text{I}$ model neuron, the voltage is the only measured variable, an expected experimental limitation on what can be measured in real neurons. To our synthetic data, Gaussian noise is added at each time point. We then run the data assimilation algorithm and compare the output, estimations of unknown parameters and unobserved states, to their true values. Following this, the true values or trajectories are compared to the model predicted trajectories outside of the estimation window.

\subsection{Data Assimilation on Real Data}

A twin experiment informs us about what measurements are necessary in order for data assimilation on a real system to succeed. Our twin experiments on a single compartment $\text{HVC}_\text{I}$ neuron, to be described in a later section, demonstrate the conditions that a time series measurement of only the voltage will suffice to produce accurate predictions. Bounds, such as those on kinetic parameters associated with $I_{CaT}$, were chosen to be around values reviewed in the literature \citep{huguenard1996low}. Experimental voltage data was sampled at either 50,000 or 10,000 Hz. Lower time resolution has the disadvantage that less subthreshold information is available. However, because simulations are run on single computing nodes and therefore larger problems take a long time to run, an advantage of coarse grained time resolution is that currents stimulating more degrees of freedom in the model can be chosen. Bounds and estimates of the parameters are given in table~\ref{table:real_estimated_parameters}. $E_{Na}$, $E_K$, $E_H$, $V_T$, and $[Ca]_{ext}$ are known for these neurons, so were fixed during data assimilation.

\! \begin{table}
\centering
\begin{tabular}{|c|c|c|c|}
\hline 
Parameter & Lower Bound & Upper Bound & Estimate \\ 
\hline 
$C_{m} (nF)$ & 0.01 & 0.033 & 0.0317 \\ 
\hline 
$g_{Na} (\mu \text{S})$ & 0.01 & 10.0 & 0.63 \\ 
\hline 
$g_K (\mu \text{S})$ & 0.01 & 15.0 & 2.15 \\ 
\hline 
$g_{H} (\mu \text{S})$ & 0.0001 & 0.01 & 0.0032 \\ 
\hline 
$g_{Ca_T} (\mu \text{S} \mu \text{M}^{-1})$ & 0.00001 & 0.01 & 0.0064 \\ 
\hline 
$g_L (\mu \text{S})$ & 0.0001 & 0.01 & 0.0052 \\ 
\hline 
$E_{Na}$ (mV) & - & - & 55.0 \\ 
\hline
$E_K$ (mV)& - & - & -90.0 \\ 
\hline
$E_H$ (mV)& - & - & -40.0 \\ 
\hline
$E_L$ (mV)& -90.0 & -30.0 & -66.32 \\ 
\hline
$\theta_H$ (mV)& -85 & -55 & -81.62 \\ 
\hline 
$\sigma_H$ (mV)& -62.5 & -5.0 & -9.80\\ 
\hline 
$t_{H_1}$ (ms)& 1.0 & 1000.0 & 214.39 \\ 
\hline 
$t_{H_2}$ (ms)& 10.0 & 2000.0 & 157.80 \\ 
\hline 
$\theta_{H_t}$ (mV)& -80.0 & -40.0 & -59.70 \\ 
\hline 
$\sigma_{H_t}$ (mV)& -62.5 & -5.0 & -5.52 \\ 
\hline 
$\theta_a$ (mV)& -80.0 & -30.0 & -30.0 \\ 
\hline 
$\sigma_a$ (mV)& 5.0 & 62.5 & 32.9\\ 
\hline 
$t_{a_1}$ (ms)& 0.01 & 5.0 & 4.44 \\ 
\hline 
$t_{a_2}$ (ms)& 1.0 & 20.0 & 4.24\\ 
\hline 
$\theta_{a_t}$ (mV)& -80.0 & -40.0 & -55.12 \\ 
\hline 
$\sigma_{a_t}$ (mV)& 5.0 & 62.5 & 5.0 \\ 
\hline 
$\theta_b$ (mV)& -90.0 & -60.0 & -61.98 \\ 
\hline 
$\sigma_b$ (mV)& -62.5 & -5.0 & -62.5 \\ 
\hline 
$t_{b_1}$ (ms)& 0.01 & 10.0 & 2.90 \\ 
\hline 
$t_{b_2}$ (ms)& 1.0 & 100.0 & 7.57 \\ 
\hline 
$\theta_{b_t}$ (mV)& -90.0 & -50.0 & -59.6 \\ 
\hline 
$\sigma_{b_t}$ (mV)& -62.5 & -5.0 & -15.1 \\ 
\hline 
$\theta_m$ (mV)& -50.0 & -30.0 & -32.304 \\ 
\hline 
$\sigma_{m}$ (mV)& 5.0 & 62.5 & 32.4 \\ 
\hline 
$t_{m_1}$ (ms)& 0.001 & 1.0 & 0.001 \\ 
\hline 
$\theta_h$ (mV)& -60.0 & -20.0 & -58.54 \\ 
\hline 
$\sigma_h$ (mV)& -62.5 & -5.0 & -59.2 \\ 
\hline 
$t_{h_1}$ (ms)& 0.01 & 1.0 & 0.42 \\ 
\hline 
$t_{h_2}$ (ms)& 1.0 & 10.0 & 4.44 \\ 
\hline 
$\theta_{h_t}$ (mV)& -60.0 & -20.0 & -60.0 \\ 
\hline 
$\sigma_{h_t}$ (mV)& -100.0 & -5.0 & -12.5 \\ 
\hline  
$\theta_n$ (mV)& -60.0 & -20.0 & -30.01 \\ 
\hline 
$\sigma_n$ (mV)& 5.0 & 62.5 & 62.5 \\ 
\hline 
$t_{n_1}$ (ms)& 0.01 & 1.0 & 0.01 \\ 
\hline 
$t_{n_2}$ (ms)& 0.1 & 10.0 & 10.0 \\ 
\hline 
$\theta_{n_t}$ (mV)& -60 & -20 & -30.79 \\ 
\hline 
$\sigma_{n_t}$ (mV)& -100.0 & -5.0 & -37.7 \\ 
\hline 
$\phi (\mu \text{M} \text{nA}^{-1})$ & 0.01 & 10.0 & 3.88 \\ 
\hline 
$\tau_{Ca}$ (ms)& 0.1 & 100.0 & 0.143 \\ 
\hline 
$V_T$ (mV)& - & - & 12.5 \\ 
\hline 
$[Ca]_{ext} (\mu \text{M})$ & - & - & 2500.0 \\ 
\hline 
$[Ca]_0 (\mu \text{M})$ & 0.01 & 5.0 & 1.11 \\ 
\hline 
\end{tabular}
\caption{Parameter estimates after 28 annealing steps using a voltage recording of an actual $\text{HVC}_{\text{I}}$ neuron as input into the data assimilation algorithm. A few of the parameters are at or near the bounds, but most are somewhere in between. The predictions are good, so there may be a few degeneracies in the model with respect to several parameters for the currents used. A `-' here denotes that a parameter was set at the indicated value because it is either fixed by experiment or well known (The various $E_x$'s and the extracellular calcium concentration $[Ca]_{ext}$), for example.}
\label{table:real_estimated_parameters}
\end{table}

\section{Results}
\label{sec:results}

\subsection{Data Assimilation for a Real $\text{HVC}_\text{I}$ Neuron}
\label{sec:RealDataAssimilation}

\begin{figure}
\includegraphics[scale=0.175]{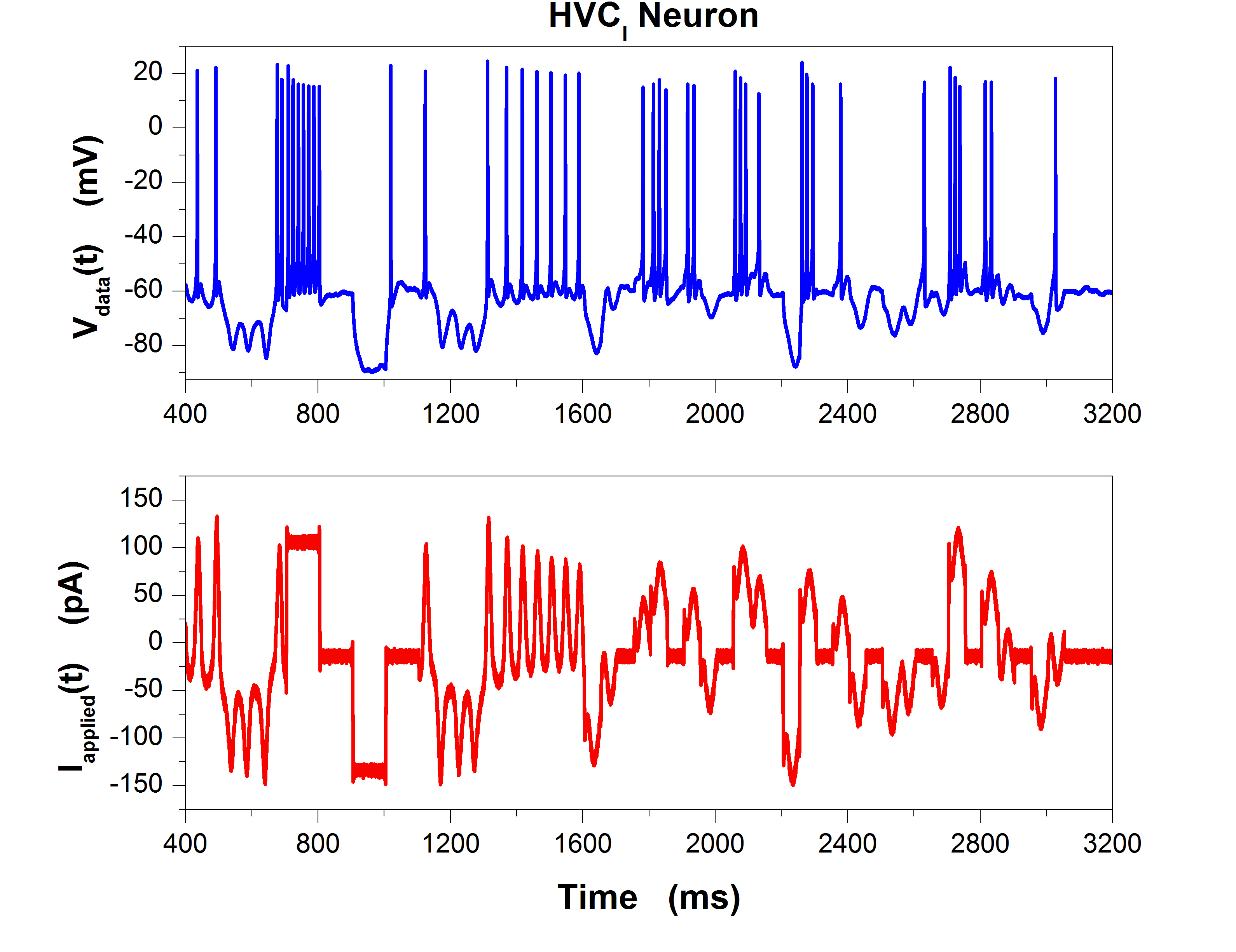}
\includegraphics[scale=0.175]{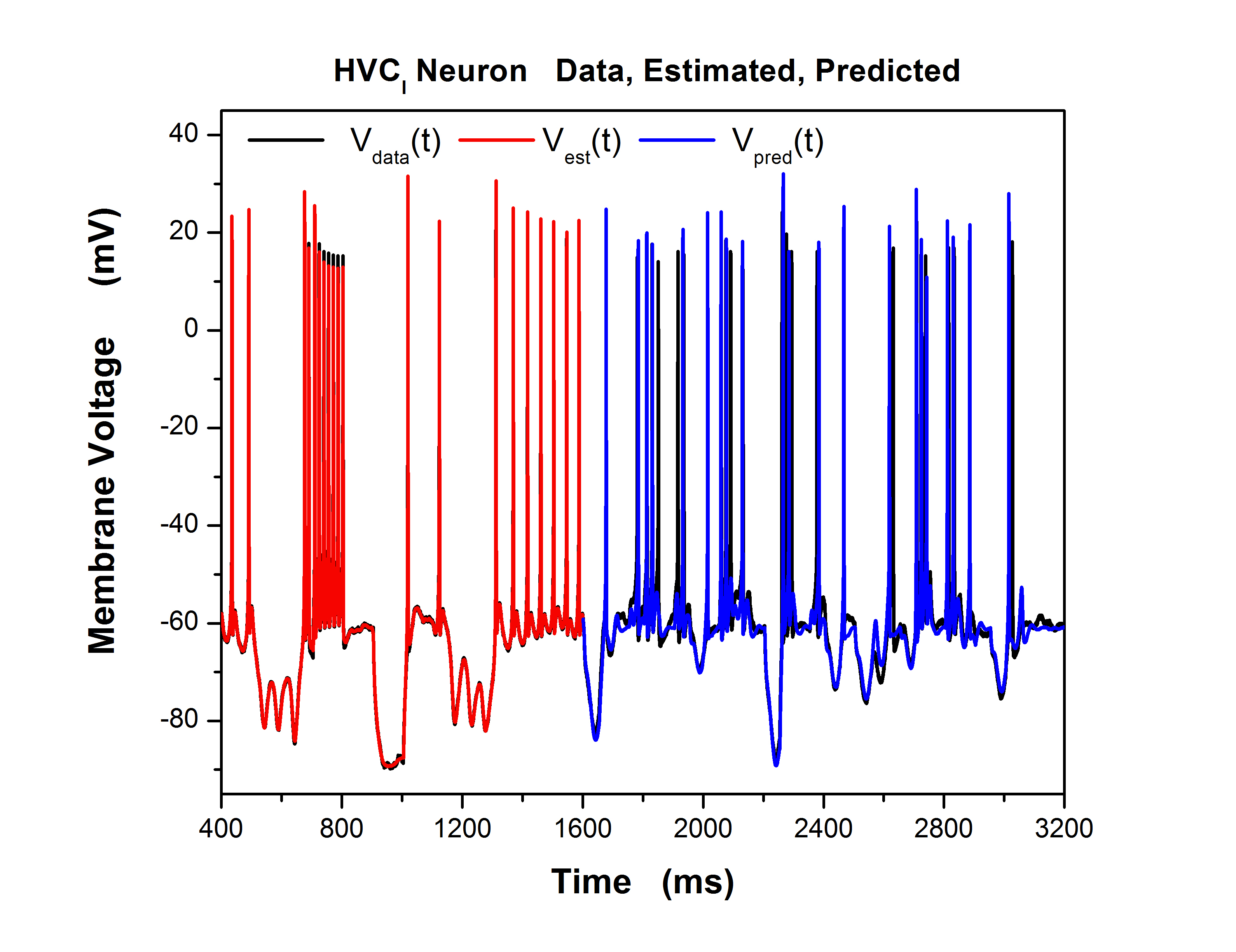}
\caption{Top: Voltage trace (blue) and injected current (red) used in data assimilation, obtained from an $\text{HVC}_{\text{I}}$ recorded \textit{in vitro}.
\newline
Bottom: Comparison of estimated voltage (red) and predicted voltage for times $t > t_M$ with the recording (black). The prediction is obtained by integrating the configuration of the system at the end of the estimation window $\mathbf{x}(t_M)$ forward with the estimated parameters. The agreement between prediction and observed data is excellent.}
\label{figure:real_hvci_assimilation}
\end{figure}

As a preliminary analysis we attempted data assimilation with voltage recordings obtained from real $\text{HVC}_\text{I}$ neurons \textit{in vitro}. The estimates for parameters during this analysis were used to later generate synthetic data used in twin experiments to better understand the extent that our method of data assimilation is able to recover dynamical properties of single neurons. Our estimation window consisted of 16001 time points sampled at 10000 Hz. These data - the resulting estimation of the voltage, the comparison of the obtained prediction of the voltage with the actual recording outside of the training window, and the stimulating current used - are plotted in Figure~\ref{figure:real_hvci_assimilation}. The parameter set obtained from data assimilation is included in Table \ref{table:real_estimated_parameters}. The results of prediction agree well with the data. As discussed in Section~\ref{sec:qualitativefeatures}, qualitative features of $\text{HVC}_\text{I}$ neurons including a sag in the voltage in response to hyperpolarizing current injection and rebound spiking upon termination of a hyperpolarizing stimulus were reproduced by the estimated parameter set. When $I_{H}$ is blocked by setting $g_H = 0$ in the model and simulating the behavior of the  neuron, the sag is eliminated and weaker rebound spiking is observed. When $I_{CaT}$ is blocked in addition to $I_{H}$, neither the sag nor rebound spiking is observed, and the resting membrane voltage is depressed by about 10 mV, all in accordance with experimental observation \citep{kubota1998electrophysiological, daou2013electrophysiological}.

\subsection{Qualitative Behavior of $\text{HVC}_\text{I}$ Model}
\label{sec:qualitativefeatures}

Previously performed experiments on $\text{HVC}_\text{I}$ neurons in vitro show that a sag in the voltage appears when negative DC injected current is applied. When this negative DC current is removed, rebound spiking is observed ~\citep{kubota1998electrophysiological, daou2013electrophysiological}. \cite{daou2013electrophysiological} also shows through pharmacological manipulation and computer modeling that the biological mechanisms responsible for this are a T-type calcium current and a hyperpolarization activated cation current. \cite{huguenard1996low} describes features of T-type calcium currents in the central nervous system. Channels gating these currents are open only when their simultaneous activation and deinactivation are achieved. When the neuron is depolarized, $I_{CaT}$ becomes increasingly inactivated and deinactivation requires a duration of quiescence or hyperpolarization. During hyperpolarization, $I_H$ causes a sag in the voltage waveform to appear. Deinactivated $I_{CaT}$ acts in conjunction with $I_H$ when a neuron is released from hyperpolarization to cause rebound spiking.

$\text{HVC}_\text{I}$ neurons are highly excitable, firing with high frequency in response to depolarizing current pulses. This may be in part due to the depolarizing influence of $I_H$ and $I_{CaT}$ which bring the neuron close to firing threshold. Large maximal conductances $g_{Na}$ and $g_K$ corresponding to the currents $I_{Na}$ and $I_K$ may also contribute to high excitability.

\begin{figure}
\includegraphics[scale=0.3]{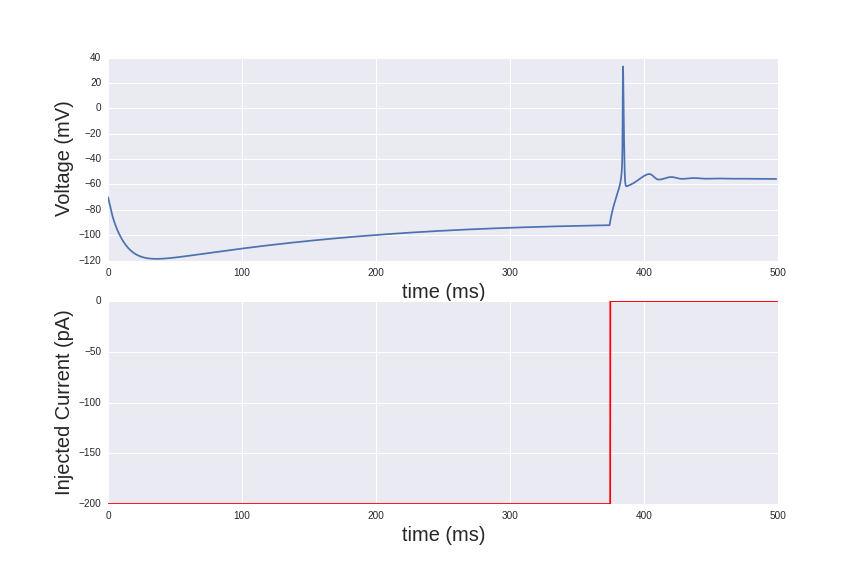}
\includegraphics[scale=0.3]{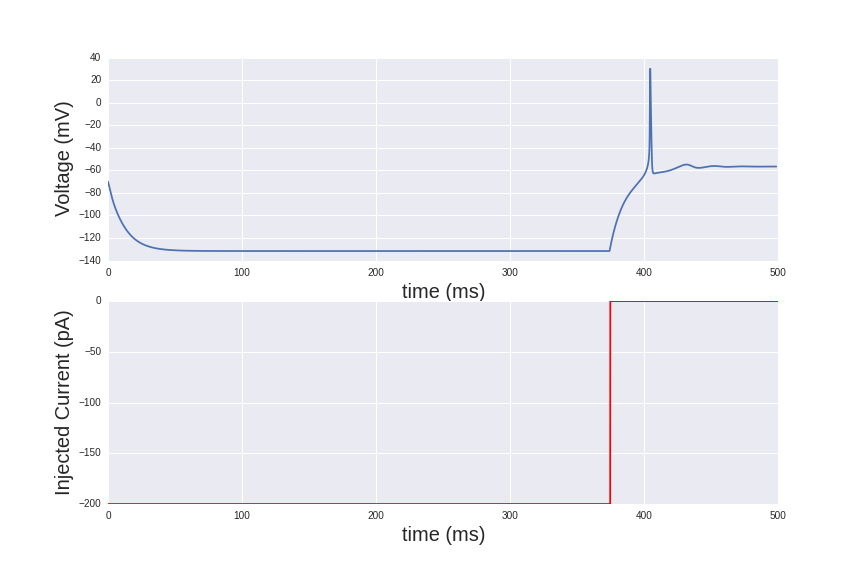}
\includegraphics[scale=0.3]{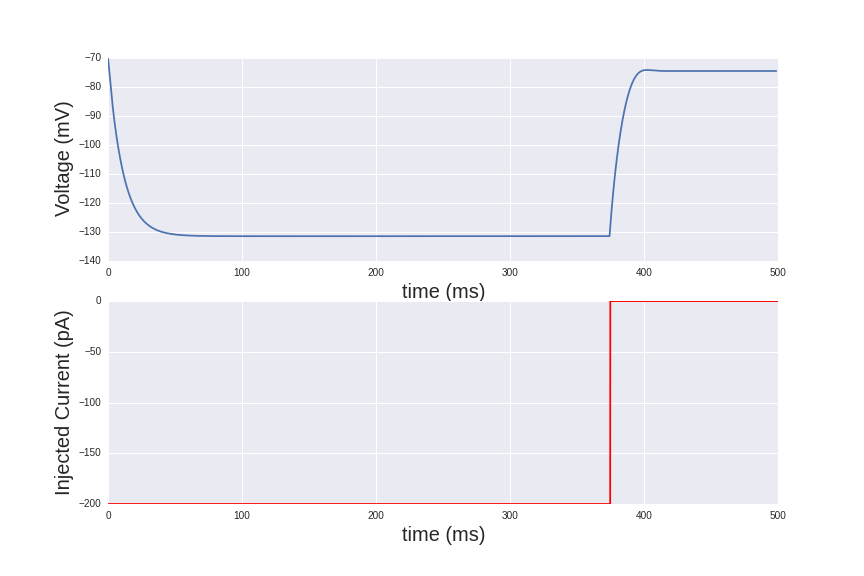}
\caption{Qualitative features of the $\text{HVC}_\text{I}$ model neuron with a parameter set (Table~\ref{table:real_estimated_parameters}) obtained from data assimilation on a real $\text{HVC}_\text{I}$ neuron. When $I_H$ and $I_{CaT}$ are not blocked, the model produces rebound spiking when released from hyperpolarizing current injection. When $I_H$ is blocked but $I_{CaT}$ is not blocked, the model produces weak rebound spiking while the sag disappears (top). When $I_H$ and $I_{CaT}$ are both blocked, neither rebound spiking nor a sag are seen (bottom). Additionally, the resting membrane voltage is depressed by about 10 mV, in accordance with experimental observations~\citep{daou2013electrophysiological}.}
\label{fig:threefigs}
\end{figure}

Table~\ref{table:real_estimated_parameters} shows the set of parameters that were used in generating Fig.~\ref{fig:threefigs}. When developing a model, a central idea to data assimilation is that the form of the model must be determined by the modeler. However, once the model is specified, the values of the unknown parameters are to be determined by the algorithm.

$I_H$ and $I_{CaT}$ have the following form:

\begin{equation}
\centering
\begin{aligned}
I_H(V(t)) =& g_HH(t)^2(E_H - V(t)) \\
I_{CaT}(V(t),[Ca](t)) =& g_{Ca_T}a(t)^3b(t)^3\Phi_{GHK}(V(t),[Ca](t))
\end{aligned}
\end{equation}

In \autoref{fig:threefigs} we verified that our model produces the expected results: $I_H$ induces sag, while $I_H$ and $I_{CaT}$ both contribute to the rebound spiking. When $g_H$ is set to zero, the sag is eliminated while weaker rebound spiking remains. When $g_{Ca_T}$ is also set to zero, both the sag and rebound spiking are eliminated (Figure~\ref{fig:threefigs}). This verifies the results of pharmacologically blocking $I_H$ and $I_{CaT}$~\citep{daou2013electrophysiological}. 

\subsection{Real Data: Dependence of Data Assimilation Results on Stimulating Currents and Sample Rate}

We contrasted the effects of different stimulating current protocols on the results of data assimilation and examined the structure of model 'symmetries' given assimilated data collected from $\text{HVC}_\text{I}$ neurons in vitro. A 'symmetry' is the presence of multiple sets of parameters and initial conditions that produce similar voltage waveform behavior given a stimulating current. More precisely, integrating forward our $\text{HVC}_\text{I}$ model using a number of estimated parameter sets produces indistinguishable (or nearly so) time evolution in the voltage given a stimulating current protocol. A number of measures, to be enumerated below, were used to analyze the structure in the model producing these symmetries given the assimilated data. To explore the tradeoffs given finite computational power between stimulating extra degrees of freedom and losing resolution of the measurements in time by downsizing the measured data, the same voltage traces were sampled at different frequencies. One set was sampled at 50 kHz, while the other set was sampled at 10 kHz. We used 24001 data points (480.02 ms) when analyzing the 50 kHz voltage trace and 10001 data points (1000.1) data points when analyzing the 10 kHz voltage trace. For either the 10 kHz or 50 kHz sampling rate conditions, 3 different stimulating current protocols were used to drive the voltage; a step current, a high frequency chaotic current, and a low frequency chaotic current. The quality of the resulting estimates were based on the predictive capabilities of the model estimations. In each case the lower frequency chaotic current protocol producing a voltage waveform with the highest quality predictions. The step current produced the lowest quality predictions, while the high frequency chaotic current performed in the middle. This is in contrast to the situation in twin experiments in section \ref{sec:twinexperimentsDependence}, where the step current and low frequency chaotic current stimulus fared the best, with the high frequency chaotic current giving the most mediocre results. This discrepency can be accounted for by the fact that the voltage waveform of the model neuron in twin experiments did not become driven to the subthreshold regime, while the voltage of the \textit{in vitro} $\text{HVC}_\text{I}$ neuron did become driven to this regime given the same stimulating current. Since $I_{CaT}$ and $I_H$ are both active in the subthreshold regime, the assimilation failed to produce a model which captured the effects of these currents when the synthetic voltage waveform was presented as data. Another possible factor is that due to the low pass filter properties of the differential equation giving the time evolution of the voltage, the added Gaussian noise in the synthetic voltage waveform in section \ref{sec:twinexperimentsDependence} was so large that the signal to noise ratio was too small to inform the model in the case of a high frequency stimulating protocol. The step current protocol likely performed better with synthetic data due to the fact that the system generating the data is identical to the assimilated model, a simplification not available in the case of real physical systems. We now turn to the use of data assimilation in the analysis of real neurons.

When multiple excitatory ionic currents such as $I_{CaT}$ and $I_{Na}$ with similarly fast activation times are simultaneously present in a model, different combinations of $I_{CaT}$ and $I_{Na}$ can lead to voltage behavior which appears the same.
There are a few other ways to characterize how estimates which produce similar and accurate predictions might be structured.
\begin{enumerate}
\item Plot the max attained amplitude and time averaged magnitude of individual ionic currents for each set of parameters as compared to the true value. 
\item Compare the relative probability of each estimate producing accurate predictions by examining the distribution of the cost function.
\end{enumerate}

Analysis of data will be provided in the same format for each stimulating current protocol and each of the sampling rate conditions as in experiments with synthetic data in section \ref{sec:twinexperiments} below. In each section, the stimulating current and driven voltage waveform, the action level plot, box plots showing the distribution of the estimated average value and maximum magnitude of each of the theoretical ionic currents, and exemplar plots of the estimation and prediction beyond $t = t_M$ of the time evolution of the voltage waveform and theoretical ionic currents will be presented for each stimulating current protocol.


\subsection{Real Data: Step Current}

\begin{figure}[h!]
  \centering
  \begin{subfigure}{0.49\textwidth}
  \includegraphics[width = \textwidth]{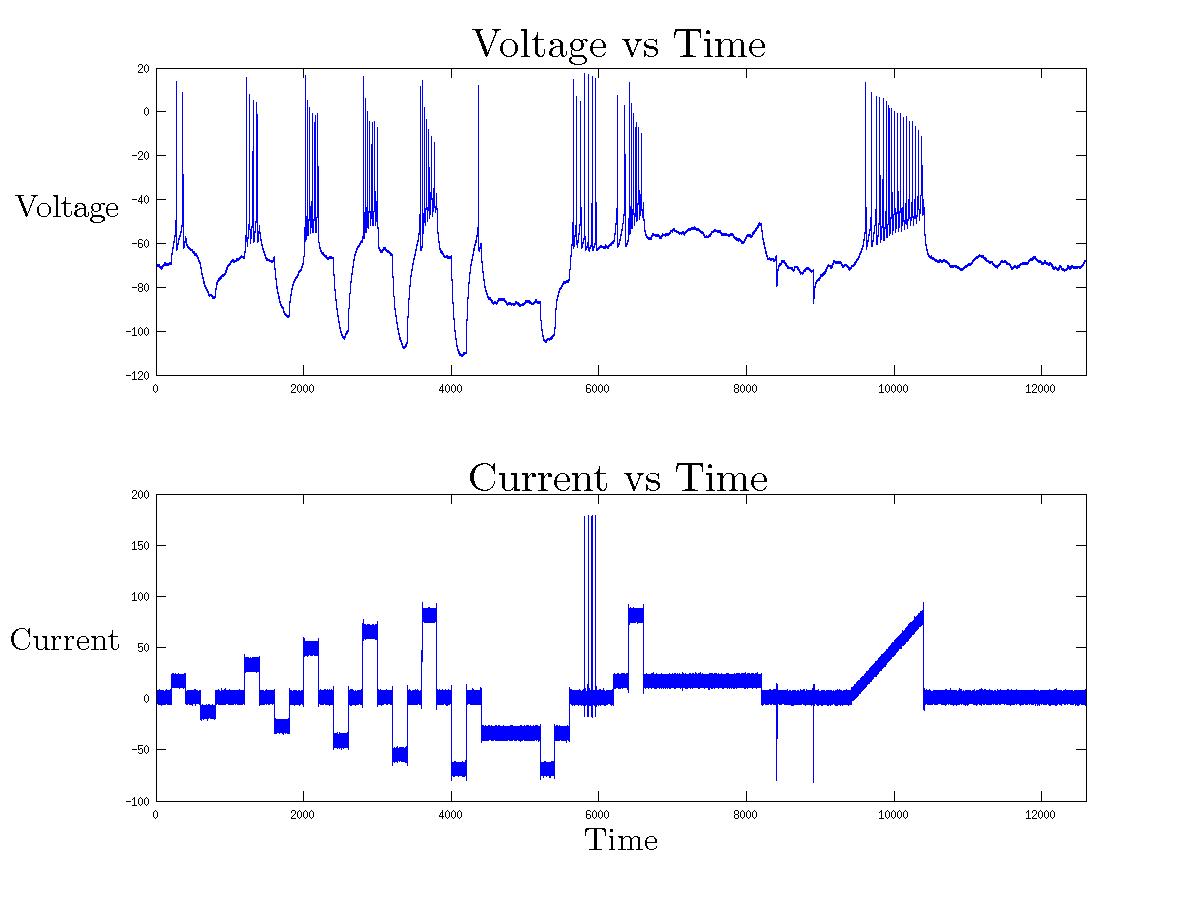}
  \captionsetup{width = 0.9\textwidth}
  \subcaption{Voltage and current traces for the step current protocol}
 \end{subfigure}
 ~
  \begin{subfigure}{0.49\textwidth}
  \includegraphics[width = \textwidth]{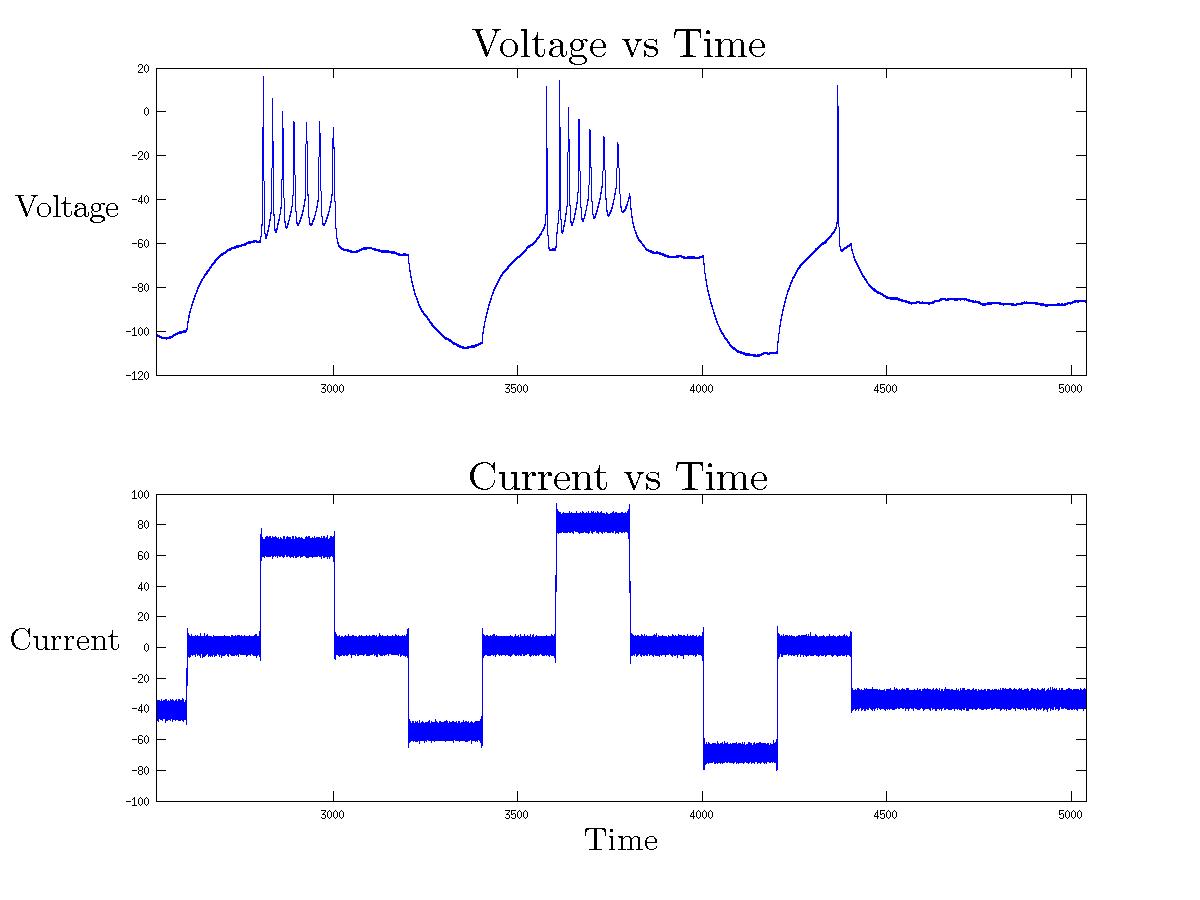}  
  \captionsetup{width = 0.9\textwidth}
  \subcaption{A closer display of the waveform information in the step current protocol}
  \end{subfigure}
  \captionsetup{width = 0.5 \textwidth}
  \caption{Step current shaped current waveform and elicited voltage. This is the first of three stimulating current protocols used.}
  \label{input0}
 \end{figure}

This section analyzes the data produced by the step current protocol. This current is presented in \autoref{input0}
 \subsubsection{50kHz Sampling Rate}
In this section, the maximum number of time steps used for the estimation window given constraints on computational resources was not long enough to obtain good predictions. Below is the action plot for this estimation procedure.
 
 \begin{figure}[h!]
 \centering
  \includegraphics[width = 0.5 \textwidth]{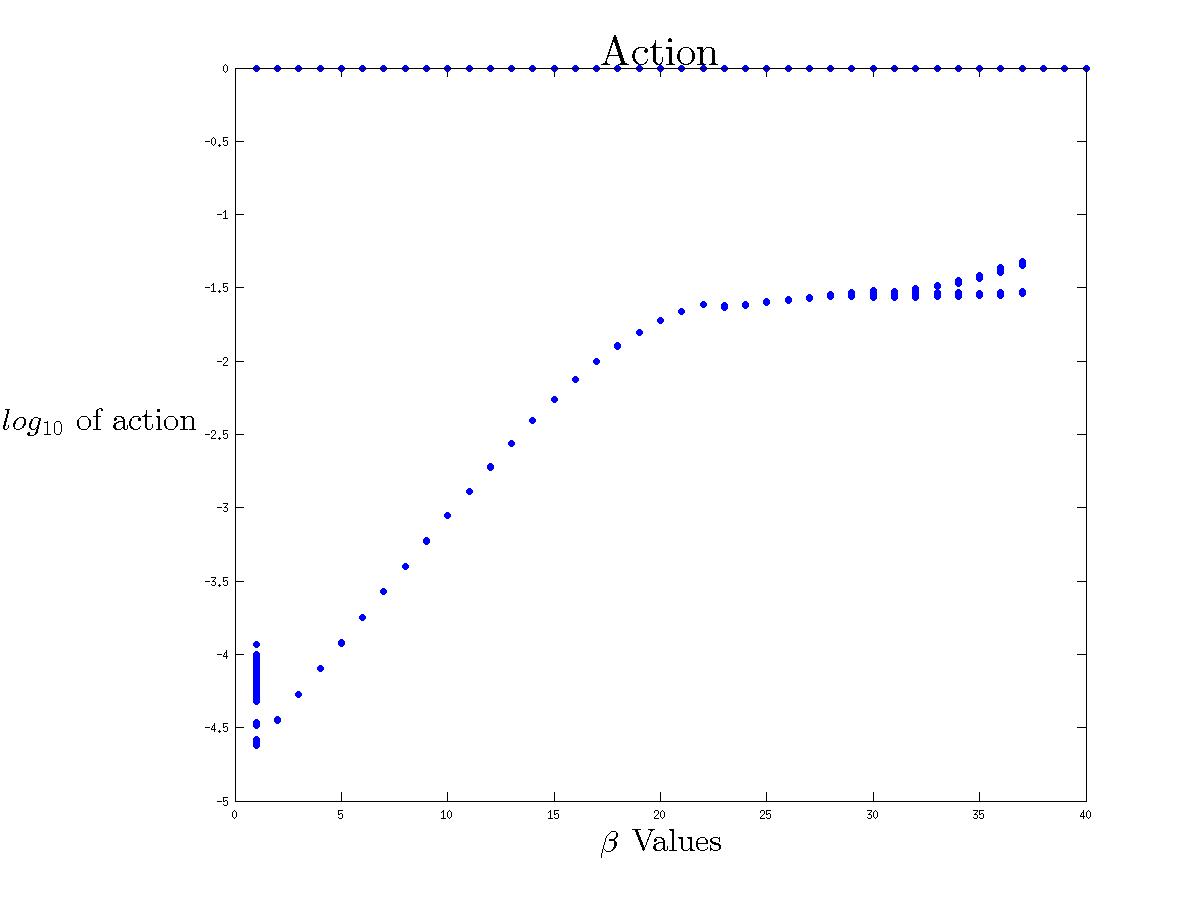}
  \captionsetup{width=.5\textwidth}
  \caption{Action level plot resulting from DA on the step current protocol of \autoref{input0} sampled at 50 kHz. Here $R_f$ is increased by a factor $\alpha = 1.5$ and splits into two distinct minima at high $\beta$. Neither of the paths yielding these values of the action resulted in accurate predictions, suggesting that densely sampled data resulting from step current protocols do not provide enough information to resolve unobserved processes in the model.}
 \label{Action50-0}
 \end{figure}

In \autoref{Action50-0}, the action rises as a function of beta and levels off to a constant value, suggesting a minimum in the action has been found. There appear two distinct minima, neither of which produced successful predictions. With this information we conclude that our 480 ms window at 50 kHz does not provide enough information to produce an accurate characterization of an $\text{HVC}_\text{I}$ neuron. This is likely due to the fact that the ionic currents are not well sampled as a function of voltage. A step current drives the fast ionic currents to equilibrium, while slow ionic currents vary in time and alter the baseline state of the system. Regardless of the sign, size, and duration of a step current, when a step current protocol is used, the ionic currents of a neuron tend to relax to dynamical equilibrium - spike trains, bursts, or quiescence - in only a few ways. This undesirable feature of step currents makes it difficult to tease apart the contributions from different ionic currents, resulting in the model symmetries seen in twin experiments in the sections preceding. Evidence for model symmetries, even when high quality predictions appear to validate the model, is a higher variance in the shape of theoretical I-V curves (section \ref{sec:twinexperimentsDependence}) which tend to deviate substantially from their true shape. This tends to make step current protocols unsuitable for use in data assimilation. 

\subsubsection{10kHz Sampling Rate}

Using the downsampled (10kHz) data resulting from the step current protocol as an input into the data assimilation achieved better results than the 50kHz data. This analysis was able to predict regions in which spiking occurred, but was still unable to properly predict the frequency or timing of those spikes. The results of data assimilation are displayed in \ref{10kHz-current}.
 
 \begin{figure}[h!]
     \centering 
            \includegraphics[width=0.5\textwidth]{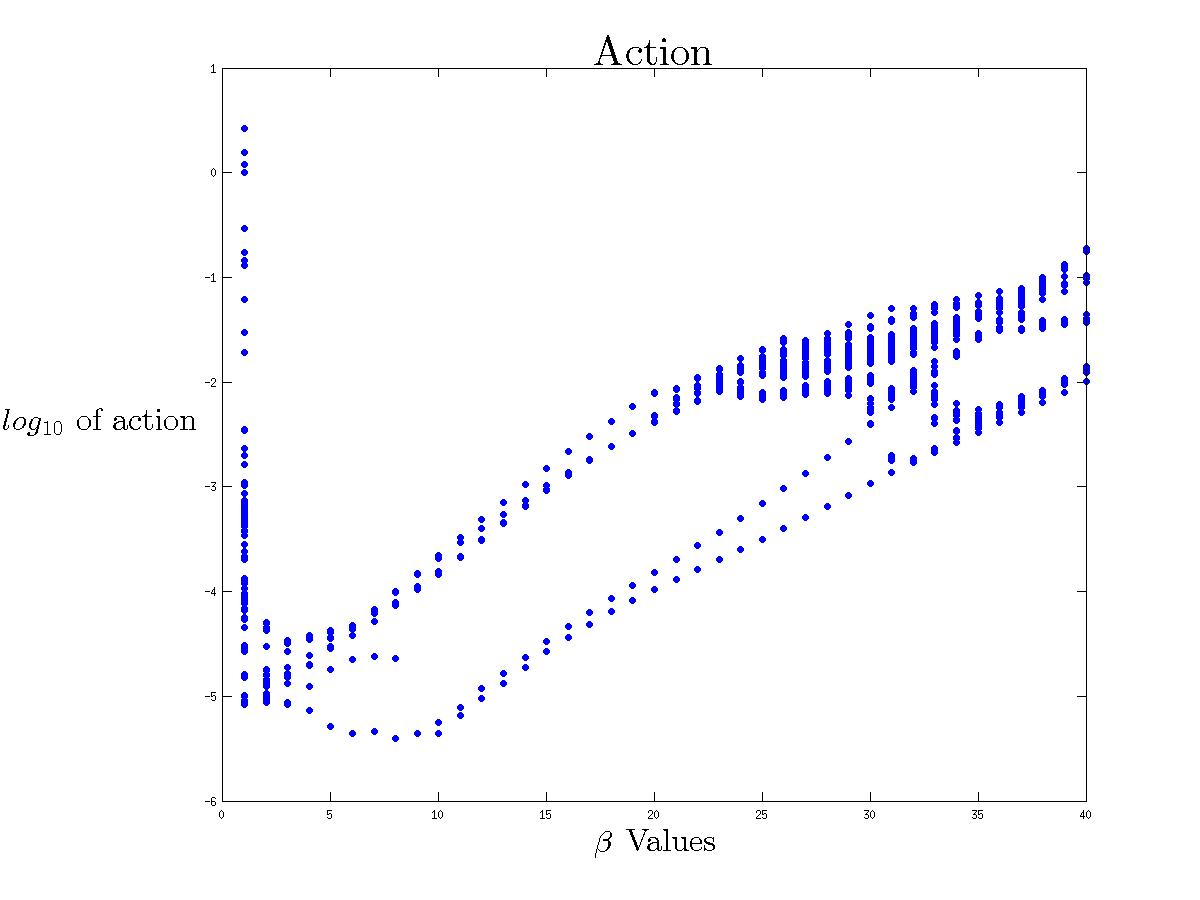} 
            \captionsetup{width = 0.5\textwidth}
            \caption{Action level plot resulting from data resulting from the step current protocol of \autoref{input0} is fed into the DA algorithm when sampled at 10kHz instead of 50 kHz. $R_f$ is increased by a factor of $\alpha = 1.5$, as in \autoref{Action50-0}. The number of levels for this plot is much higher than in the case of the 50kHz data, where limitations on computational power force a shorter overall time duration of the estimation window because of the higher sampling. A large number of action levels are present in the graph, reflecting the presence of a distribution of parameters producing similar time evolution in the voltage trace when integrated forward. Additionally, none of the paths corresponding to the plotted action values yielded predictions matching both the waveform information and spike times of \autoref{input0}.}
            \label{10kHz-action}
     \end{figure}
      \begin{figure}[h!]
     \centering 
      \begin{subfigure}{0.49\textwidth}
      \centering 
            \includegraphics[width=\textwidth]{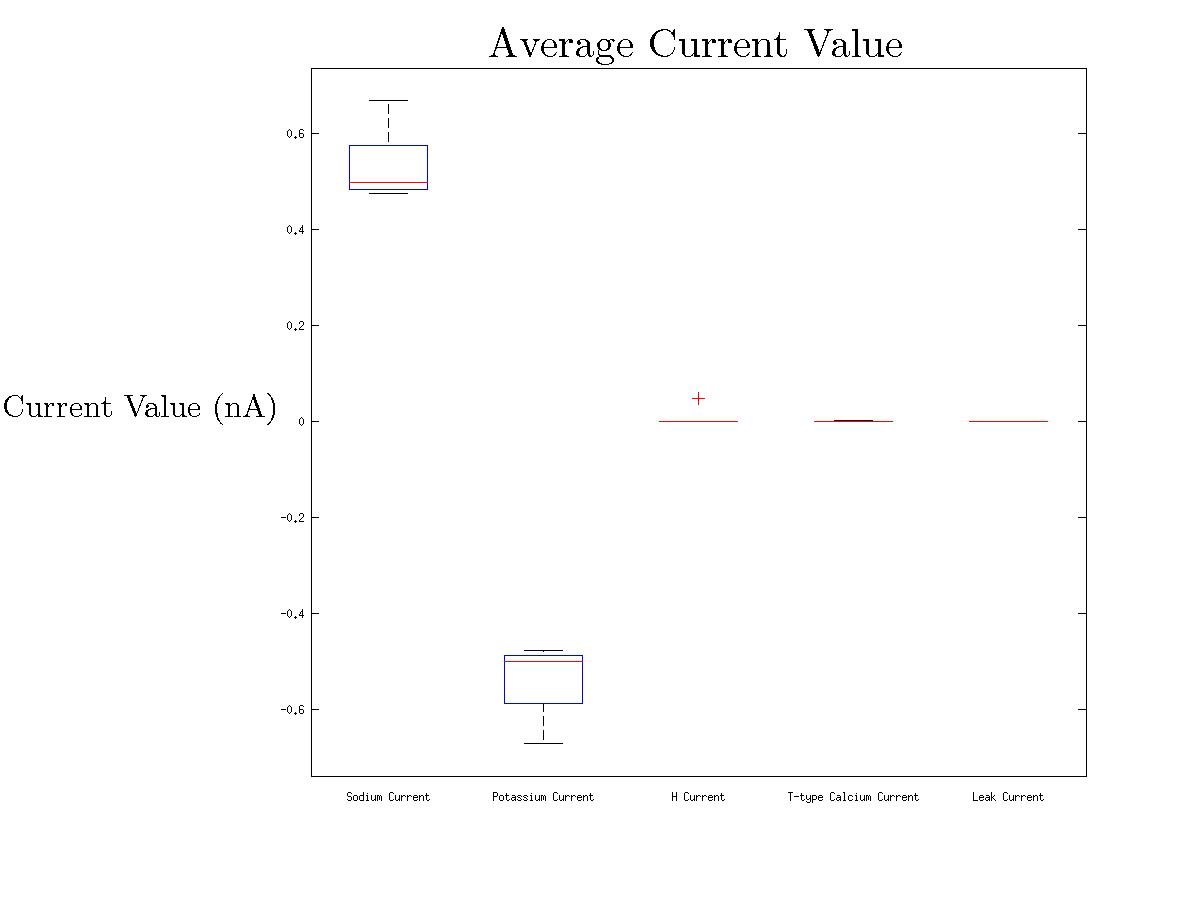}
            \captionsetup{width = 0.9\textwidth}
            \subcaption{A box plot comparing the time averaged value of each ionic currents in the model estimated by DA.}
            \label{10-0mean}
      \end{subfigure}
      ~
      \begin{subfigure}{0.49\textwidth}
      \centering 
            \includegraphics[width=\textwidth]{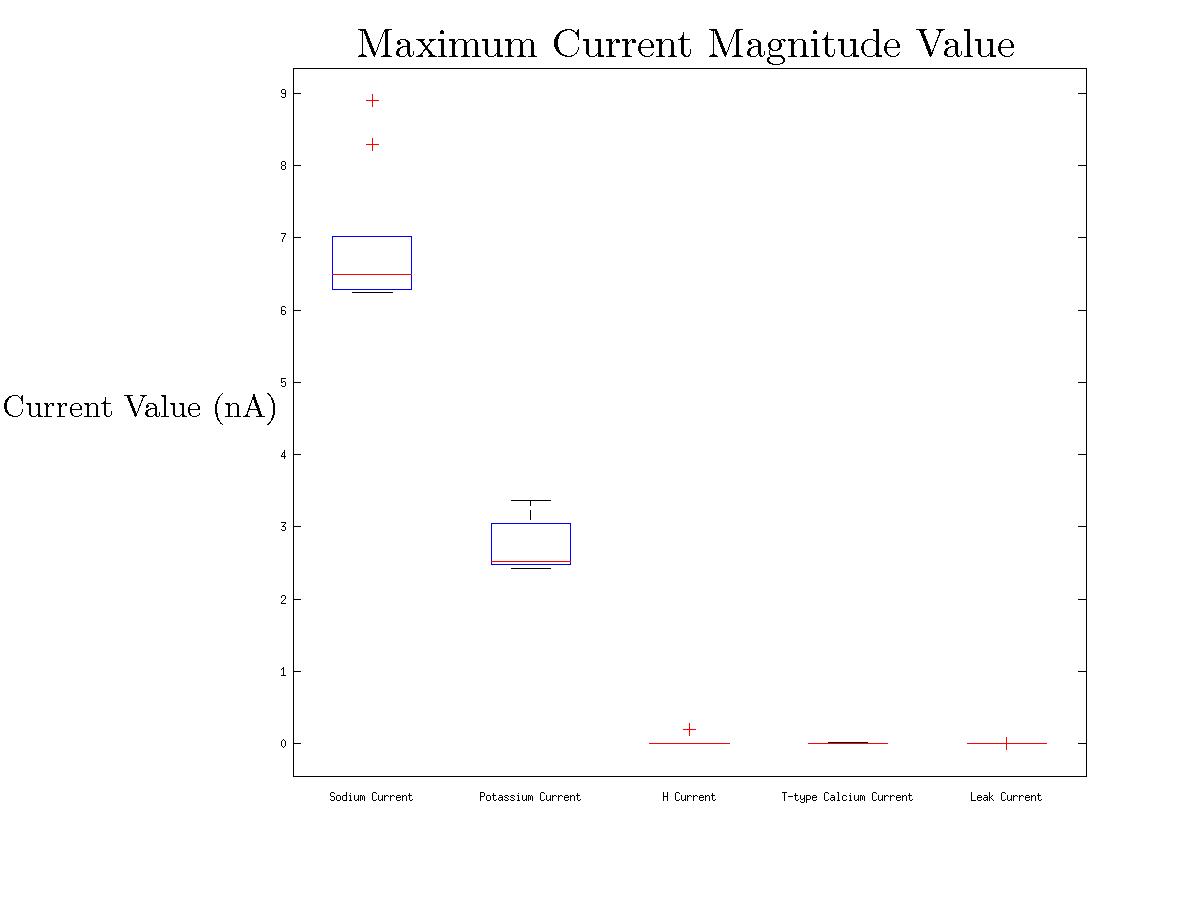}
            \captionsetup{width = 0.9\textwidth}
            \subcaption{A box plot comparing the magnitude of the maximum value over time of each ionic current in the model estimated by DA.}
     \end{subfigure}
     \caption{Box plots showing a wide distribution of features resulting from DA on data and step current of \autoref{input0} sampled at 10 kHz. The red data points in these plots are features of sets of parameters in the model producing 'good' predictions. For the case of the step current protocol, these only match the periods of spiking and quiescence but neither the precise spike times nor the waveform information.}
     \end{figure}
     
     \begin{figure}[h!]
      \begin{subfigure}{0.5\textwidth}
       \centering
       \includegraphics[width = \textwidth]{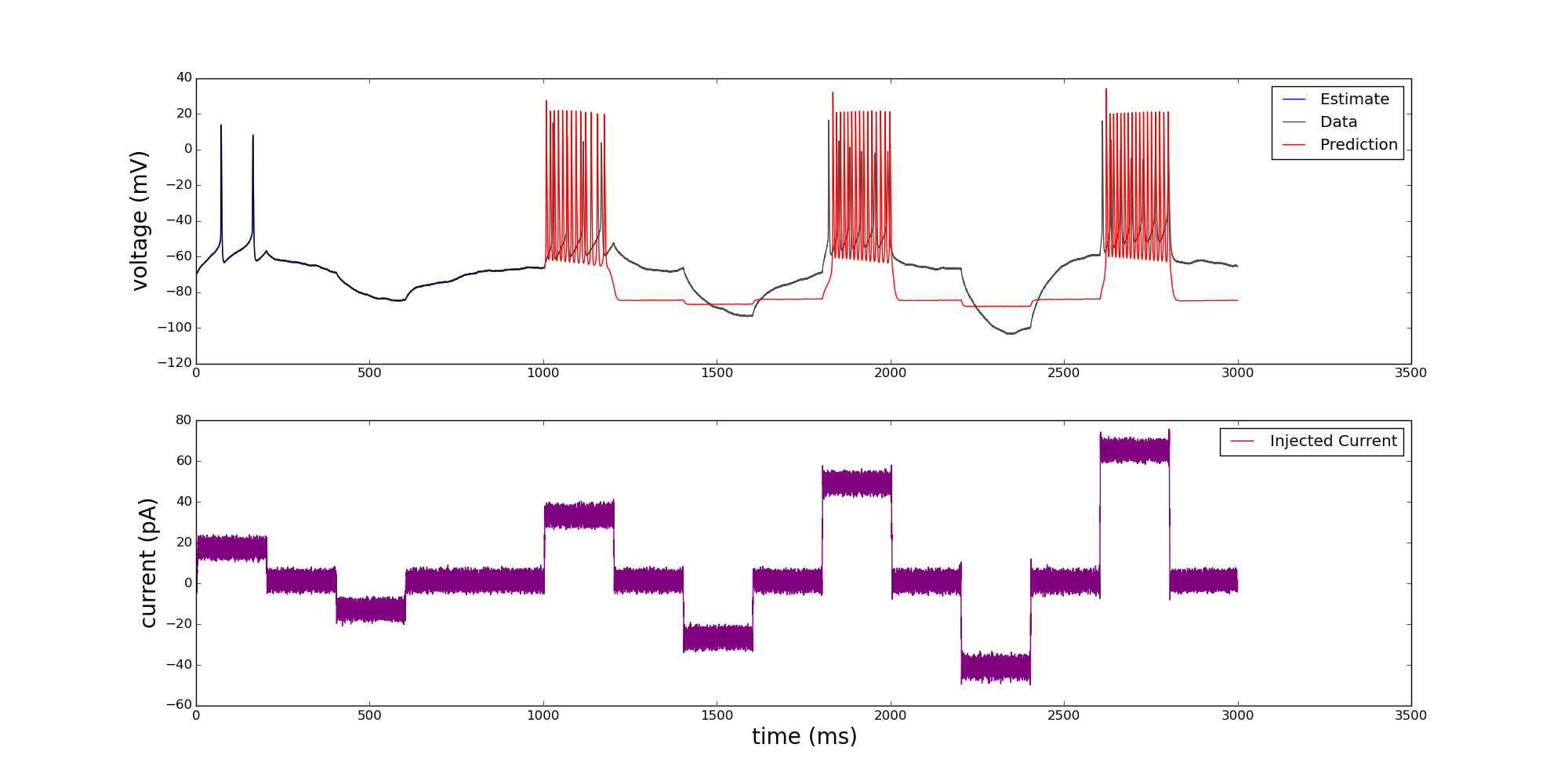}
       \captionsetup{width = 0.9\textwidth}
       \subcaption{A representative prediction (red) plotted against the data (black) and stimulating current protocol (purple) using the data of \autoref{input0} producing the action level plot of \autoref{10kHz-action}.}
      \end{subfigure}
      ~
      \begin{subfigure}{0.4 \textwidth}
       \centering
       \includegraphics[width = \textwidth]{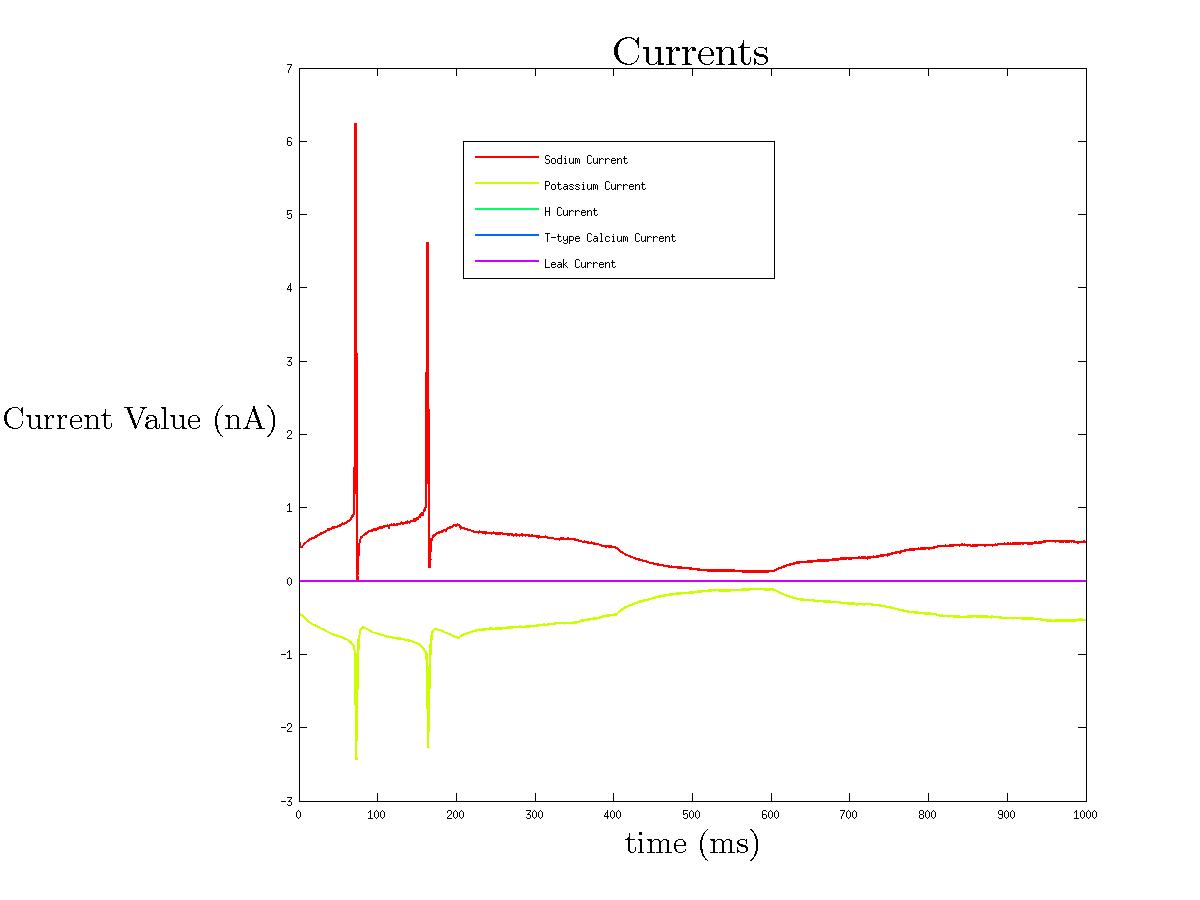}
       \captionsetup{width = 0.9\textwidth}
       \subcaption{Estimated ionic currents in the estimation window.}
       \label{10kHz-current}
      \end{subfigure}
      \captionsetup{width = 0.5\textwidth}
      \caption{The above are a representative example of the predictions and features of the model obtained using the step current of \autoref{input0} sampled at 10kHz (estimation window of 1000.1 ms) as input into data assimilation.}
      \label{predict0down}
     \end{figure}
     
The action levels in \autoref{10kHz-action} appear to cross each other and present a large number of closely spaced values of the action for a number of different paths. This is an undesirable feature because if a large number of closely spaced action levels producing 'good' predictions is found, the expected value of parameters cannot be approximated by the parameters producing the lowest value of the action. In the case of a large number of closely spaced values of the action, a weighted average must be computed following equation \ref{eq:prob_dis} in section \ref{sec:path_integral_methods}. This weighted average cannot be computed if all significant contributions to the integral of \ref{eq:prob_dis} are not accounted for. This pattern of closely spaced action level plots is present in all of the stimulating current protocol conditions.    

In \autoref{predict0down}, an example of predictions and theoretical ionic currents at times $t > t_M$ after the estimation window are plotted. As can be seen in \autoref{predict0down}, timing of the spike bursts is well predicted, as is the increase in burst frequency. The actual spike frequency, spike timing, and subthreshold behavior is poorly predicted. The most likely explanation for the poor prediction quality can be deduced from \autoref{10kHz-current}. In that plot $I_H$ and $I_{CaT}$ are not activated during the estimation window and thus estimated as negligible in magnitude. This will lead to a poor estimation of any parameters related to these currents and poor performance of the model in reproducing any behavior that relies on them. It is likely that in the prediction window these currents are subsequently activated in the real system, partially explaining the poor predictions of the model.
     
\subsection{Real Data: High Frequency Complex Current}

\begin{figure}[h!]
  \centering
  \begin{subfigure}{0.49\textwidth}
  \includegraphics[width = \textwidth]{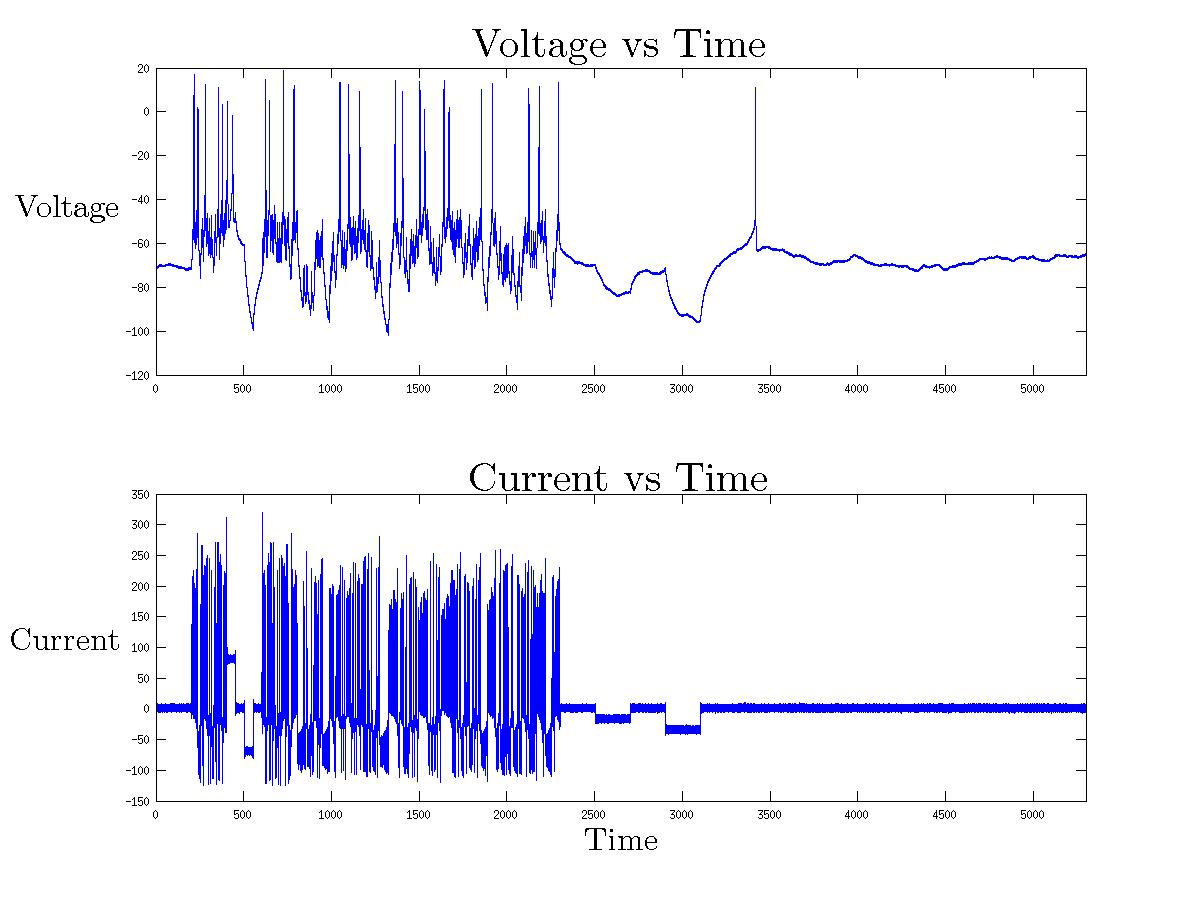}
  \captionsetup{width = 0.9\textwidth}
  \subcaption{Voltage and stimulating current protocol for the high frequency, complex waveform condition}
 \end{subfigure}
 ~
  \begin{subfigure}{0.49\textwidth}
 \includegraphics[width = \textwidth]{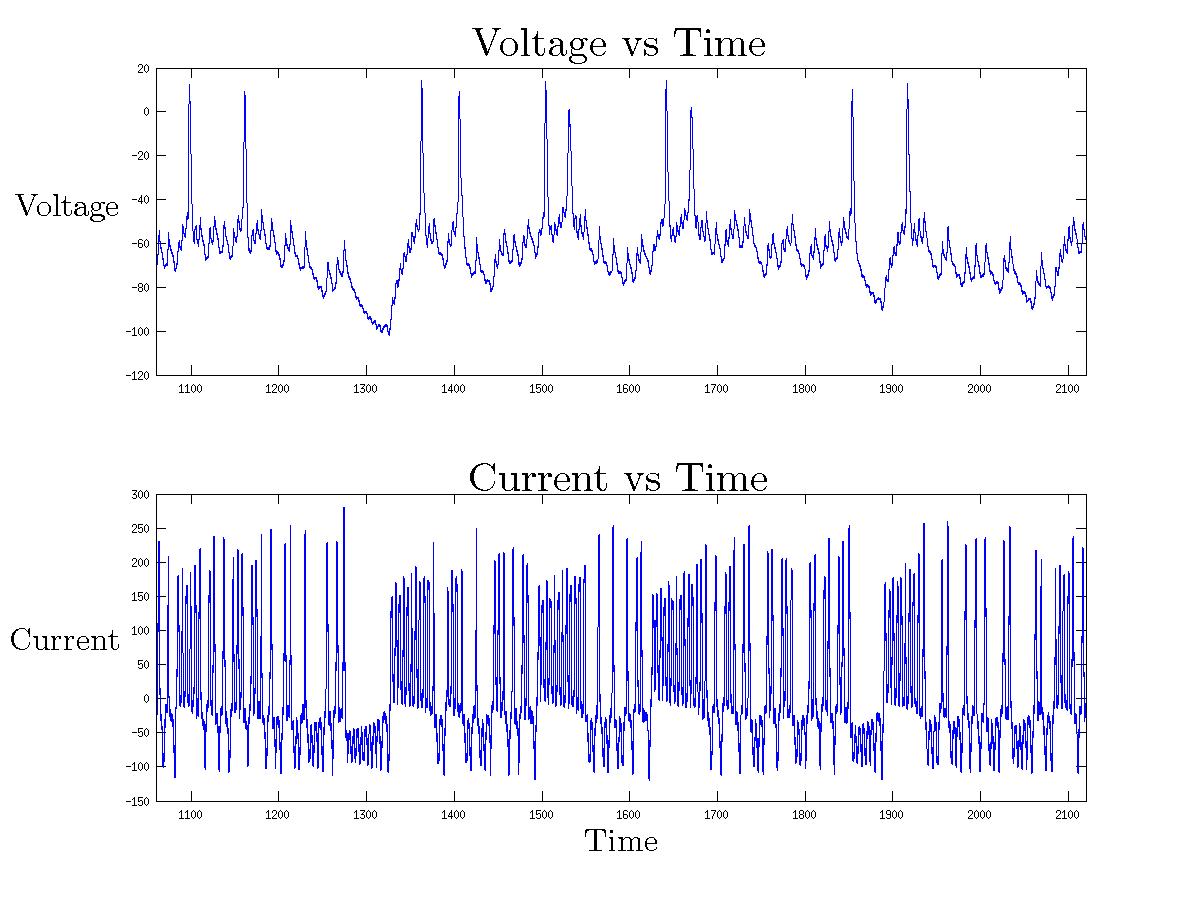}  
  \captionsetup{width = 0.9\textwidth}
  \subcaption{Voltage and current trace of a small section of the epoch to show more of the waveform detail}
  \end{subfigure}
  \captionsetup{width = 0.5 \textwidth}
  \caption{Different views of the high frequency, complex shaped current waveform and voltage elicited. This is the second of three stimulating current protocols used.}
  \label{input1}
 \end{figure}

The stimulating current with a high frequency and complex shape is identical to the waveform used in this condition in twin experiments in section \ref{sec:twinexperimentsDependence}. The response voltage waveform elicited and from a real HVC$_{\text{I}}$ neuron is displayed in \autoref{input1}. Notably, the high frequency oscillations in the current trace oscillate about as rapidly as elicited voltage spikes.
 
 \subsubsection{50kHz Sampling Rate}
 
\begin{figure}[h!]
    \centering
  \includegraphics[width = 0.5\textwidth]{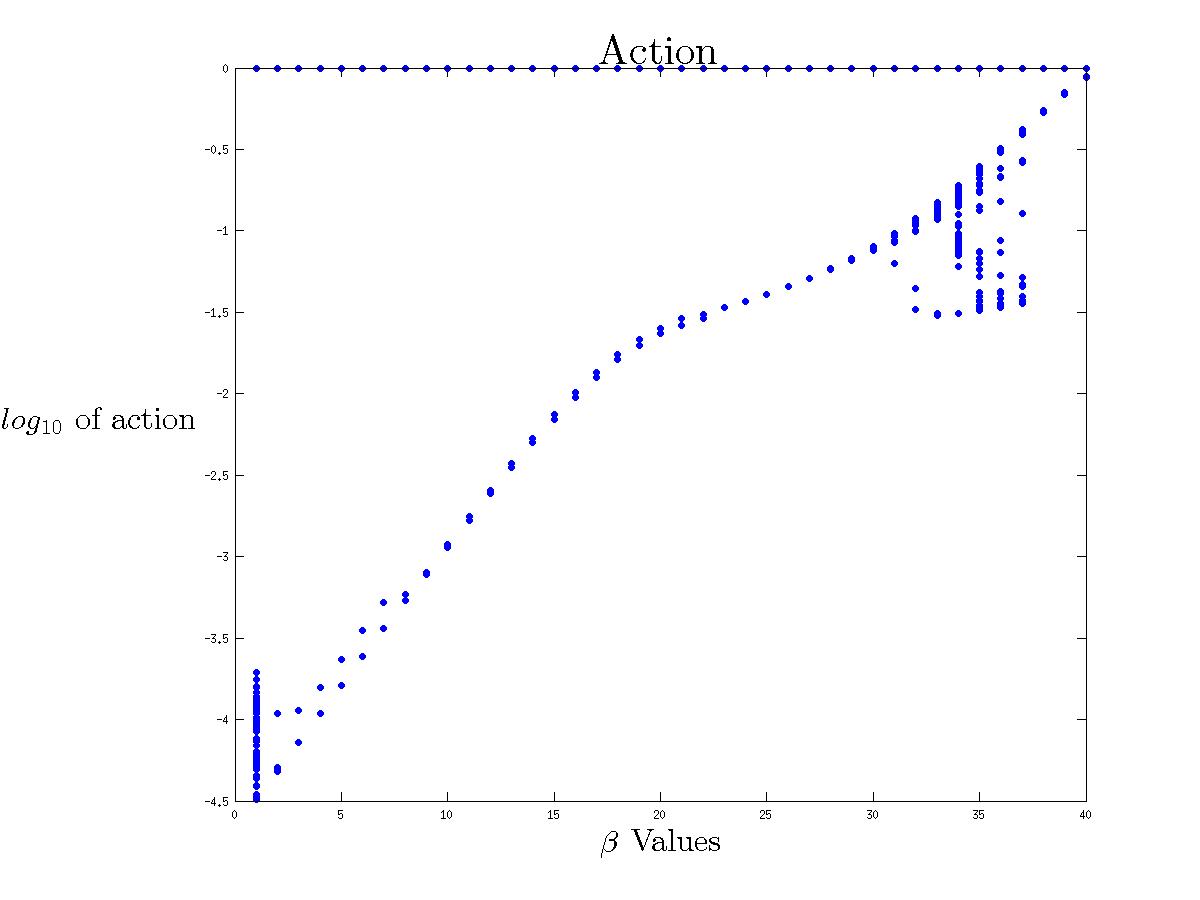}
  \captionsetup{width=.5\textwidth}
  \caption{Action level plot using voltage data and the high frequency complex current protocol of \autoref{input1} fed into the DA algorithm at 50 kHz. $R_f$ is increased by a factor of $\alpha = 1.5$ at each annealing step. The number of levels for this plot is much higher than in the case of the step current protocol of \autoref{Action50-0}. As with other action level plots, a large number of action levels are present in the graph, reflecting the presence of a distribution of parameters producing similar time evolution in the voltage trace when integrated forward. Some of the paths produced predictions matching much of the waveform information \autoref{input1}, though information such as the spike frequency was not accurately reproduced.}
 \label{Action50-1}
\end{figure}

At 50kHz (480.2 ms), only 2 of the initial 100 paths were able to successfully predict, at multiple anneal steps, the time evolution of the voltage waveform. The action plot as a function of $\beta$ for all of the initial paths is plotted in \autoref{Action50-1}. The levels of the action are not as distinct as in the step-current 50kHz action plot, and the splitting of the action into two distinct levels is not observed here. This could be a reflection of the fact that the complex current protocol causes the action surface to become more complicated. Despite the action surface potentially being more complicated, we find that the resulting sets of parameters produce better predictions than with the step current protocol.
 
Of the quality measures used in previous sections, we find only plots of the predictions and ionic currents informative to display here, as the other measures are statistical and only 2 of the initial 100 paths gave acceptable predictions.

\begin{figure}[h!]
\centering
      \begin{subfigure}{0.5\textwidth}
       \centering
       \includegraphics[width = \textwidth]{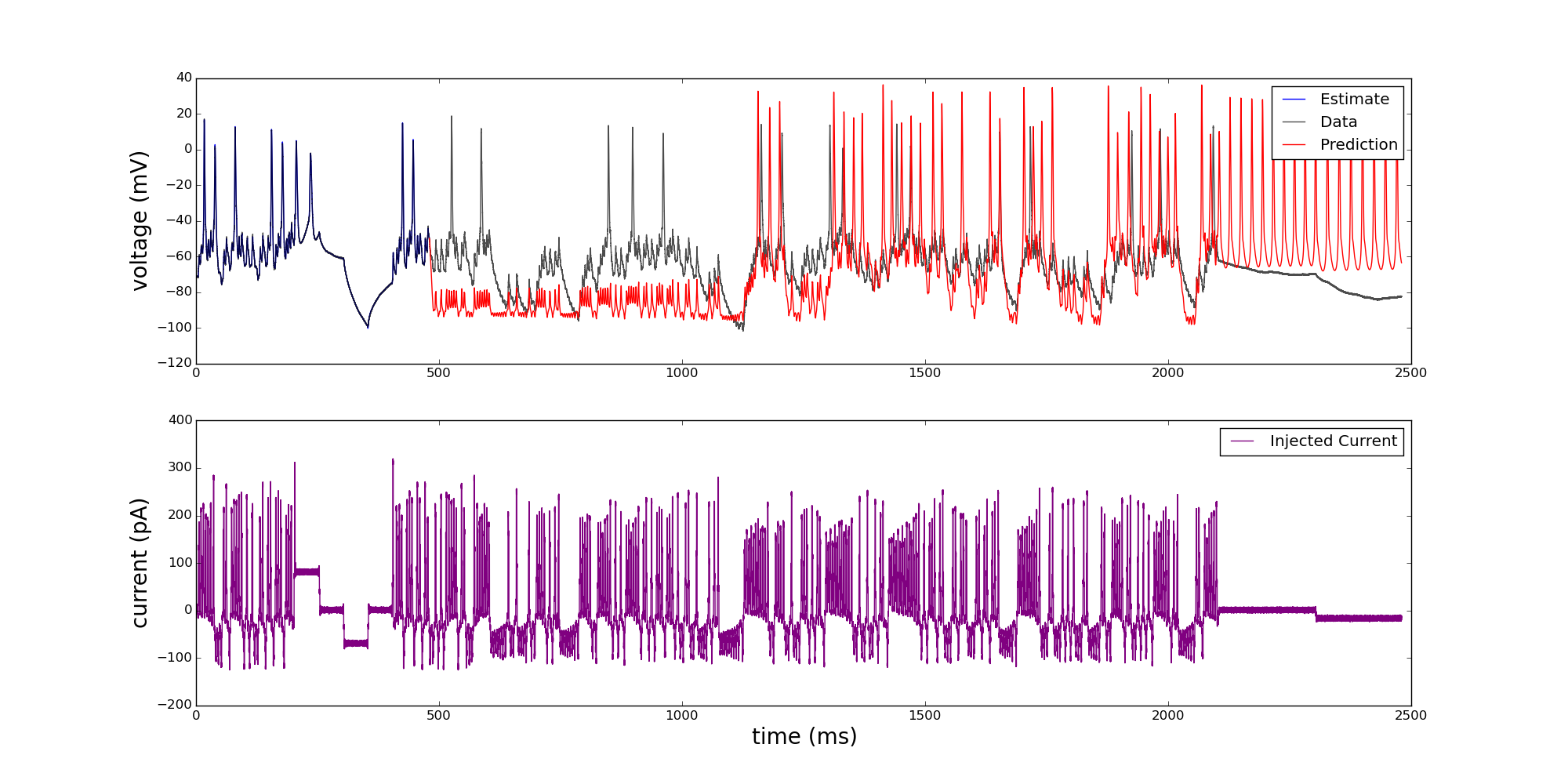}
       \captionsetup{width = 0.9\textwidth}
       \subcaption{This is one of the predictions that was classified as `acceptable' from the action plot in \autoref{Action50-1}.}
       \label{predict50-1}
      \end{subfigure}
      ~
      \begin{subfigure}{0.4 \textwidth}
       \centering
       \includegraphics[width = \textwidth]{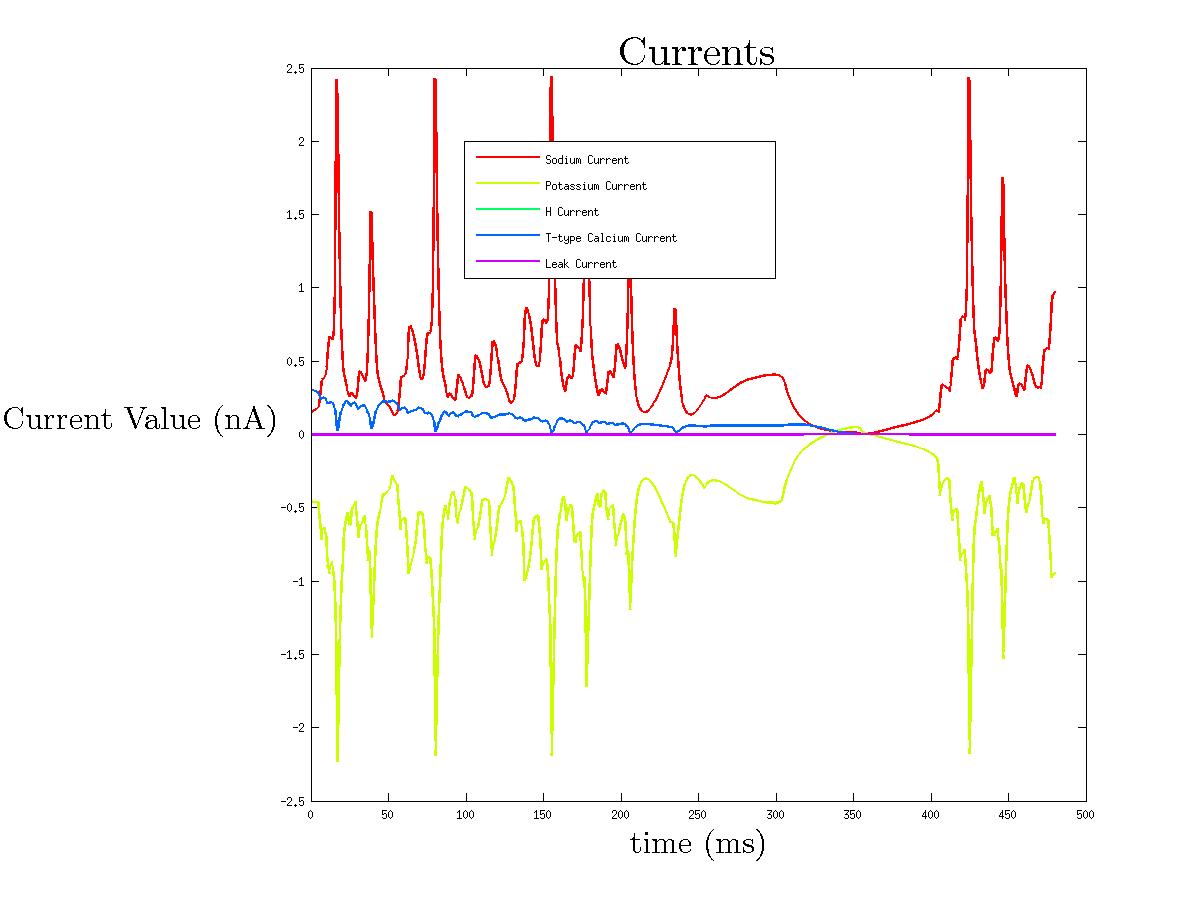}
       \captionsetup{width = 0.9\textwidth}
       \subcaption{Estimated ionic currents in the estimation window.}
       \label{50kHz1-current}
      \end{subfigure}
      \captionsetup{width = 0.5\textwidth}
      \caption{The above are an example of predictions and features of the model obtained with data assimilation using the high frequency complex current protocol of \autoref{input1} sampled at 50 kHz (480.2 ms estimation window)}
     \end{figure}

It is clear in \autoref{predict50-1} that the prediction rapidly improves at approximately $t=1100$, when the general structure of the current changes. Previous to this time, many spikes in the predicted voltage trace are missed. Subsequently, the subthreshold behavior appears to be adequate, but the simulated neuron produces many spikes not present in the data. One possibility for the improvement in the subthreshold behavior is the region of time after $t>t_M$ (times after the estimation window) and before the section of hyperpolarizing current that directly precedes the improvement in prediction. During this time, all of the states might be reset to the correct value, allowing the appropriate behavior to follow. If this is the case, then some of  the properties of the estimated model could be accurate, while the state of the system, particularly the unmeasured states, are incidentally badly estimated at the end of the estimation window. Though this was the best prediction from the set of data and analysis, the estimated model is clearly inadequate as a description of an $\text{HVC}_\text{I}$.
 
\subsubsection{10kHz Sampling Rate}

We present results of a similar analysis on the corresponding downsampled (10 kHz) set of data. Out of the 100 initial paths, 66 were deemed acceptable given the quality of the model predictions.

\begin{figure}[h!]
\centering 
            \includegraphics[width=0.5\textwidth]{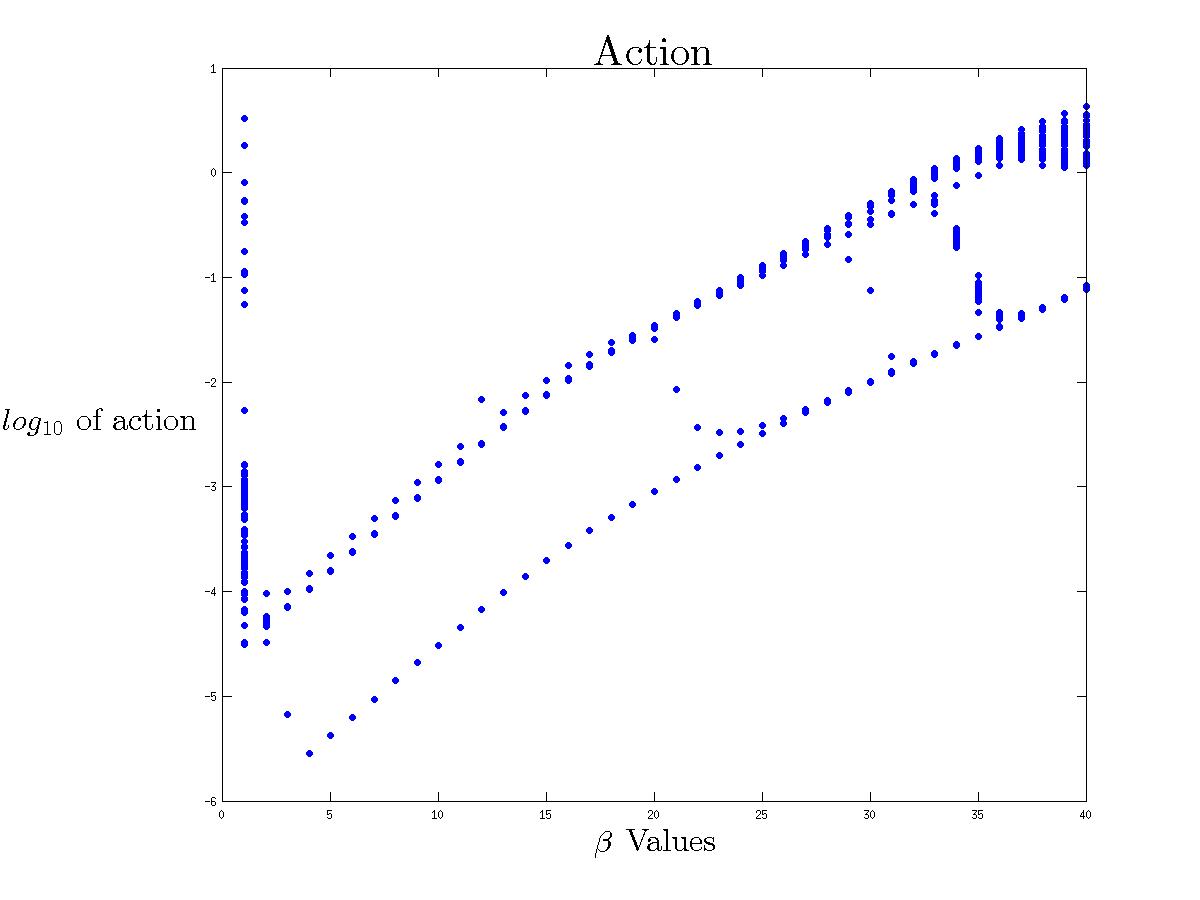} 
            \captionsetup{width = 0.5\textwidth}
            \caption{Action level plot using voltage data and the high frequency complex current protocol of \autoref{input1} fed into the DA algorithm at 10 kHz. $R_f$ is increased by a factor of $\alpha = 1.5$ at each annealing step. The number of levels for this plot is much higher than in the case of the step current protocol of \autoref{Action50-0}. As with other action level plots, a large number of action levels are present in the graph, reflecting the presence of a distribution of parameters producing similar time evolution in the voltage trace when integrated forward. Some of the paths produced predictions matching much of the waveform information \autoref{input1}, though information such as the spike frequency was not accurately reproduced.}
            \label{10kHz1-action}
\end{figure}

The action plot for this analysis is presented in \autoref{10kHz1-action}. There is significantly more separation between upper and lower action levels than in \autoref{Action50-1}. The analysis with this current and sampling rate condition also appears to demonstrate paths 'jumping' from one minimum in the action to another over successive annealing steps, which is manifested in the abrupt drop in the value of the action at around $\beta = 30$. The quality of prediction results for this set of paths is reflected in the larger separation between the lowest and higher levels in the action level plot. In other words, the estimated parameter sets and unmeasured states producing the best predictions also produced the lowest action levels.

\begin{figure}[h!]
\centering
\begin{subfigure}{0.49\textwidth}
      \centering 
            \includegraphics[width=\textwidth]{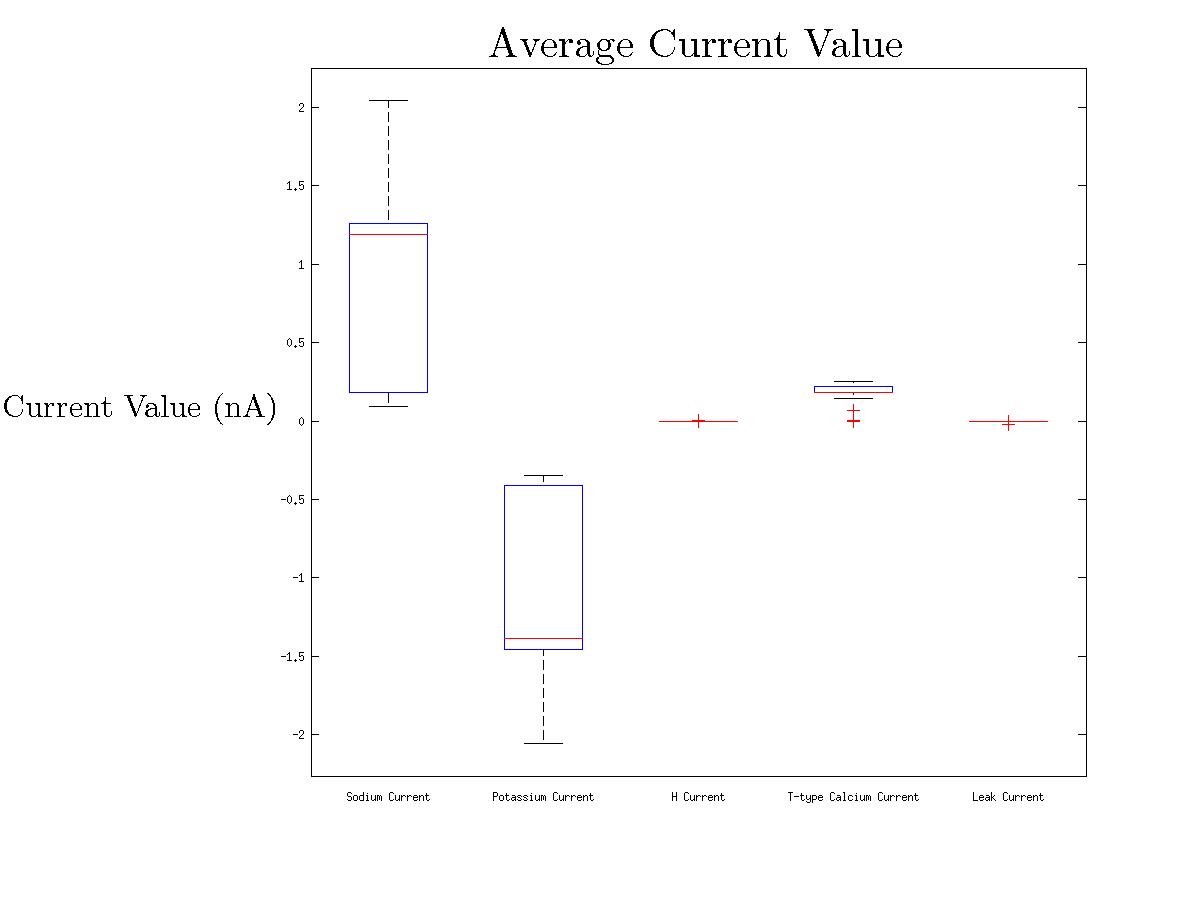}
            \captionsetup{width = 0.9\textwidth}
            \subcaption{Time averaged magnitude of each of the ionic currents.}
      \end{subfigure}
      ~
      \begin{subfigure}{0.49\textwidth}
      \centering 
            \includegraphics[width=\textwidth]{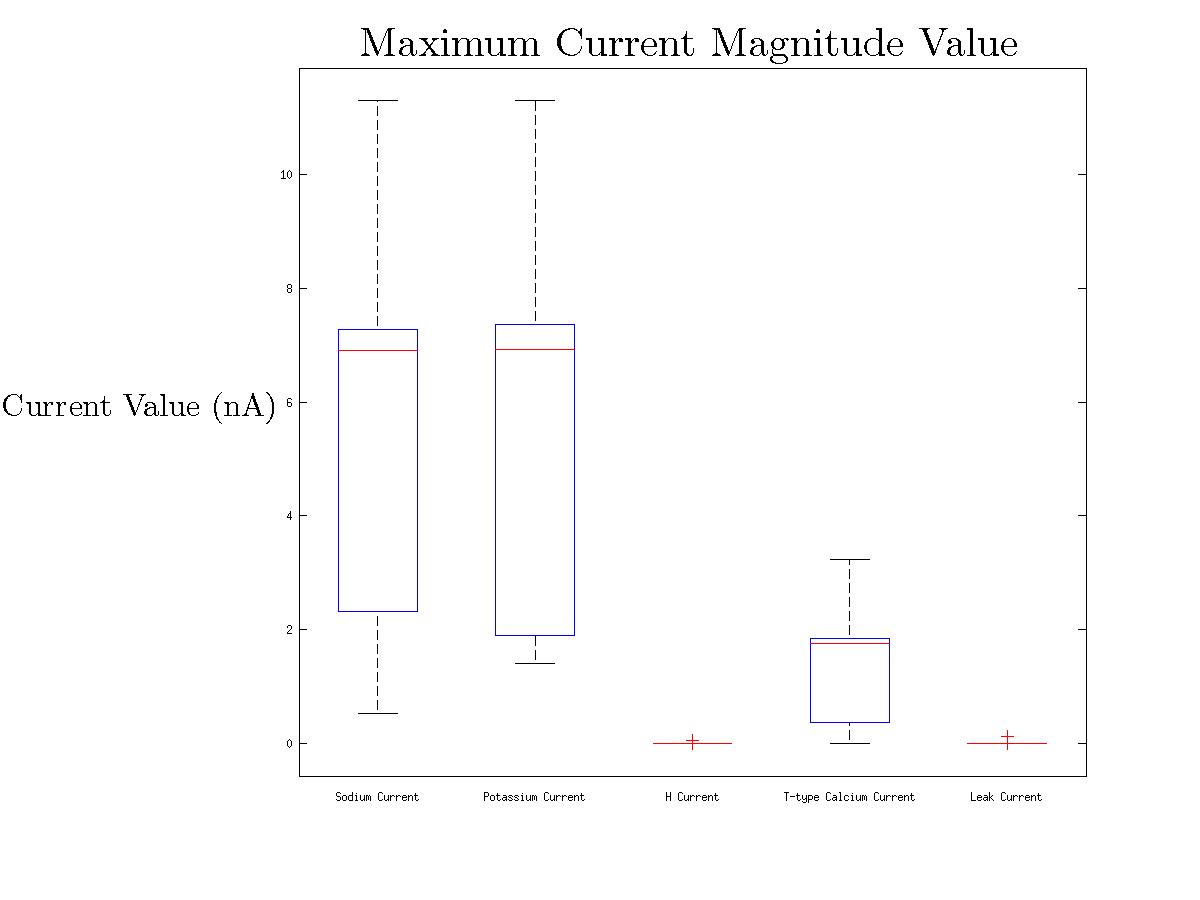}
            \captionsetup{width = 0.9\textwidth}
            \subcaption{Magnitude of the maximum value of each ionic current in time.}
     \end{subfigure}
     \caption{Box plots showing a wide distribution of features resulting from DA on voltage data produced by the high frequency complex stimulating current of \autoref{input1} sampled at 10 kHz. The red data points in these plots are features of sets of parameters in the model producing 'good' predictions. For this stimulating current, good predictions tend to match the waveform information and spike times well.}
     \label{10-1-box}
     \end{figure}

The box plots in \autoref{10-1-box} show more spread in $I_{Na}$ and $I_K$ than have been seen until now. This could be attributed to the two different types of predictions seen. 

\begin{figure}[h!]
\centering
      \begin{subfigure}{0.5\textwidth}
       \centering
       \includegraphics[width = \textwidth]{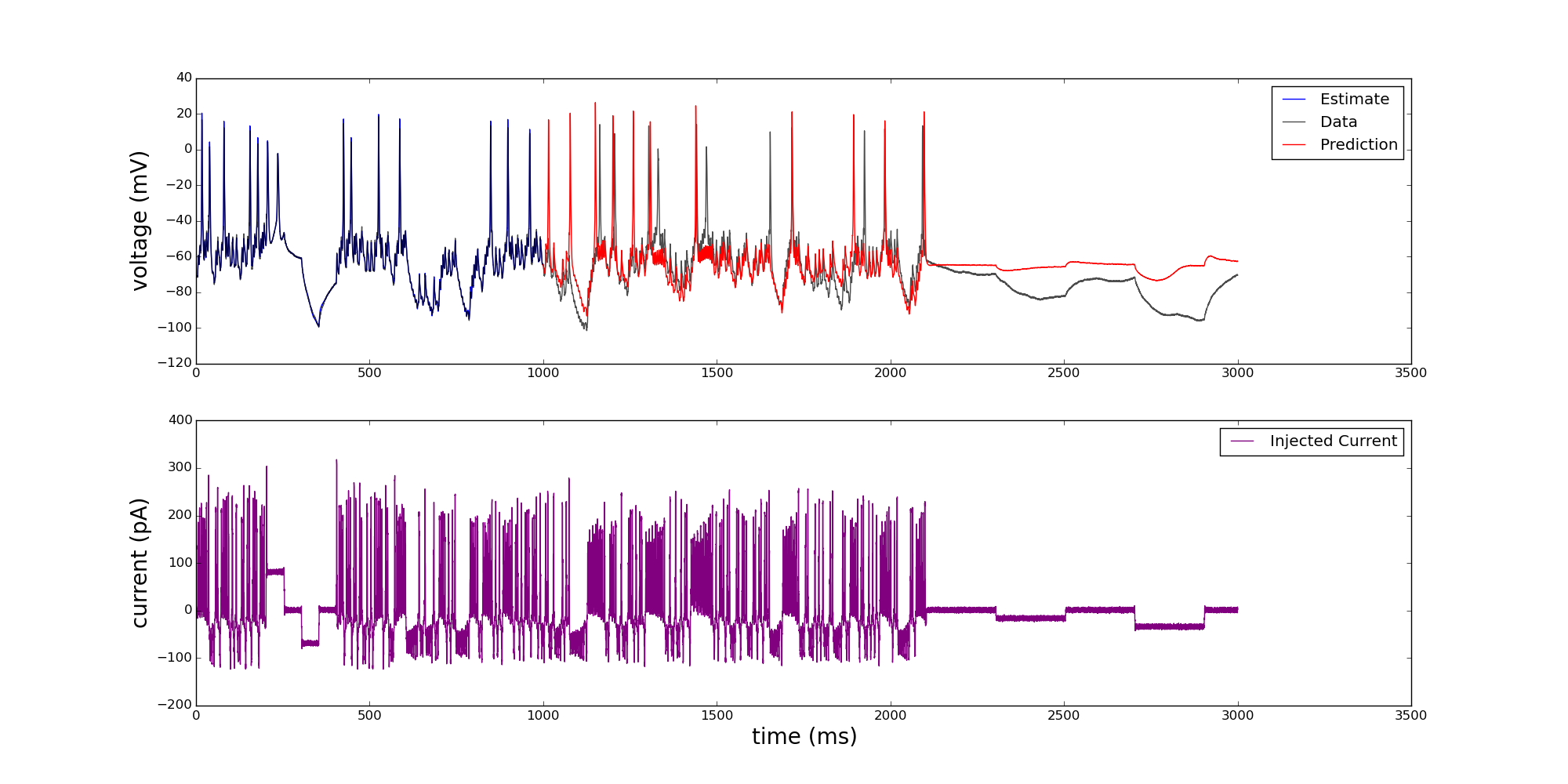}
       \captionsetup{width = 0.9\textwidth}
       \subcaption{First class of prediction. Very good subthreshold behavior, with approximately accurate spike timing behavior.}
      \end{subfigure}
      ~
      \begin{subfigure}{0.4 \textwidth}
       \centering
       \includegraphics[width = \textwidth]{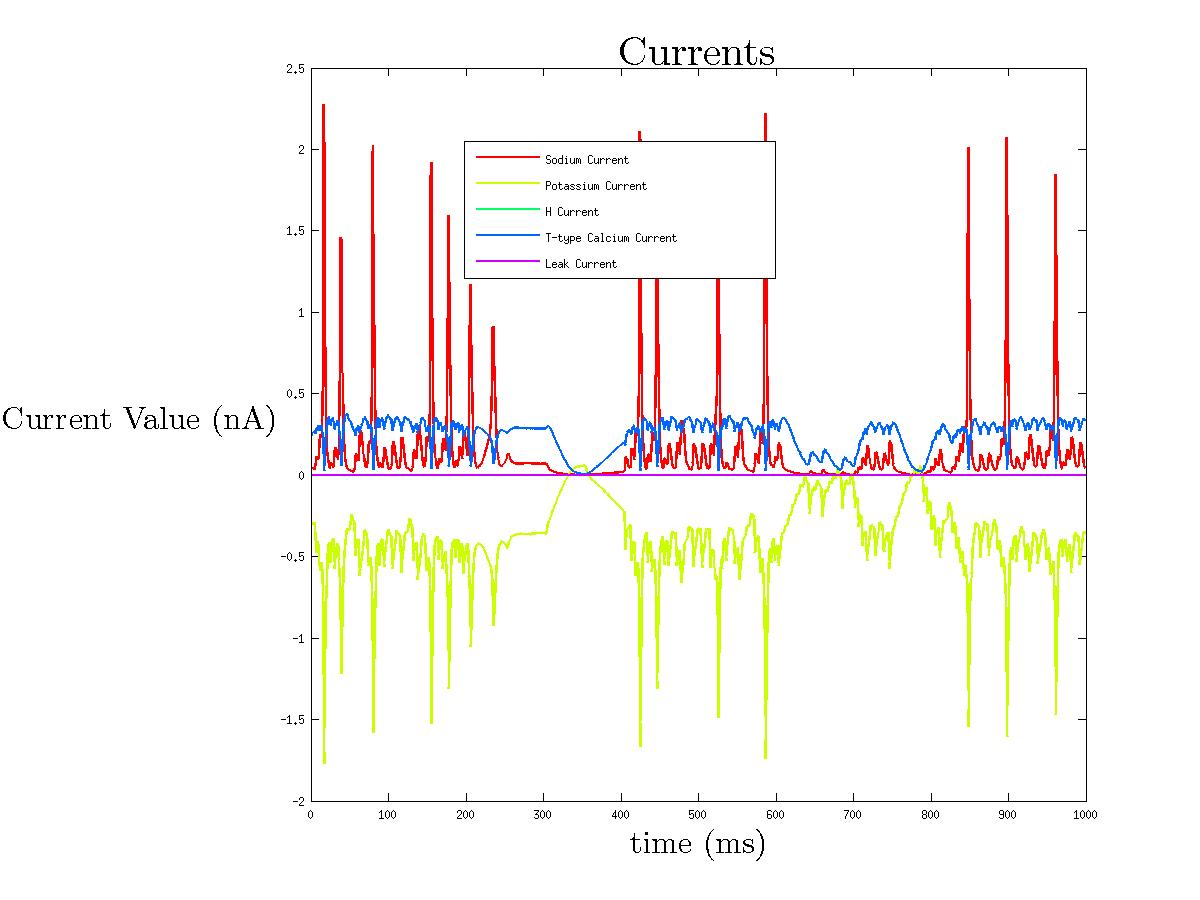}
       \captionsetup{width = 0.9\textwidth}
       \subcaption{Estimated ionic currents in the estimation window. The significant increase in $I_{CaT}$ estimation compared to step current estimates suggests that more accurate predictions require the involvement of this current. This may be due to the fact that $I_{CaT}$ is biologically typically active at subthreshold voltages.}
       \label{10-1-current1}
      \end{subfigure}
      \captionsetup{width = 0.5\textwidth}
      \caption{Predictions and estimated ionic currents of the 10kHz sampling rate condition of the data in \autoref{input1}. The sets of parameters that produced this type of prediction occurred at $\beta \approx 24$ and $\beta \approx 40$, around the abrupt drops in the action levels indicating where the DA procedure jumps from minimum to minimum over an annealing step.}
       \label{10kHz1-output1}
     \end{figure}

\begin{figure}[h!]
\centering
      \begin{subfigure}{0.5\textwidth}
       \centering
       \includegraphics[width = \textwidth]{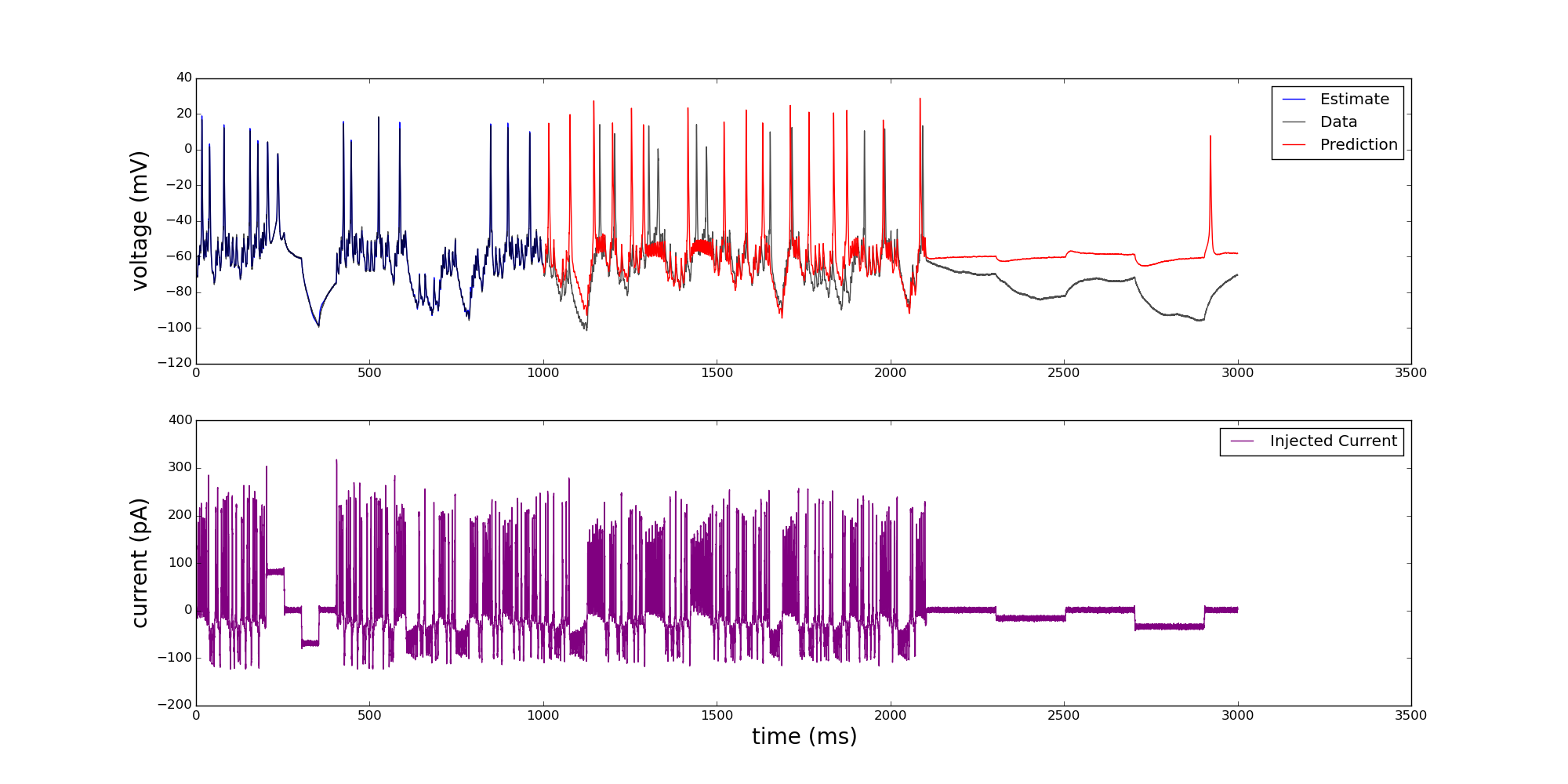}
       \captionsetup{width = 0.9\textwidth}
       \subcaption{Second class of good predictions. Very good subthreshold behavior with overly active spiking behavior. }
      \end{subfigure}
      ~
      \begin{subfigure}{0.4 \textwidth}
       \centering
       \includegraphics[width = \textwidth]{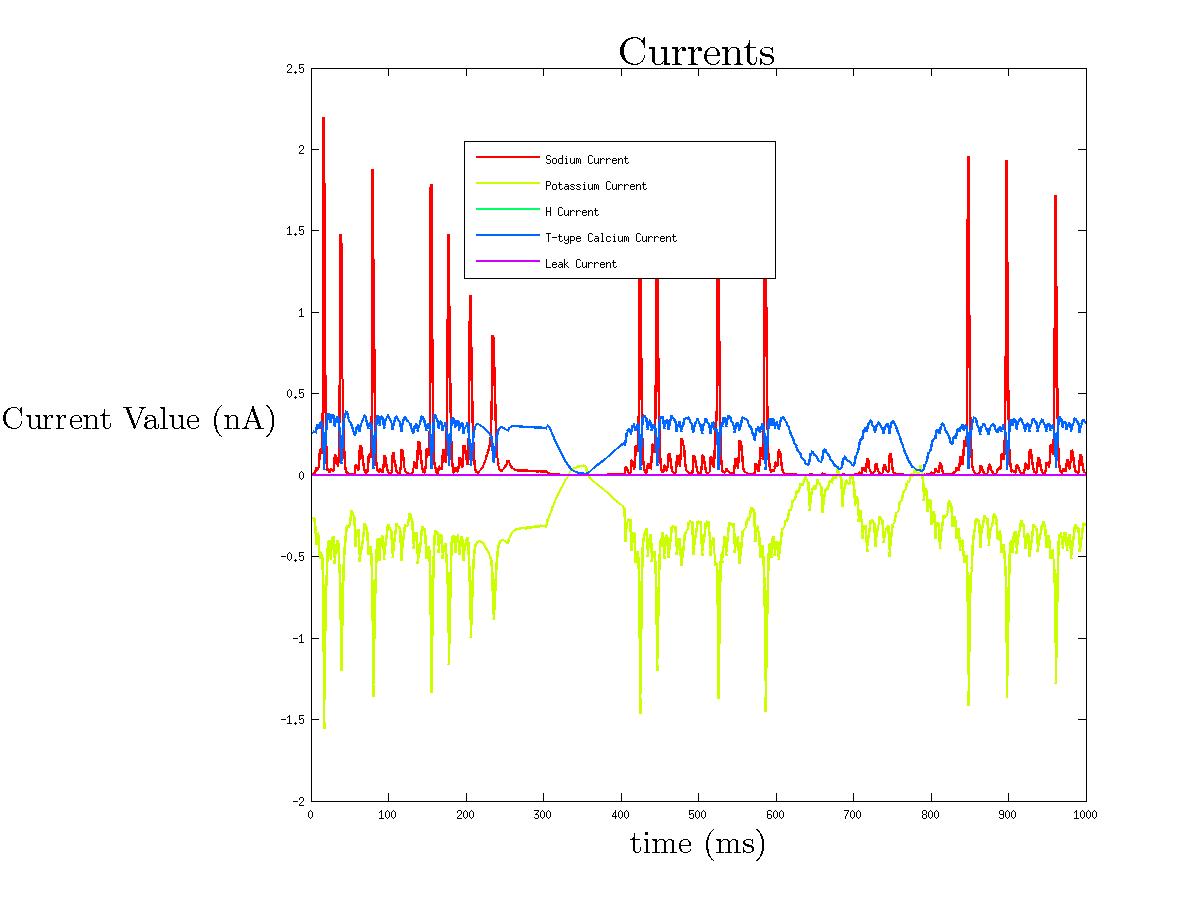}
       \captionsetup{width = 0.9\textwidth}
       \subcaption{Estimated ionic currents in the estimation window. There does not appear to be a significant difference between this plot and the current plot in \autoref{10kHz1-output1}}
       \label{10-1-current2}
      \end{subfigure}
      \captionsetup{width = 0.5\textwidth}
      \caption{A second class of predictions and estimated ionic currents of the 10kHz sampling rate condition of \autoref{input1}. The set of parameters that produced this type of prediction typically occurred at $\beta \approx 38-40$.}
       \label{10kHz1-output2}
     \end{figure}

Among the acceptable predictions, the many different paths can be grouped into two types. Both types represent the subthreshold very well. One of the types spikes far more frequently than the real neuron that the data was taken from, while the other type of result spikes less frequently than the real neuron. Parameters producing these predictions correspond to levels in the action that are lower than any of the others in \autoref{10kHz1-action} when $\beta > 30$. Both of those results have been plotted below in \autoref{10kHz1-output1} and \autoref{10kHz1-output2}.

The first class of predictions plotted (\autoref{10kHz1-output1}) is the more common of the two types of predictions designated as successful. The subthreshold behavior in this prediction is almost exact, with spike timing predictions deviating slightly from the data. Estimated ionic currents in the assimilation window in this class of results demonstrates a much higher estimated magnitude for $I_{CaT}$ than the step current protocol and corresponding 50 kHz version of the present protocol. This supports the hypothesis that estimates of parameters for this data analysis produce a more accurate model than in the previous sections, as we know from \cite{daou2013electrophysiological} that $I_{CaT}$ plays a role in the behavior of $\text{HVC}_\text{I}$ neurons.
     
The second class of predictions is demonstrated in \autoref{10kHz1-output2}. This form is a less common type of prediction than the first in this analysis. Although the subthreshold prediction of the voltage trace in this class is again quite accurate, the prediction is more prone to spiking than the actual neurons that the data was recorded from. 


To the eye, there does not appear to be a significant difference between the estimated ionic current traces in \autoref{10-1-current1} and \autoref{10-1-current2}. The differences in the predicted voltage traces could be due to small differences in many parameters resulting in the significant difference in the intrinsic excitability of the model in the two cases. 


In both \autoref{10kHz1-output1} and \autoref{10kHz1-output2}, the current plot shows a larger magnitude of $I_{CaT}$ than in previous analyses. As has been discussed, it is known that $I_{CaT}$ plays a role in $\text{HVC}_\text{I}$ neuron behavior. The increase in the estimate of this current implies that the model is being more accurately estimated using this data set and sampling frequency than in previous sections. The realistically larger magnitude in the estimate of $I_{CaT}$ supports our model of $\text{HVC}_\text{I}$ neurons and our annealing methods of data assimilation as a whole.
     
\subsection{Real Data: Low Frequency Complex Current}
\begin{figure}[h!]
  \centering
  \begin{subfigure}{0.49\textwidth}
  \includegraphics[width = \textwidth]{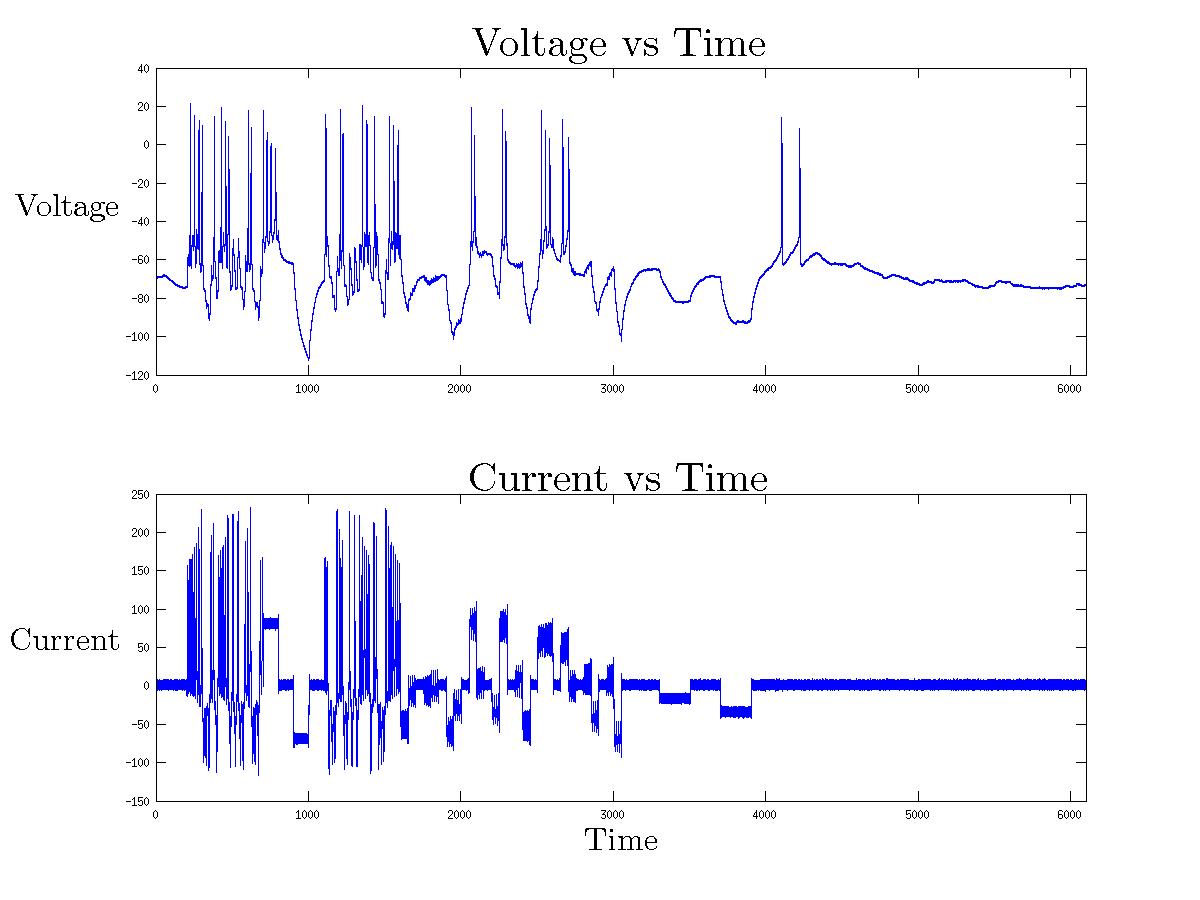}
  \captionsetup{width = 0.9\textwidth}
  \subcaption{Voltage and current traces for the full epoch}
 \end{subfigure}
 ~
  \begin{subfigure}{0.49\textwidth}
  \includegraphics[width = \textwidth]{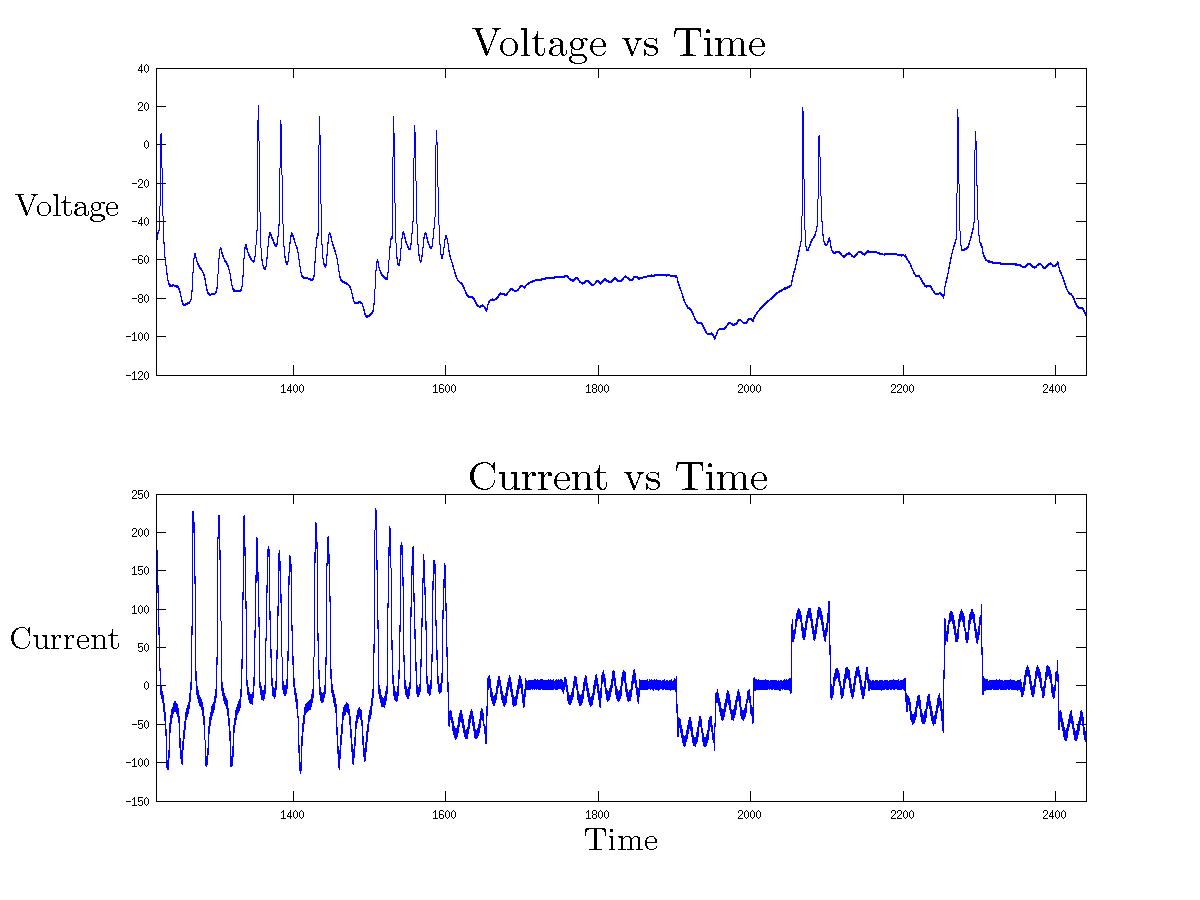}  
  \captionsetup{width = 0.9\textwidth}
  \subcaption{Voltage and current trace of a small section of the epoch to show detail of the waveform information}
  \end{subfigure}
  \captionsetup{width = 0.5 \textwidth}
  \caption{Data and low frequency complex stimulating current protocol. Comparison to \autoref{input1} shows that the frequency of oscillations is significantly lower.}
  \label{input4}
 \end{figure}

This section addresses the analysis of voltage data generated from a final stimulating current protocol. The stimulating current for this epoch is also generated partially from Lorenz '63, a system of equations capable of generating chaotic behavior. The frequency of the oscillations in the input current during this epoch is lower than in the previously analyzed chaotic current protocol. This difference in stimulating current frequency affects the quality of predictions here, which also is found in twin experiments in section \ref{sec:twinexperimentsDependence}. This epoch produces the most successful predictions when 10 kHz data is fed into the assimilation algorithm. The current and voltage data for this epoch is displayed in \autoref{input4}.
     
 \subsubsection{50kHz Sampling Rate}
\begin{figure}[h!]
 \centering
  \includegraphics[width = 0.5\textwidth]{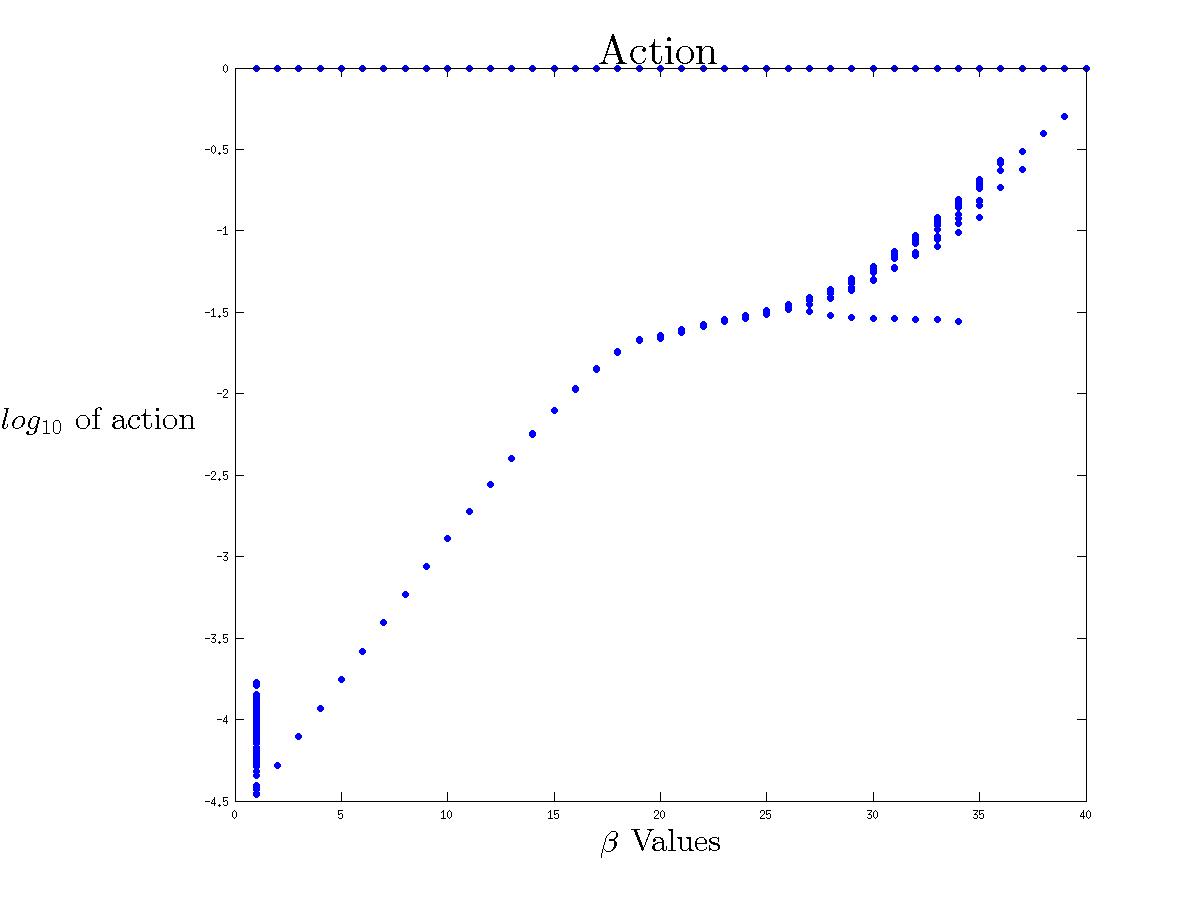}
  \captionsetup{width=.5\textwidth}
  \caption{50kHz data with $\alpha = 1.5$. Input from \autoref{input4}}
 \label{Action50-4}
\end{figure}

 \autoref{Action50-4} is the action plot which results from the analysis of 50kHz data, displayed in \autoref{input4}. It has significantly more separation between the action levels than the previous two 50kHz action plots. However, the best predictions from this stimulating current condition are not very good. The subthreshold behavior is not predicted well, though the spike timing and spike frequency is well estimated. This is due to the fact that subthreshold voltages are not well explored in the estimation window.
 
\begin{figure}[h!]
\centering
      \begin{subfigure}{0.49\textwidth}
      \centering 
            \includegraphics[width=\textwidth]{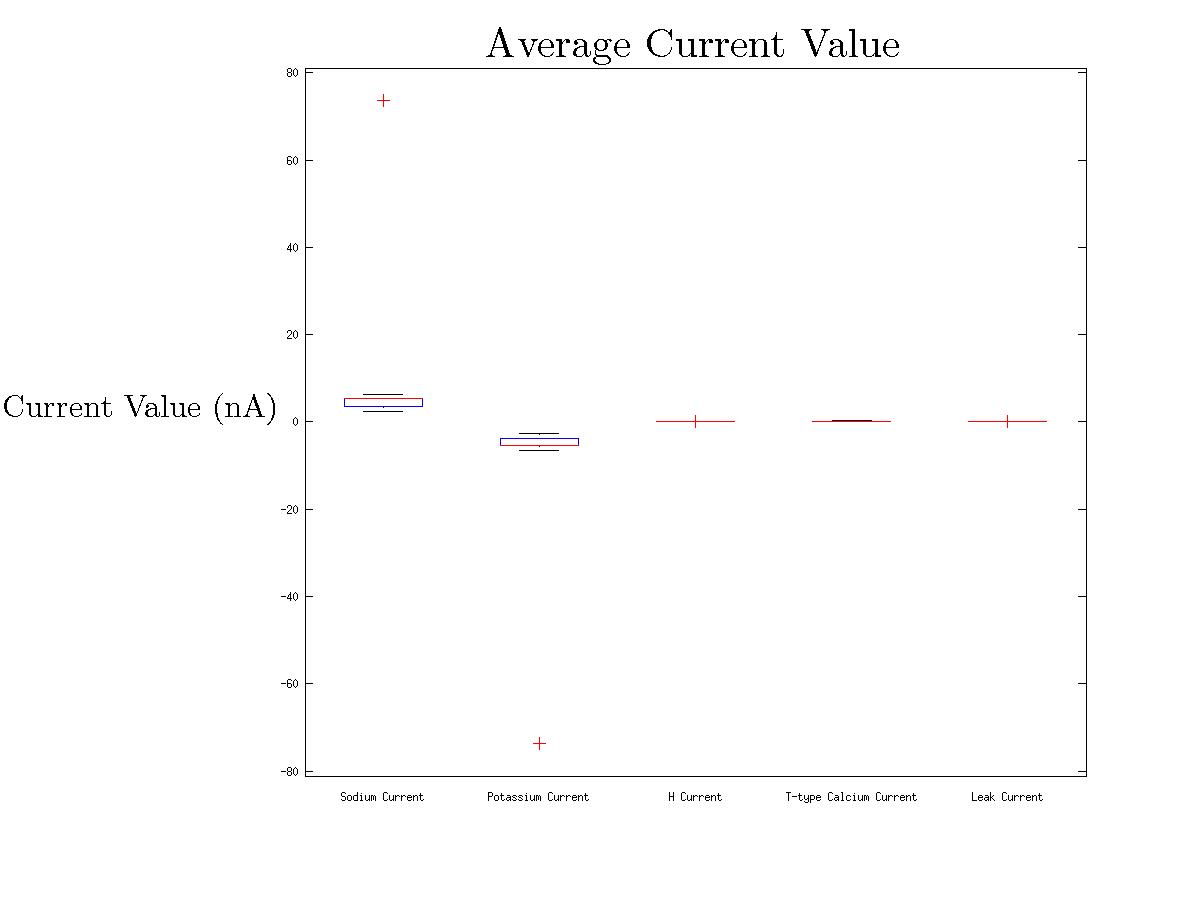}
            \captionsetup{width = 0.9\textwidth}
            \subcaption{Time averaged magnitude of the ionic currents for each estimated parameter set}
      \end{subfigure}
      ~
      \begin{subfigure}{0.49\textwidth}
      \centering 
            \includegraphics[width=\textwidth]{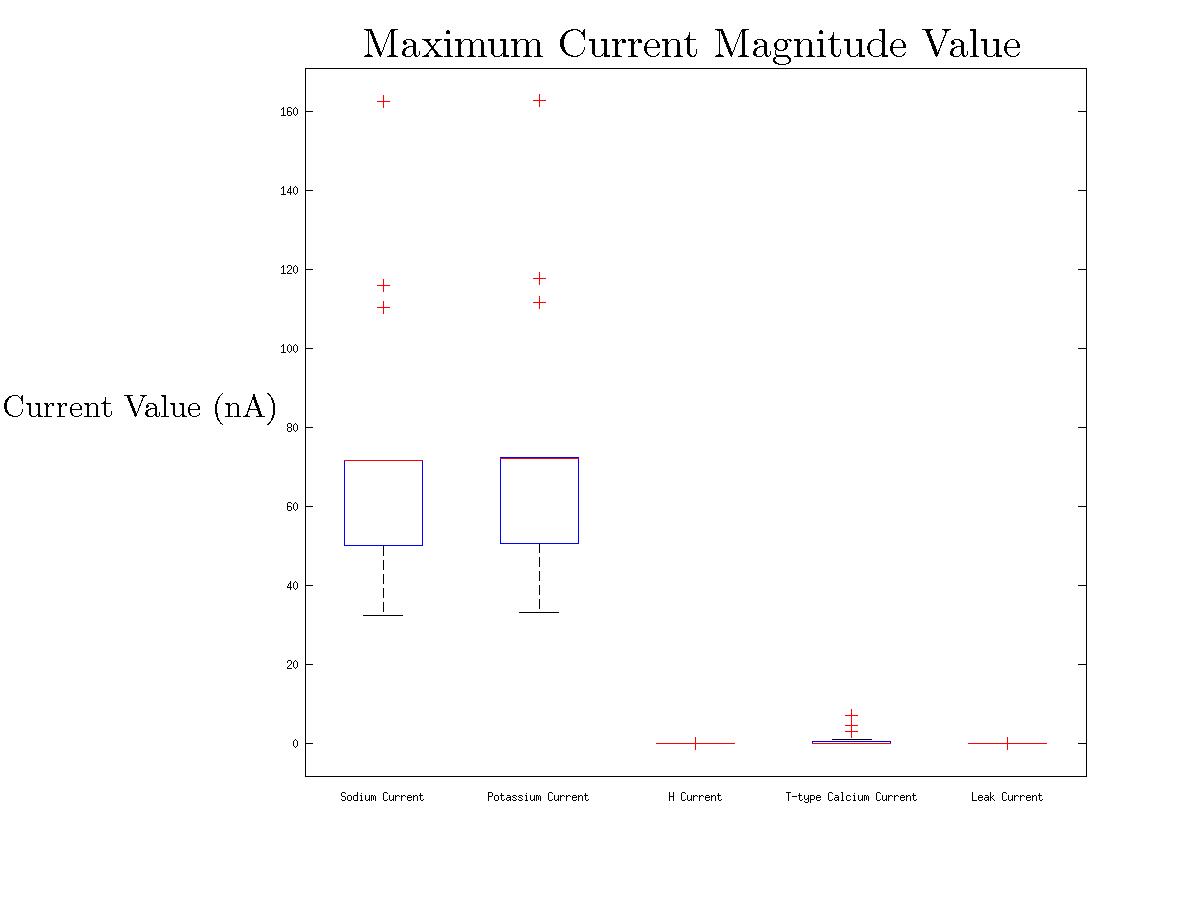}
            \captionsetup{width = 0.9\textwidth}
            \subcaption{Distribution of the maximum magnitude of each ionic current over parameter sets}
     \end{subfigure}
     \caption{Box plots showing a wide distribution of features resulting from DA on voltage data produced by the low frequency complex stimulating current of \autoref{input4} sampled at 50 kHz. The red data points in these plots are features of sets of parameters in the model producing 'good' predictions. These predictions do not accurately reproduce subthreshold behavior.}
 \end{figure}
 
 Of the three sets of data analyzed using 50kHz sampling rate, results using the stimulating current of \autoref{input4} produced the best predictions. The resulting set of estimated currents have a higher maximum and average current value for both $I_{Na}$ and $I_K$ than other currents. 
 
 
  \begin{figure}[h!]
  \centering
      \begin{subfigure}{0.5\textwidth}
       \centering
       \includegraphics[width = \textwidth]{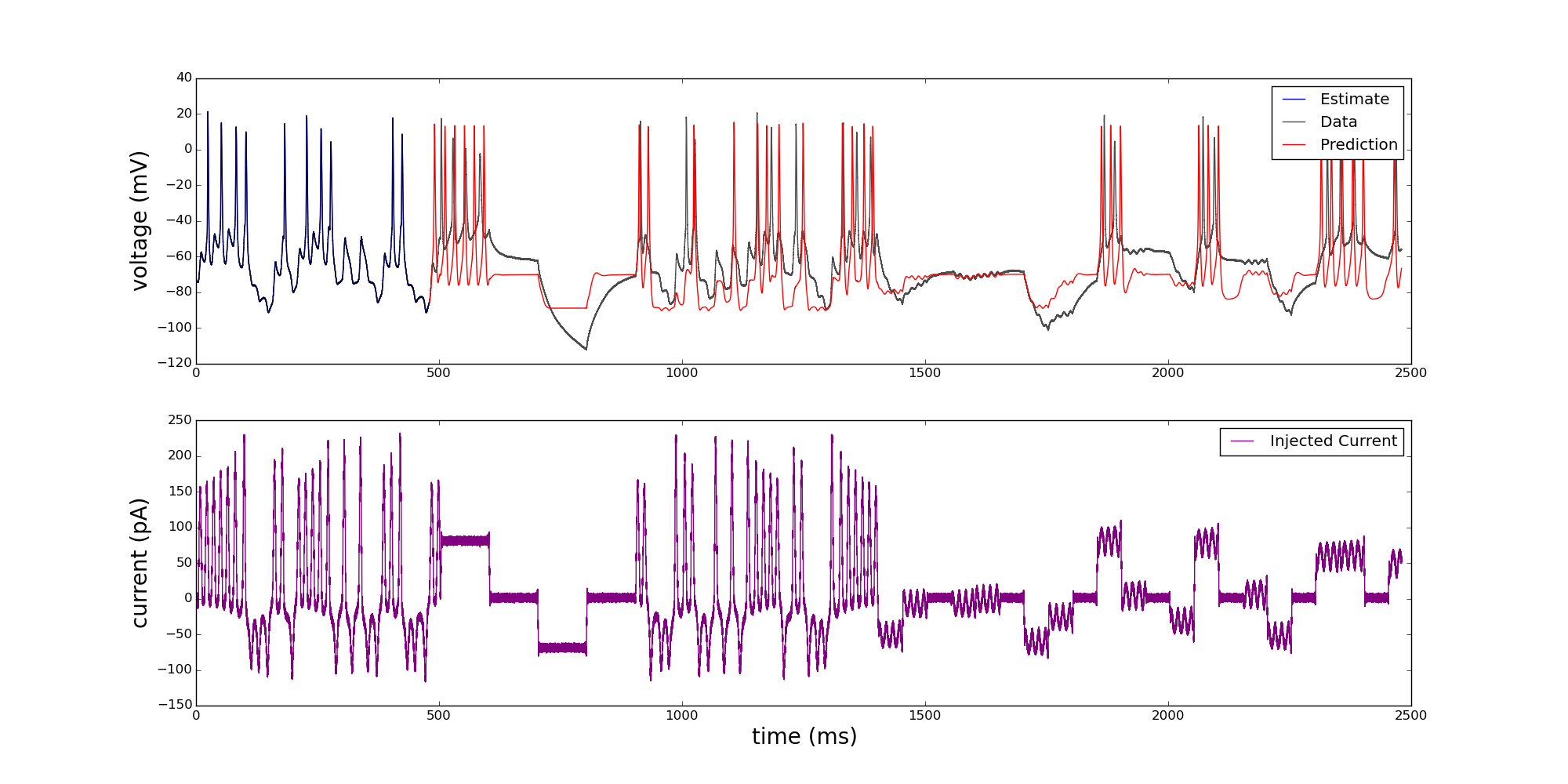}
       \captionsetup{width = 0.9\textwidth}
       \subcaption{A representative example of the good predictions from this set of initial paths using 50kHz data (480.2 ms)}
       \label{50kHz4-output}
      \end{subfigure}
      ~
      \begin{subfigure}{0.4 \textwidth}
       \centering
       \includegraphics[width = \textwidth]{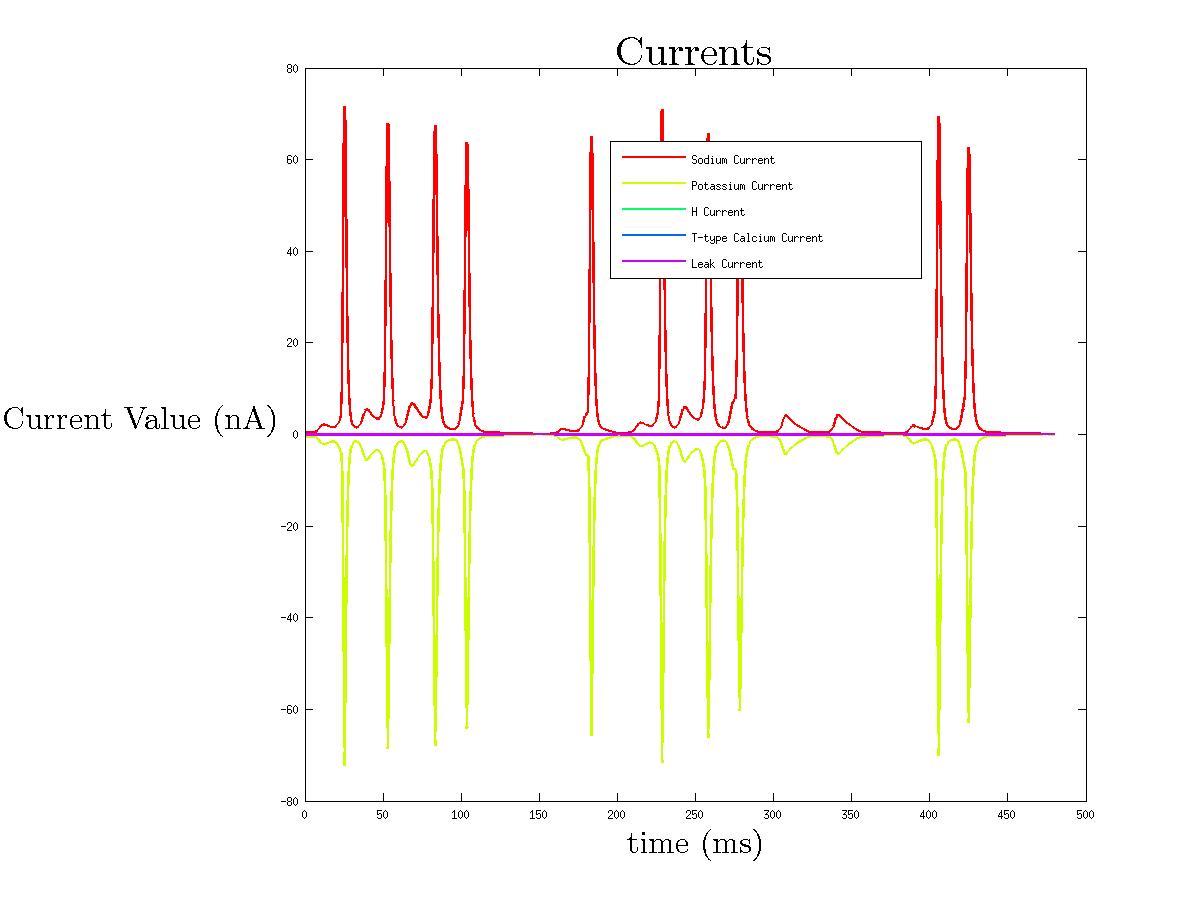}
       \captionsetup{width = 0.9\textwidth}
       \subcaption{Estimated ionic currents in the estimation window.}
      \end{subfigure}
      \captionsetup{width = 0.5\textwidth}
      \caption{Predictions and estimated ionic currents of the 50kHz sampling rate condition of the data in \autoref{input4}.}
       \label{50kHz4}
     \end{figure}
     
 The estimates of $I_{CaT}$ and $I_H$ are nearly zero. Although $I_{CaT}$ and $I_H$ are in fact smaller in magnitude than the other two currents, the estimates contradict the known importance of $I_H$ and $I_{CaT}$ \citep{daou2013electrophysiological} in $\text{HVC}_\text{I}$ neurons. 

It can be concluded that the estimate of properties of $I_H$ and $I_{CaT}$ is incorrect. 
$I_H$ and $I_{CaT}$ are poorly estimated because the input current in the estimation window causes the neuron to spend very little time in the low subthreshold regime, where the overall conductance of the neuron appears to be overestimated. This can be seen in the relatively small drop in voltage of the model accompanying the injected current steps of \autoref{50kHz4-output}.

\subsubsection{10kHz Sampling Rate}
\begin{figure}[h!]
\centering
  \includegraphics[width = 0.5\textwidth]{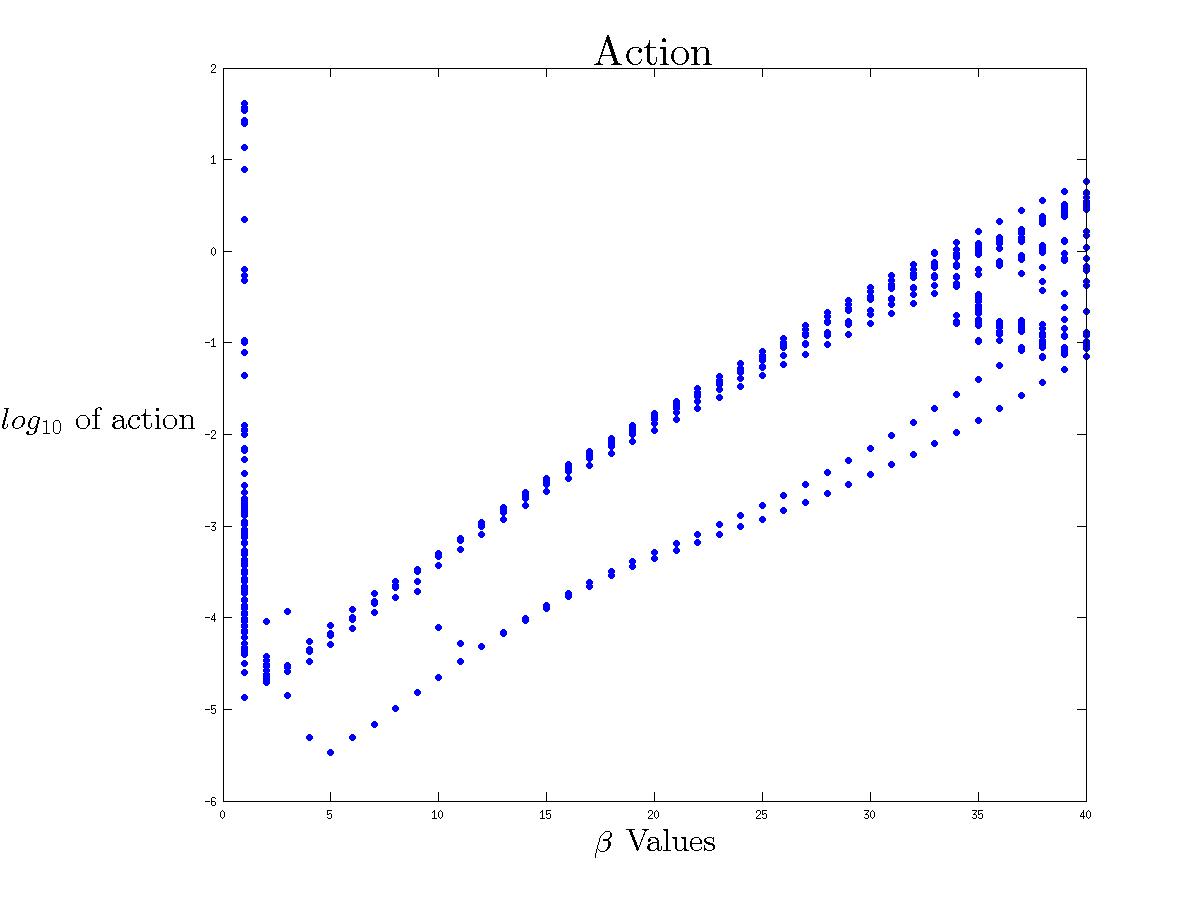}
  \captionsetup{width=.5\textwidth}
  \caption{Action level plot using voltage data and the low frequency complex current protocol of \autoref{input4} fed into the DA algorithm at 10 kHz. $R_f$ is increased by a factor of $\alpha = 1.5$ at each annealing step. As with other action level plots, a large number of action levels are present in the graph, reflecting the presence of a distribution of parameters producing similar time evolution in the voltage trace when integrated forward.}
 \label{Action10-4}
\end{figure}

The results of analysis on the data of \autoref{input4} sampled at 10 kHz produced markedly better results.
The resulting action plot is shown below in \autoref{Action10-4}. Out of 100 initial paths, 49 arrived at estimates producing sufficiently good predictions. Both the spike timing and the subthreshold behavior of the voltage traces in the prediction match the data better than any of the previous attempts. Based on the following results it is concluded that the estimated model properties are more accurate than any of the other sampling rate and stimulating current waveform conditions.

Once again when moving from a sampling rate of 50kHz to one of 10kHz with the same data, the action plot becomes significantly `messier'. Action level separation is unclear at higher $\beta$ values. Though individual levels are harder to distinguish, there is still a large range between the final action level of the lowest and highest plots. The levels themselves are less discrete.

\begin{figure}[h!]
\centering
\begin{subfigure}{0.49\textwidth}
      \centering 
            \includegraphics[width=\textwidth]{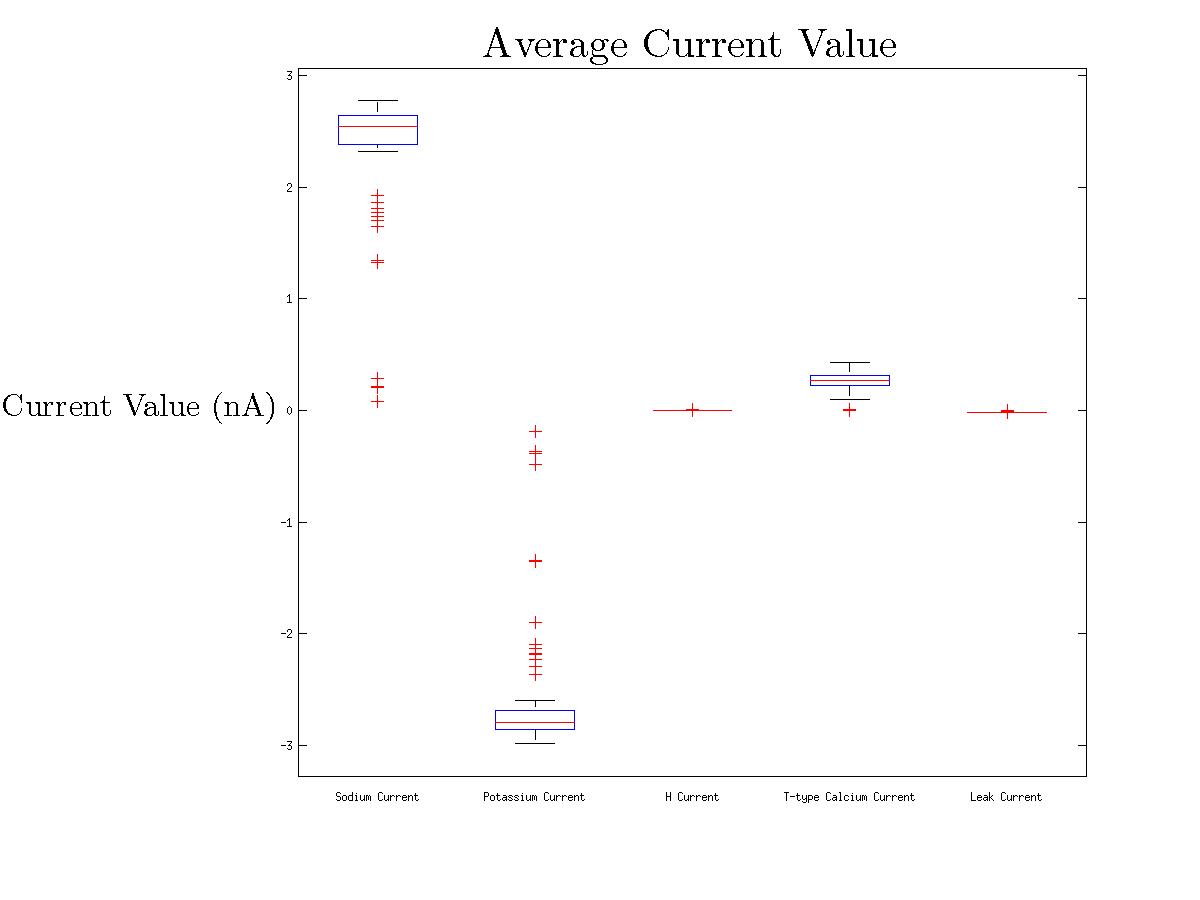}
            \captionsetup{width = 0.9\textwidth}
            \subcaption{Distribution of the time averaged magnitude of the ionic currents.}
      \end{subfigure}
      ~
      \begin{subfigure}{0.49\textwidth}
      \centering 
            \includegraphics[width=\textwidth]{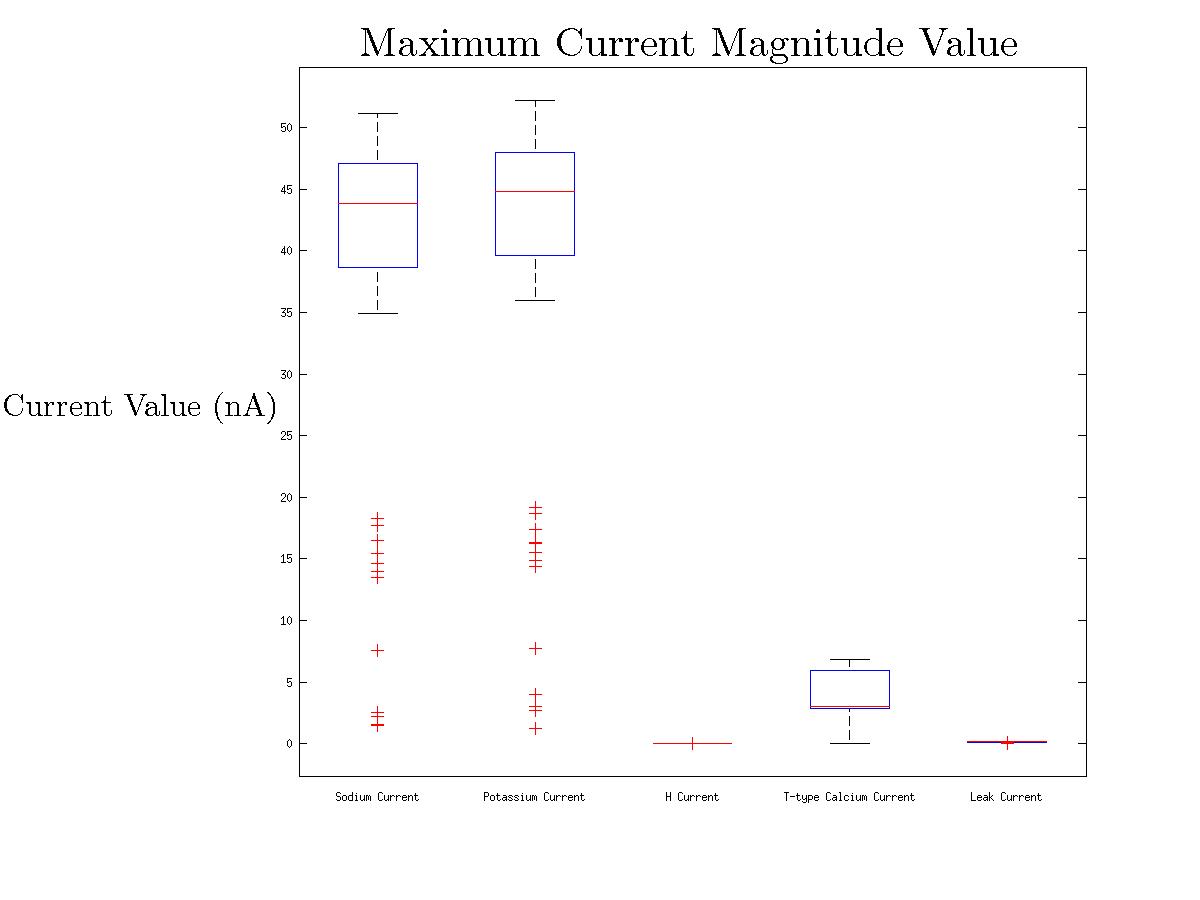}
            \captionsetup{width = 0.9\textwidth}
            \subcaption{Maximum magnitude of each of the ionic currents over time.}
     \end{subfigure}
     \caption{Box plots showing a wide distribution of features resulting from DA on voltage data produced by the low frequency complex stimulating current of \autoref{input4} sampled at 10 kHz. The red data points in these plots are features of sets of parameters in the model producing 'good' predictions. These predictions tend to match subthreshold information and the spike timing and frequency information well, and this coincides with a somewhat smaller variance in the features in the plots above.}
     \label{box4down}
\end{figure}

Though the box plots in \autoref{box4down} have more outliers than all the previous box plots, the spread of the remaining data points is tighter than in previous plots.

\begin{figure}[h!]
\centering
      \begin{subfigure}{0.5\textwidth}
       \centering
       \includegraphics[width = \textwidth]{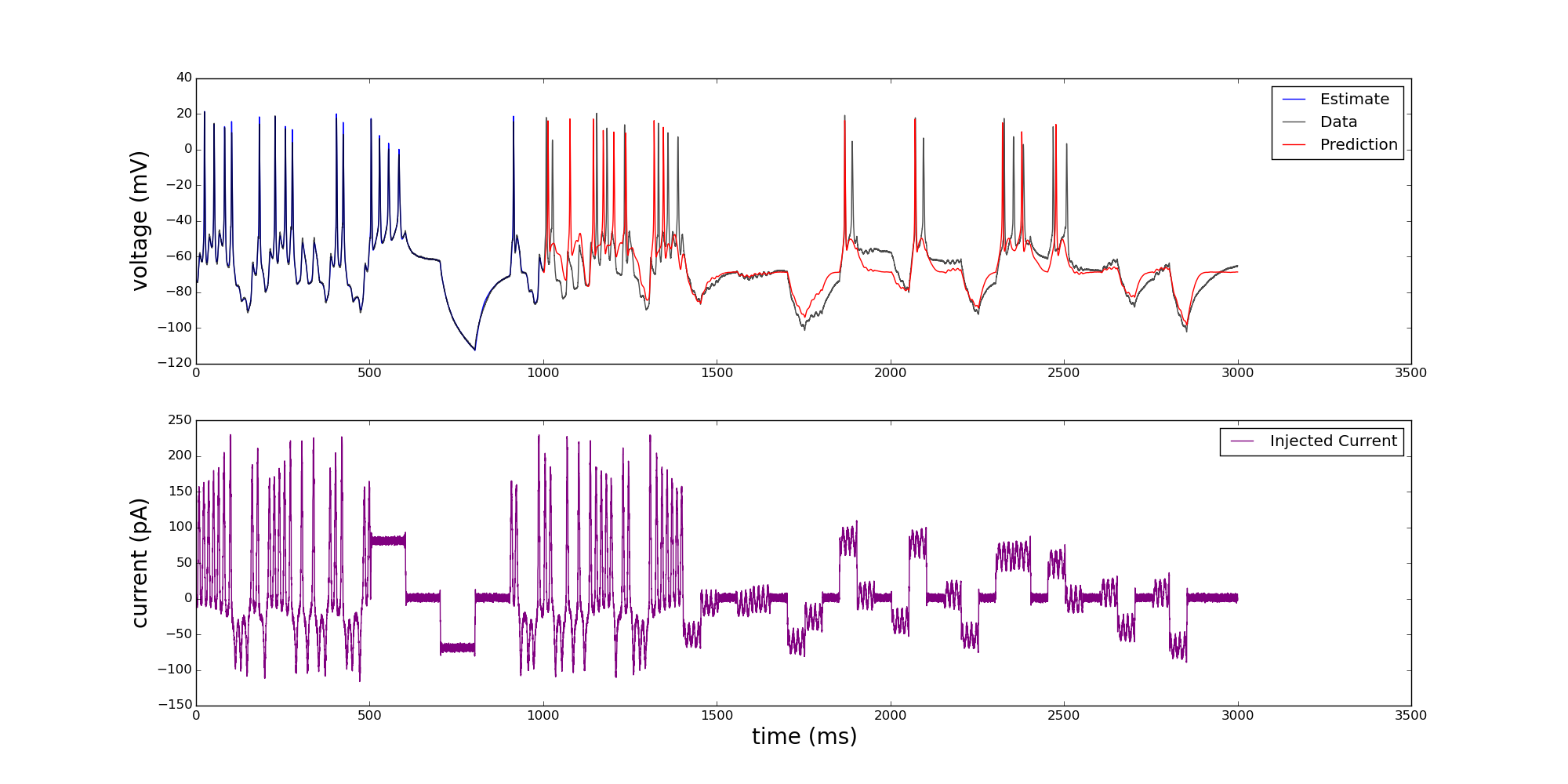}
       \captionsetup{width = 0.9\textwidth}
       \subcaption{Representative example of one of the good predictions using 10kHz (1000.1 ms) data.}
      \end{subfigure}
      ~
      \begin{subfigure}{0.4 \textwidth}
       \centering
       \includegraphics[width = \textwidth]{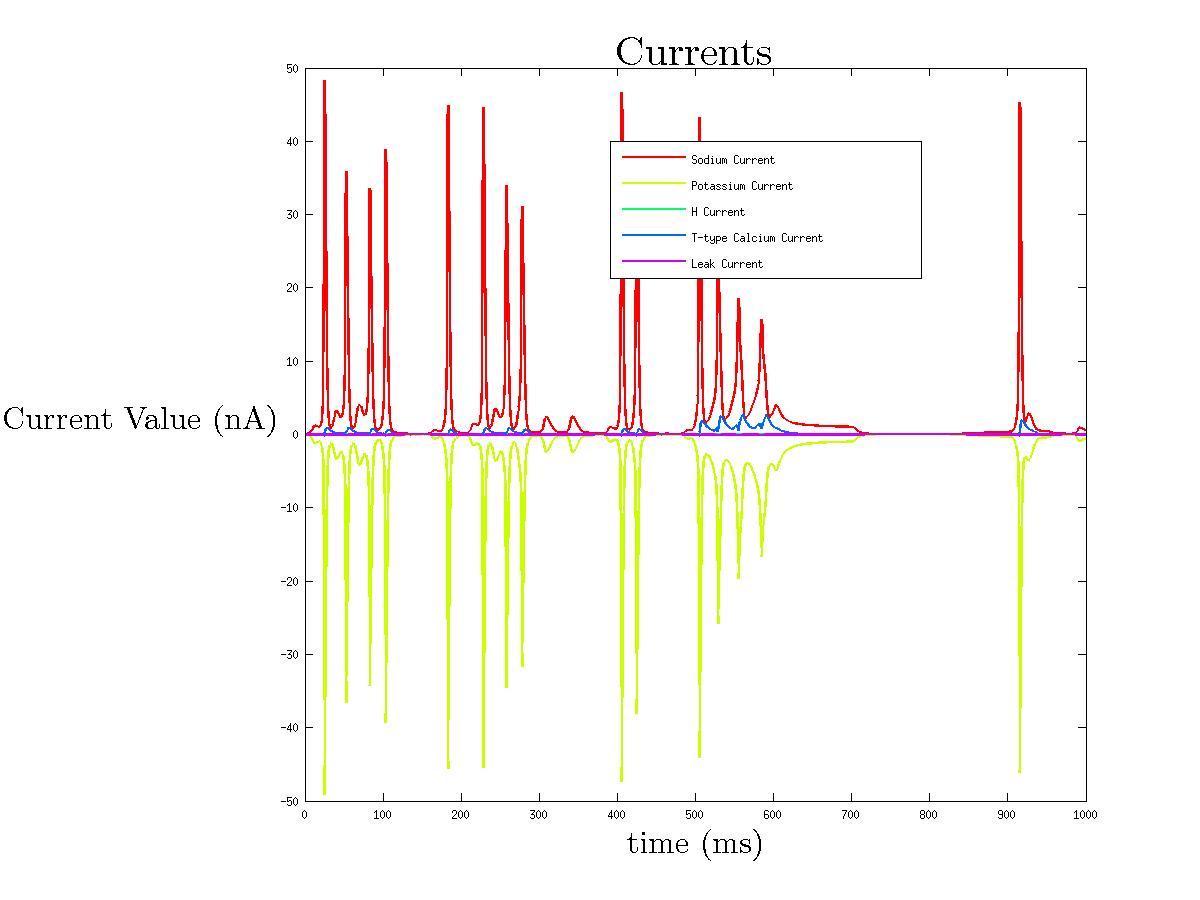}
       \captionsetup{width = 0.9\textwidth}
       \subcaption{Estimated ionic currents in the estimation window. The estimate of $I_{CaT}$ is more realistic in this example.}
      \end{subfigure}
      \captionsetup{width = 0.5\textwidth}
      \caption{10kHz sampling rate of \autoref{input4}. The set of parameter values that produced these results typically occurred at $\beta$ values of around 38-40.}
       \label{10kHz4}
     \end{figure}
     
This analysis done on \autoref{input4} at a sampling rate of 10kHz was the most successful of the three analyses done at 10kHz and was also the most successful of all analyses done regardless of sampling rate. The prediction and current plots in \autoref{10kHz4} are representative of the good estimations found in the analysis.  Other estimations with similar predictions in this analysis have slightly different maximal magnitudes of $I_{CaT}$, though the general shape and timing of this and other currents does not vary. The increase in $I_{CaT}$ in this analysis compared to the analysis in \autoref{50kHz4-output} is consistent with the improvement in subthreshold behavior prediction. The improved subthreshold behavior most likely arises because the step current sections of data now present in the new estimation window samples the relaxation properties and I-V relationships of ionic currents at several different voltages. Additionally, after the relaxation of the slow variables, stimulating the neuron to threshold and causing action potential helps to contrast the properties of each of the fast currents. This contrast may have been brought out less well with the high frequency stimulating protocol because the hyperpolarizing portion of the step current was briefer and smaller in magnitude than with the low frequency stimulating current protocol.

\subsection{Further Comments}

We were also interested in what would happen if $I_{CaL}$ were added to the model in addition to the other four ionic currents, $I_{Na}$, $I_K$, $I_{CaT}$, and $I_H$. The addition of $I_{CaL}$ did not improve the quality of the estimations and predictions. Data assimilation with addition of $I_{CaL}$ produced a variety of distinct parameter sets, but none of these parameter sets produced estimates differing significantly enough in the value of the cost function to dominate the contribution of the integral in equation \ref{eq:prob_dis}. However, dropping $I_{CaT}$ and retaining $I_{CaL}$ eliminated the ability of the assimilated model to reproduce rebound spiking. This suggests that $I_{CaL}$ is not important for reproducing observed behavior and therefore can be dropped from the model. 

We also attempted data assimilation at a variety of sampling rates higher and lower than 10 kHz. For conditions in which the sampling rate was higher than 10 kHz, we increased the dimension D of the problem so that the estimation window was the same length in time for a 50 kHz and 25 kHz condition and allowed the algorithm to run as long as necessary to complete. Even with the condition of increasing the computational resources available for the problem, we did not find any improvement in the quality of predictions or undesirable multimodality of the probability distribution.

\subsection{Twin Experiments: Dependence of Data Assimilation Results on Stimulating Currents and Sample Rate}
\label{sec:twinexperimentsDependence}

To continue our investigation into parameter set degeneracies and how successfully our data assimilation methods produced accurate predictions when the sampling rate and stimulating currents varied, we compared predictions from 6 different conditions using synthetic data generated from our model with the estimated parameters of Table \ref{table:real_estimated_parameters}. 
For this analysis, 3 different current protocols from real experiments were used for data assimilation. As before, to explore the tradeoffs given finite computational power between stimulating extra degrees of freedom and losing resolution of the measurements in time by downsizing the measured data, the same voltage traces were sampled at different frequencies. One set was sampled at 50 kHz, while the other set was sampled at 10 kHz. The same stimulating current waveforms were used here as with DA on real data to demonstrate the presence of multiple sets of parameters and initial conditions that produce similar voltage waveform behavior when the system producing the data was known. Similar measures of these degeneracies were used here as were in the analysis of real data, to be enumerated again below.
For all conditions, the length of the estimation window was 16001 time points. This is around the maximum possible number of time points which allows the estimation procedure to complete in 24 hours given constraints on our computational resources. For 10 kHz, this is a 1600.1 ms time window, while for 50 kHz, this is a 320.02 ms time window. For annealing, $\alpha = 2$ and $\beta$ is incremented by 1 at each annealing step.

\begin{figure}
\includegraphics[scale=0.18]{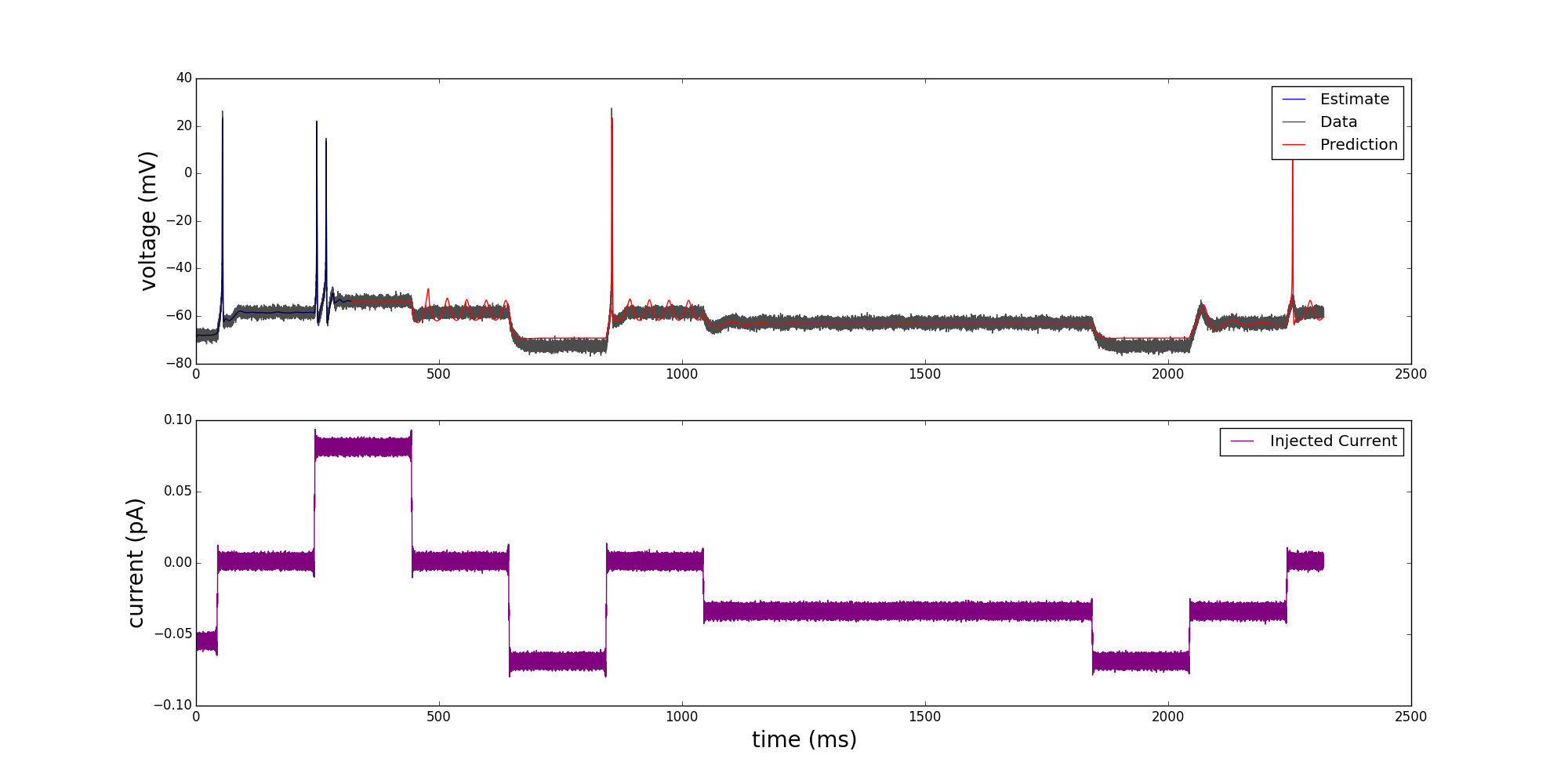}
\includegraphics[scale=0.18]
{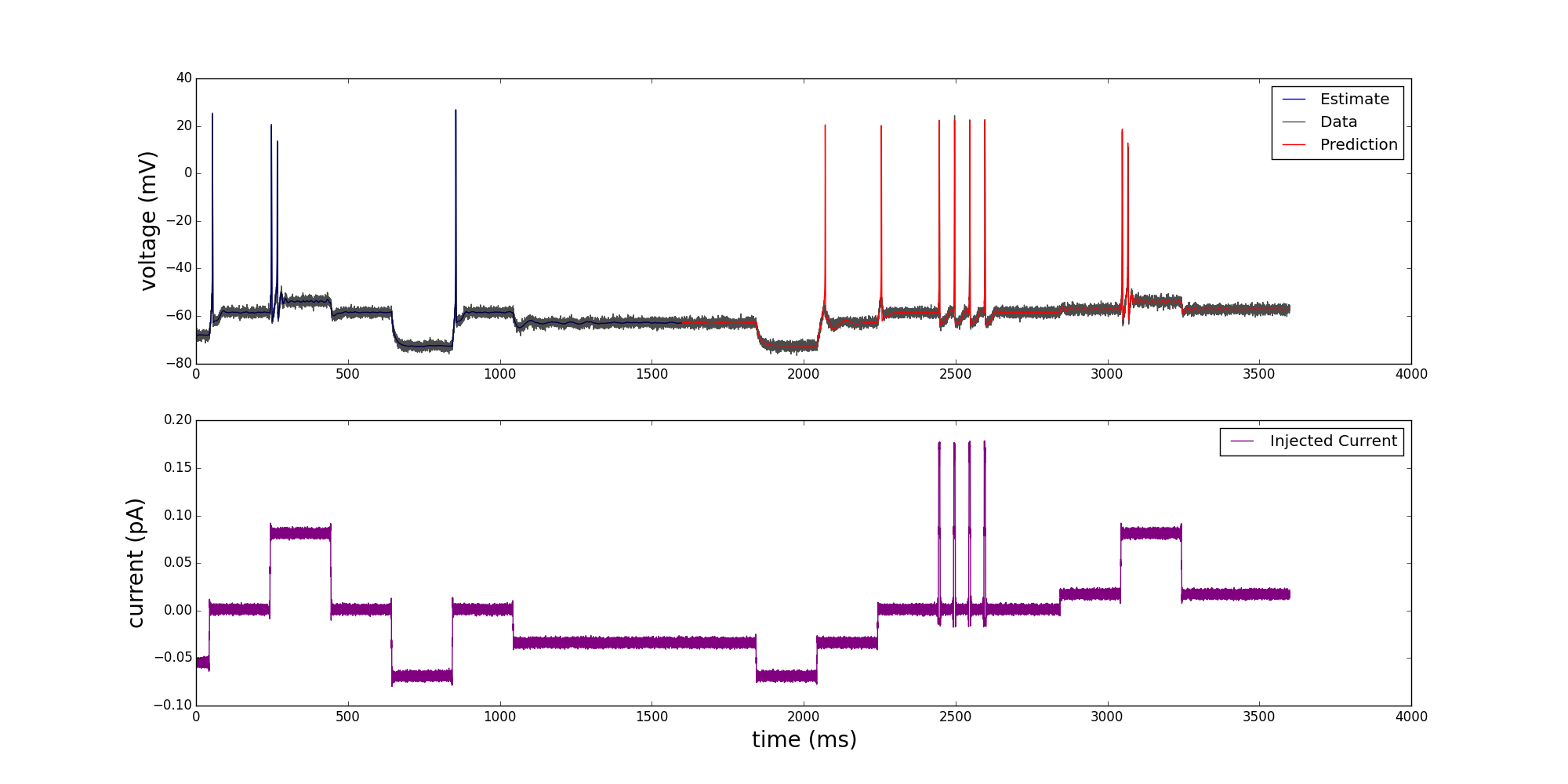}
\caption{Data assimilation window and prediction window using a step current elicited synthetic recording from a zebra finch HVC$_I$ neuron. Data and a representative estimate and prediction are plotted for the 50 kHz and 10 kHz condition, top and bottom graph respectively. The graphs show membrane voltage (top of each graph) in response to injection of a \textit{step} current waveform (bottom of each graph). The \textit{black traces} show recorded voltage, and the \textit{blue traces} show estimated voltage from the data assimilation procedure for times between 0-320 ms or 0-1,600 ms, during which all state variables and parameters of the model were estimated. The \textit{red traces} show the voltage predicted by integrating the completed model with estimated parameters and state variables forward in time beyond 320 ms or 1,600ms.}
\label{fig:stimulation_currents_0}
\end{figure}

\begin{figure}
\includegraphics[scale=0.18]{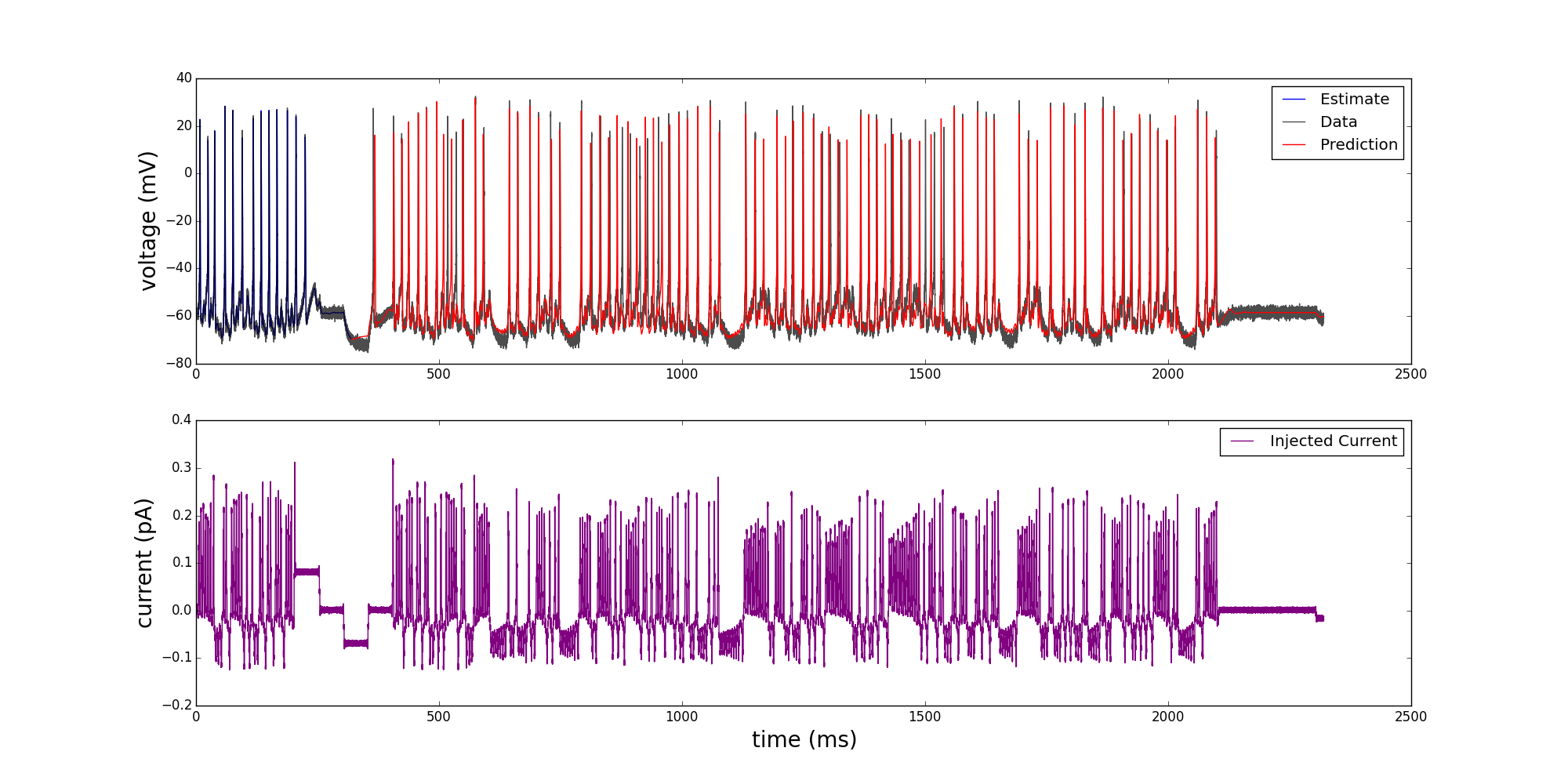}
\includegraphics[scale=0.18]{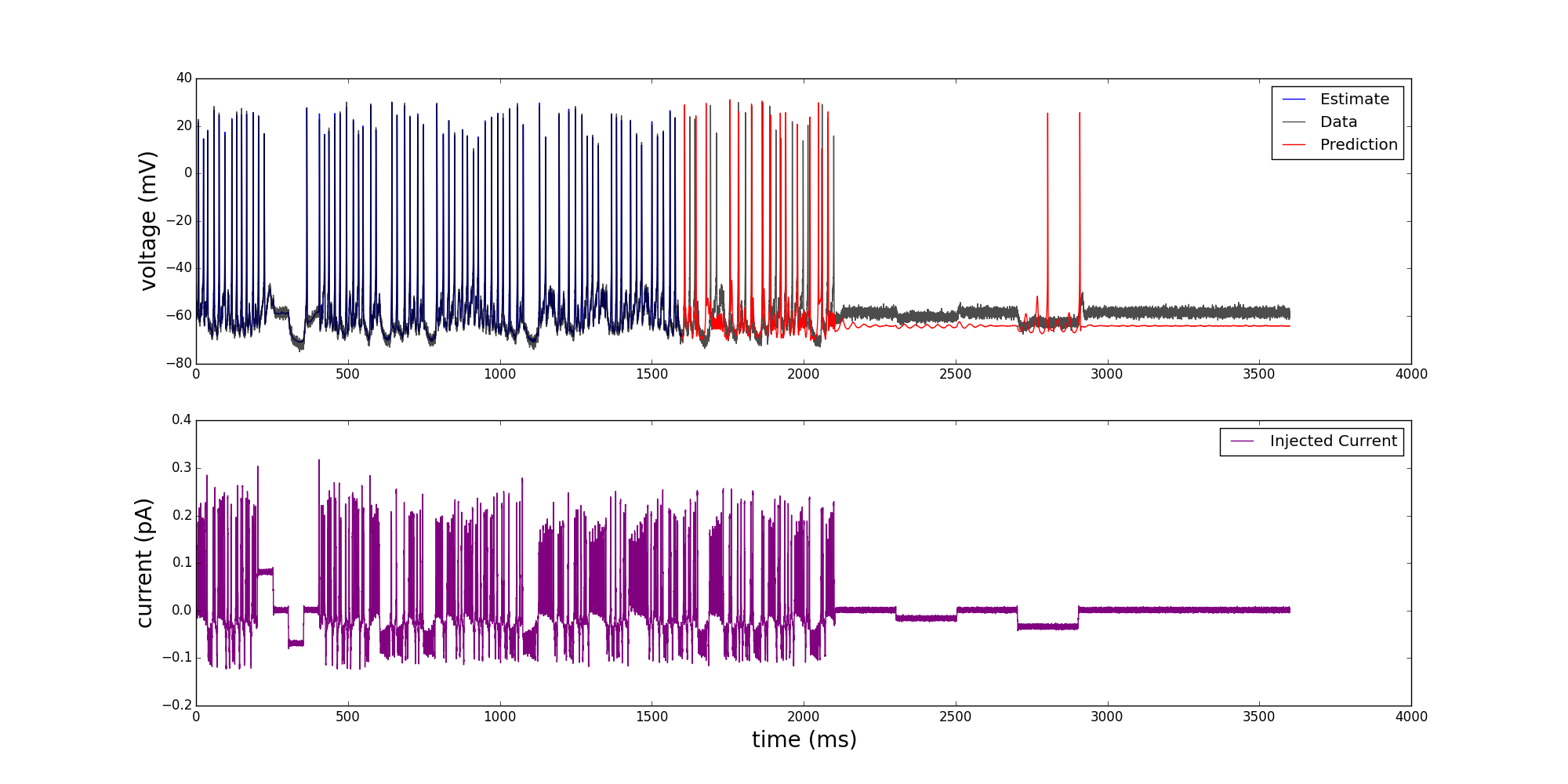}
\caption{Data assimilation window and prediction window using a high frequency complex current elicited synthetic recording from a zebra finch HVC$_I$ neuron.}
\label{fig:stimulation_currents_1}
\end{figure}

\begin{figure}
\includegraphics[scale=0.18]{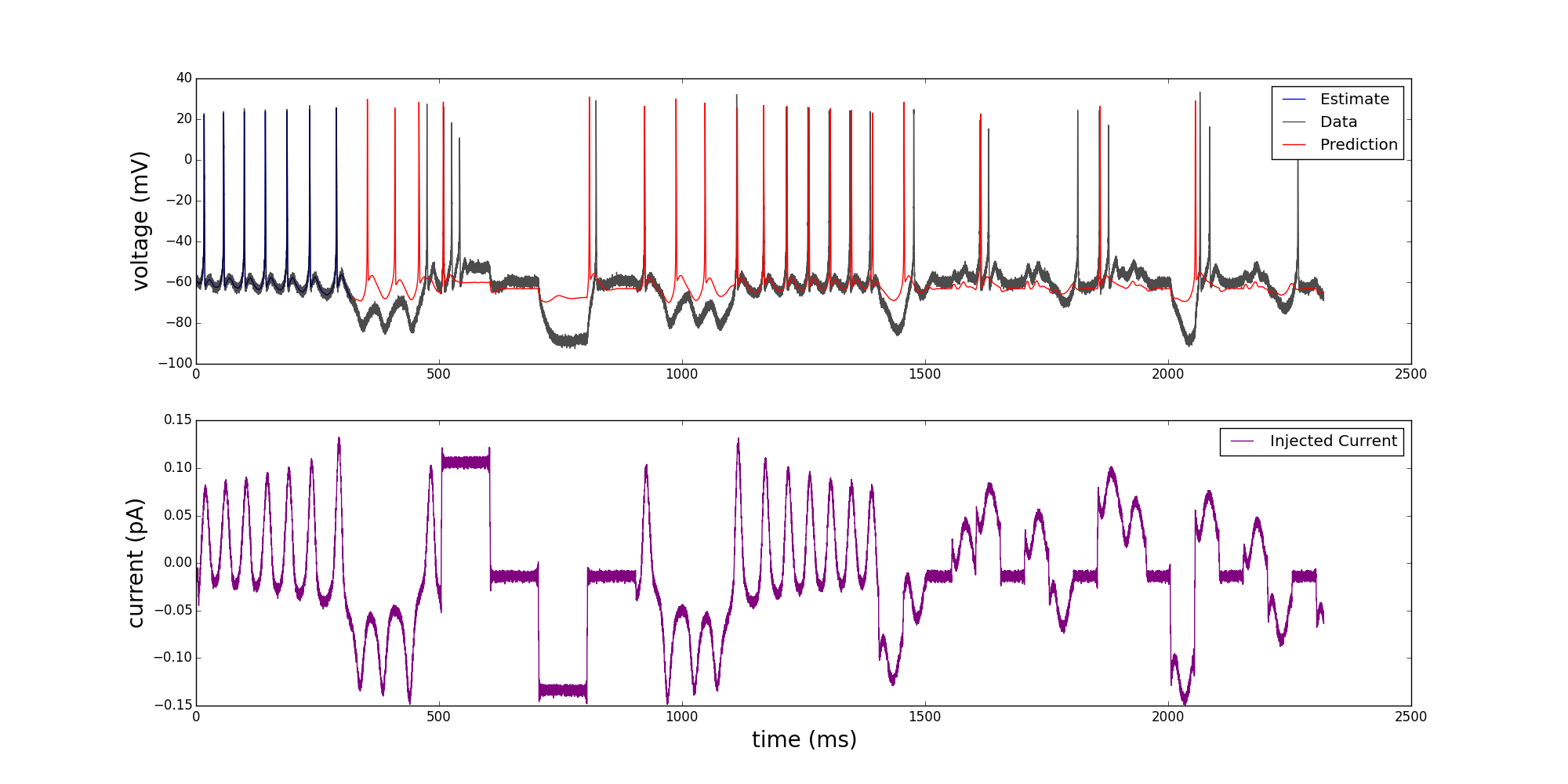}
\includegraphics[scale=0.18]{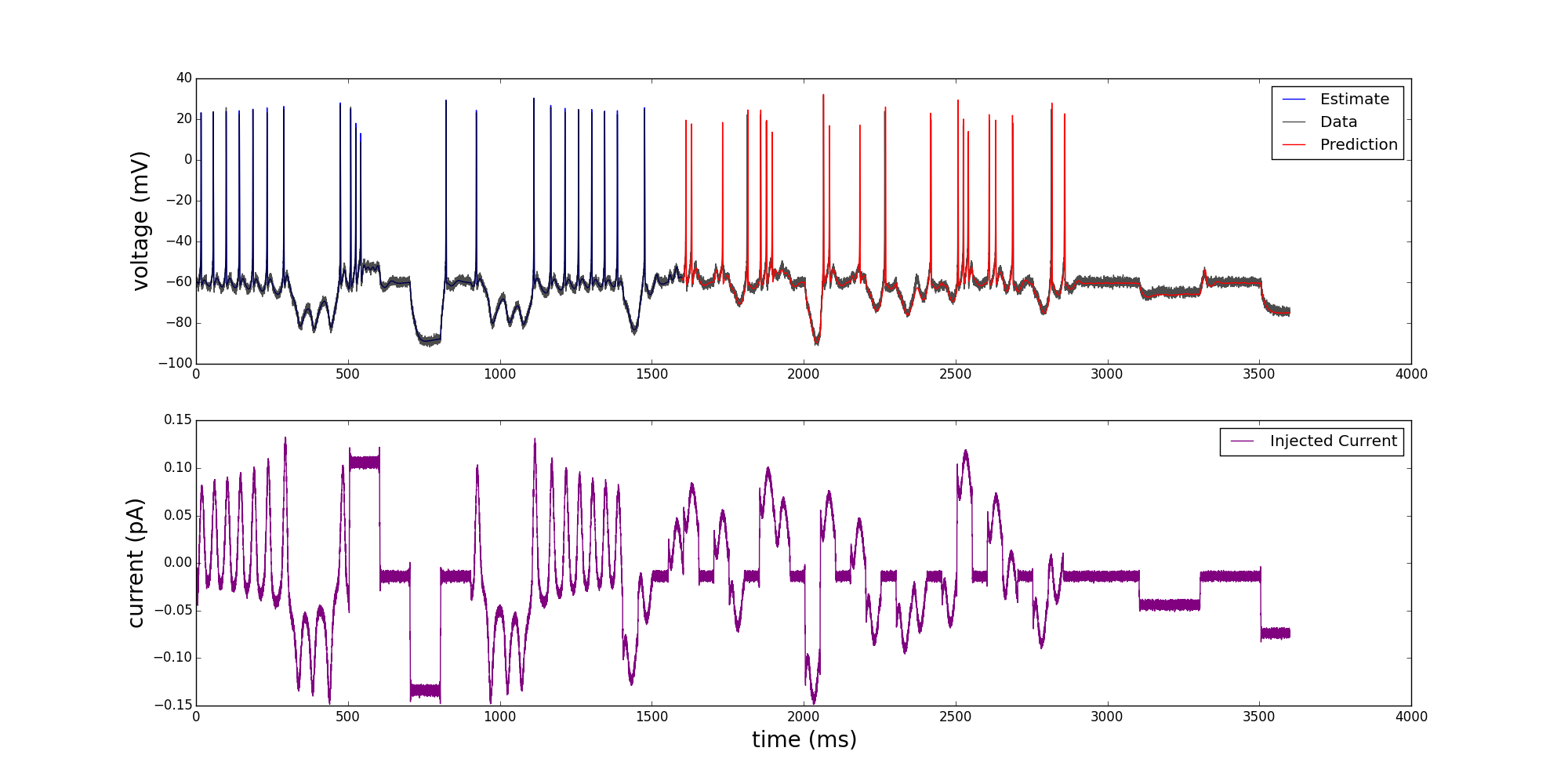}
\caption{Data assimilation window and prediction window using a low frequency complex current elicited synthetic recording from a zebra finch HVC$_I$ neuron.}
\label{fig:stimulation_currents_4}
\end{figure}

The stimulating current protocols, along with the response voltage and representative estimates and predictions for each protocol and sampling rate condition are shown in Figures~\ref{fig:stimulation_currents_0}-\ref{fig:stimulation_currents_4}.

Integrating forward our $\text{HVC}_\text{I}$ model using a number of estimated parameter sets produces indistinguishable (or nearly so) time evolution in the voltage given a stimulating current protocol.
For example, when multiple excitatory ionic currents such as $I_{CaT}$ and $I_{Na}$ with similarly fast activation times are simultaneously present in a model, different combinations of $I_{CaT}$ and $I_{Na}$ can lead to voltage behavior which appears the same. Because information about the flux of ionic currents is not typically available when taking recordings from real neurons, we first examine these potential shortcomings when estimating the accuracy of the inferred parameters in a single neuron model using synthetic data, where all the unrecorded processes are known to the experimenter but withheld from the data assimilation algorithm.

There are a few ways to characterize how estimates which produce accurate predictions might be structured.
\begin{enumerate}
\item Plot the max attained amplitude and time averaged magnitude of individual ionic currents for each set of parameters as compared to the true value. 
\item Find whether estimations producing accurate predictions have significantly different values of the cost function.
\item Plot the estimated I-V curves for individual ionic currents and compare them to the true I-V curves.
\item Compare the value of all sets of estimated parameters to the true set of values.
\end{enumerate}

The results of twin experiments with model HVC$_{\text{I}}$ neurons suggest that when data assimilation produces a set of high quality predictions in the voltage traces with different estimated parameter sets, there is clustering in all three of these measures. When the accuracy of predictions degrades, the variance of the distributions for each measure increase. Once the predictions have become sufficiently inaccurate,  usually when the subthreshold behavior is not matched well and the predicted spike timing is off, the clustering is gone and there are significant deficiencies in the estimations.

Figures \ref{fig:action_levels_0s} and \ref{fig:action_levels_4s} demonstrate that none of the sets of estimated parameters producing accurate predictions for each stimulating current protocol can be considered to be much more probable than any of the others. This is because the value of the cost function is not very different for each distinct estimated parameter set path, so the integral in the expression for the probability distribution

\begin{equation*}
P(x(t_M)|Y(t_M)) = \int dX \exp(-A_0(X,Y))
\end{equation*}

is highly multimodal and does not have any dominant contribution. An expected value for the parameters and estimated initial conditions can still be computed, but because there is no reason to believe that all the minima contributing to the integral have been found or that the expected value of parameters will correspond to the true value, there is not a sound basis for calculating this expected value.

The values of average and max currents and maximally attained values of I-V curves are also systematically observed to be biased upwards of their true values. These data support the hypothesis that large depolarizing currents can be offset by similarly large rectifying currents while maintaining an identical net membrane voltage. Some of the degeneracies in estimated parameter sets were merely due to deficiencies in the training data presented to the algorithm. In such cases, the model produced accurate predictions only when the stimulating current presented after the estimation window was sufficiently similar to the training data, but comparison of the maximally attained, 1-norm, and theoretical I-V curves between different recording epochs showed that some characteristics of the current protocols caused these measures to be closer to their true values despite superficially comparable performance in forward prediction. This demonstrates the importance of validating the model in a number of ways, such as forward integration of the model with a variety of novel stimulating current waveforms in the prediction window.

In the graphs in Figure \ref{fig:stimulation_currents_0} - \ref{fig:stimulation_currents_4}, the quality of predictions fell into roughly 3 classes of high, medium, and low quality. High quality predictions matched the voltage waveform of the data exactly. Medium quality predictions showed a close correspondence with subthreshold behavior but missed some spikes. Low quality predictions did not match the subthreshold behavior and missed many spikes.

The 3 stimulation protocols are a step current protocol, a high frequency complex waveform protocol, and a low frequency complex waveform protocol. The \textit{step} current protocol sampled at 10 kHz produced high quality predictions, while the 50 kHz step current protocol produced medium quality predictions. The \textit{high frequency complex} current protocol sampled at 10 kHz produced low quality predictions, while the 50 kHz version produced medium quality predictions. The \textit{low frequency} complex current protocol sampled at 10 kHz produced high quality predictions but produced low quality predictions when sampled at 50 kHz. These patterns can be seen in Figures \ref{fig:stimulation_currents_0} - \ref{fig:stimulation_currents_4}.

However, when the plots of the theoretical I-V curves and maximal/1-norm currents were analyzed, the 10kHz-sampled \textit{step} current protocol which produced high quality predictions did less well than the 10kHz-sampled \textit{low frequency complex} current protocol. This shows that complex current protocols better sample the available degrees of freedom in the model than step current protocols. The resulting parameter sets are then likely to generalize better to other stimulating current waveforms.

The \textit{high frequency complex} current waveform may have generated parameter sets less able to generalize well to new driving current waveforms because the subthreshold behavior of the neuron was not adequately explored during the estimation window. Although the neuron was highly excited during the estimation window, it spent most of its time in similar regions of phase space, along the limit cycle trajectory of a spike. This supports the idea that driving the voltage of a neuron into as many regions of phase space available to the model as possible is crucial when attempting to estimate the parameters and unknown states. The poor performance is likely also due to the fact that the $RC$ time constant of the membrane responded only weakly to the rapidly oscillating stimulating current waveform. This information in the voltage waveform was erased by the addition of $\approx$ 1 mV measurement noise, resulting in bad model estimates.

\subsection{Twin Experiment: Step Current, Sparse}
\label{sec:twinexperimentsstepsparse}

\begin{figure}[h]
\includegraphics[scale=0.5]{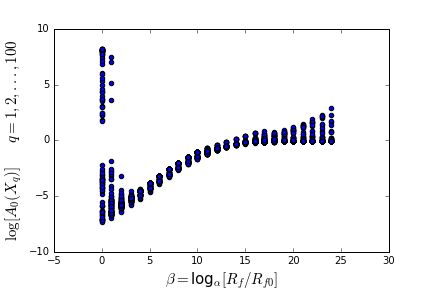}
\caption{Action level (cost function) plot for the sparsely sampled step current protocol. The natural logarithm of the action is plotted against $\beta$ for $\alpha = 2$. None of the estimated paths produce dominant contributions to the integral of equation \ref{eq:prob_dis} because the corresponding values of the cost function are not appreciatively different. The action flattens out with increasing $\beta$, and in contrast to the situation in Figure \ref{fig:action_levels_4s} in the low frequency complex current protocol, does not begin to rise again. Although these models described the data in the estimation and prediction window well, comparison of the estimated theoretical I-V curves with their true value shows that these estimated models are inferior to the models corresponding to the lowest action levels of Figure \ref{fig:action_levels_4s}. This demonstrates the importance of validating an estimated model in many ways, such as with a variety of novel stimulating currents in the prediction window.}
\label{fig:action_levels_0s}
\end{figure}

The step current protocol sampled at 10 kHz produced high quality predictions, also doing well on the clustering measures above. Out of 25 paths examined, 21 (84\%) produced high quality predictions. In each path examined, all anneal steps producing good predictions were retained. This is likely to be a confounding source of clustering in the measures used because of the similarity of the cost function landscape between annealing steps in the numerical procedure used to locate the lowest minima. A subset of these high quality predictions produced the best predictions, but in order to explore the distribution of the clustering methods above and to simulate conditions in a real experiment where unresolved processes are not known and low error on the training and validation data does not necessarily imply good generalization ability of the model, all accurate predictions were retained. One strategy to evaluate the ability of the algorithm to resolve unknown processes would be to choose only parameter sets producing accurate predictions that correspond to the lowest value of the action in the action level plot of Figure \ref{fig:action_levels_0s}. This strategy is not taken, however, because in real experiments, parameter sets corresponding to the lowest action values could be overfit to the training and validation data and generalize poorly to different stimulating currents.

Estimated parameters were symmetrically distributed around their true values in many cases with relatively small variance, or were clustered in a region close to their true values. The step current protocol produced the best quality of parameter estimates, in terms of the squared deviation of the estimates from their true values. Yet, the theoretical I-V curves were of lower quality than their low frequency chaotic current protocol counterparts. Among the very best of these predictions, the variance of the clustering measures would likely be smaller. The quality of predictions increased as the number of anneal steps increased, with no degradation in their quality in the range of beta completed in 24 hours.

The primary difference between this set of predictions and the other set of high quality predictions in the next section is that the action level plot plateaued (Figure \ref{fig:action_levels_0s}) and did not begin to increase again (Figure \ref{fig:action_levels_4s}).

\begin{figure}
\includegraphics[scale=0.3]{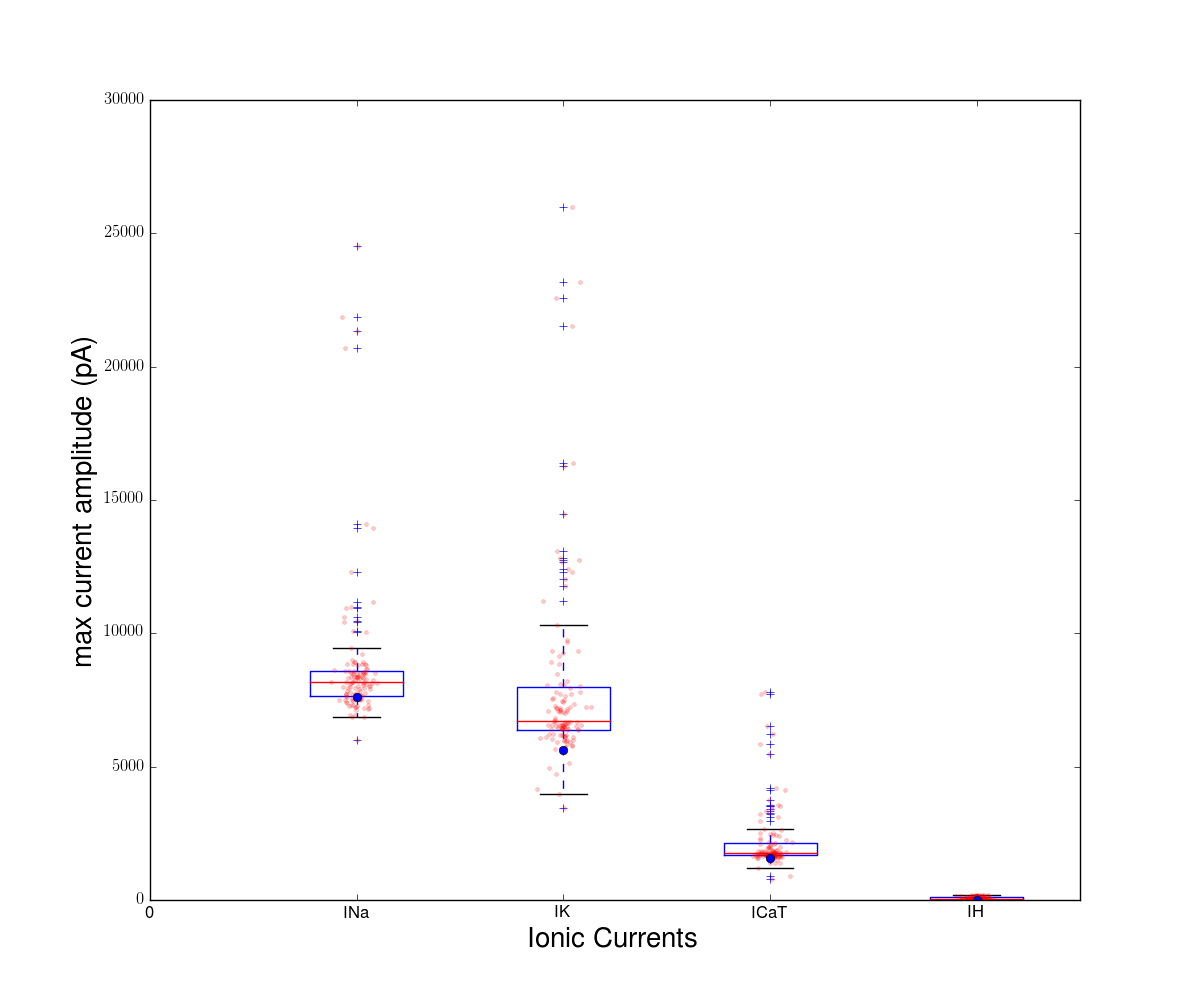}
\includegraphics[scale=0.3]{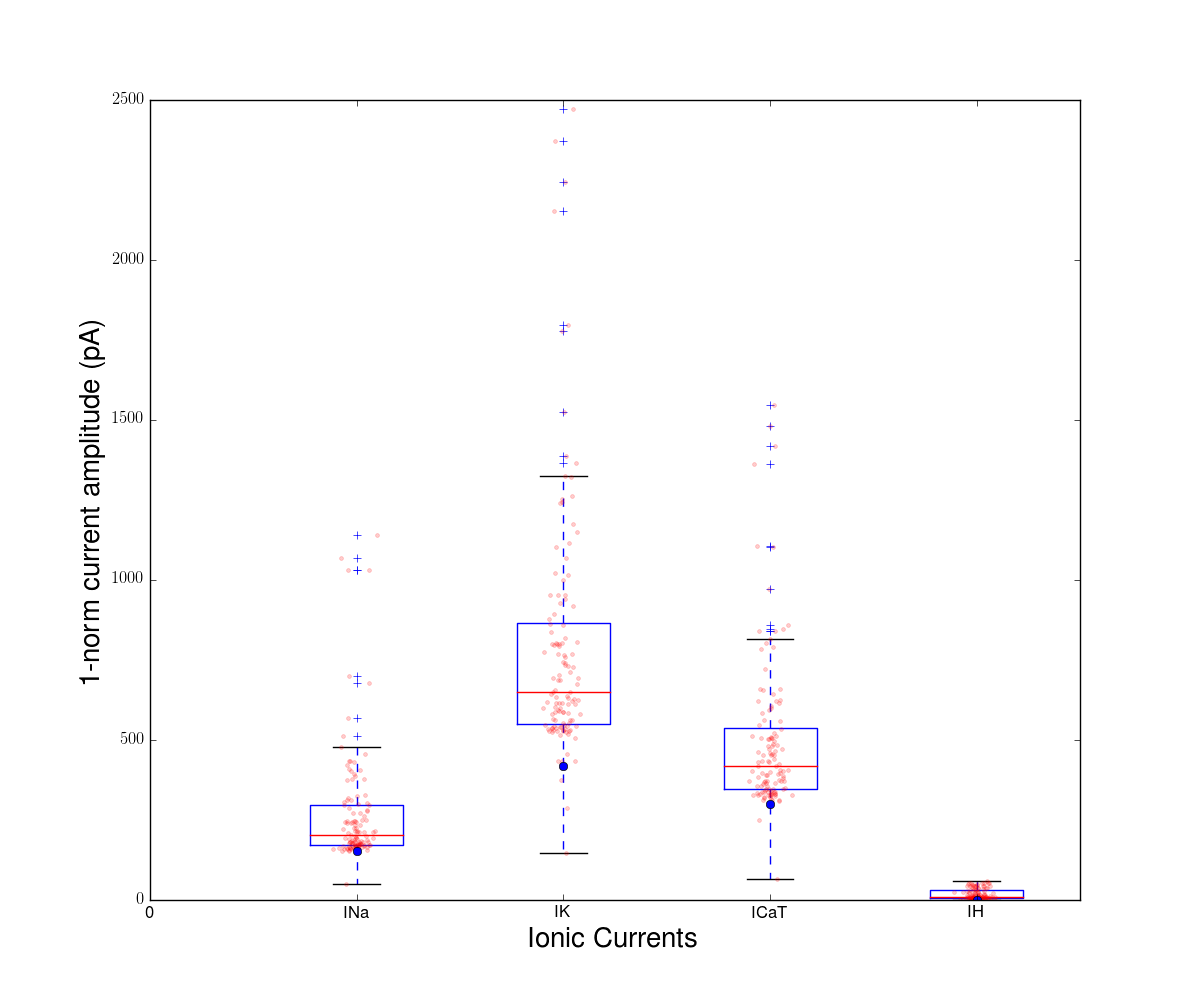}
\caption{Maximum values (top) and time averaged magnitude (bottom) of individual ionic currents over all parameter sets. The red dots are the values calculated from estimated parameter sets. The blue dots are the values calculated from the true parameter set used to generate the data from the sparse step current stimulating protocol.}
\label{fig:max_currents_0s}
\end{figure}

The maximally attained current amplitude for individual ionic currents over all parameters sets, plotted against their true values is shown in Figure \ref{fig:max_currents_0s}. On average, the estimated values are above the true value. This could be because the upper bounds on the values of the maximal conductances were usually significantly larger than the true values in order to simulate the ignorance likely to be present in a real experiment, and the model is likely to be degenerate in these parameters given the data. This can be, for example, because excessively large estimated depolarizing currents can sometimes be offset by excessively large rectifying currents. Other possible symmetries involve an increase in the maximal conductances and/or larger widths in the steady state activation functions for each ionic current offset by an increased value of the thresholds. In fact, this was often observed to be the case.

The time averaged magnitude of individual ionic currents over all parameter sets is also plotted against the corresponding true value in \ref{fig:max_currents_0s}. As with the max current amplitude measure, the estimated values are on average somewhat above the true value.

\begin{figure}
\includegraphics[scale=0.5]{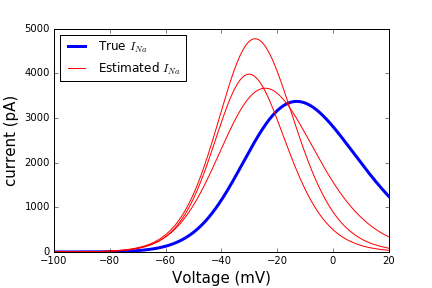}
\includegraphics[scale=0.5]{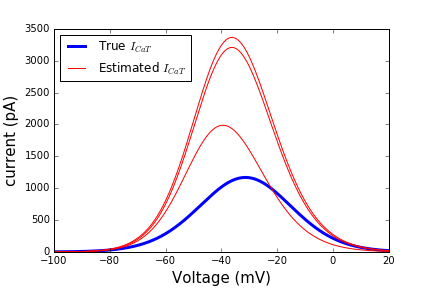}
\includegraphics[scale=0.5]{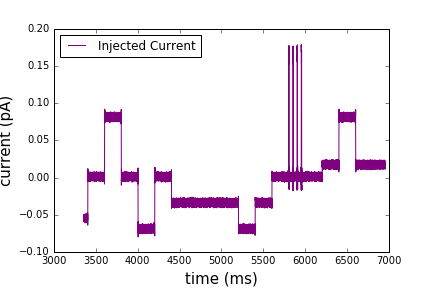}
\caption{Steady state activation functions for $I_{Na}$ (top) and $I_{CaT}$ (bottom) for the sparse step current stimulating protocol. The red curves are sample curves calculated from the estimated parameter sets, and the blue from the parameter sets used to generate the data. The estimates for $I_{CaT}$ appear to cluster around the true value, while for $I_{Na}$ are shifted a bit to the left due to $\sigma_m$ being substantially overestimated (not shown) with little compensation in other kinetic parameters. Overall, the shape of the estimated theoretical I-V curves using the step current protocol are less like the true theoretical I-V curves than those estimated using the low frequency complex current protocol of Figure \ref{fig:excitatory_currents_4s}}
\label{fig:excitatory_currents_0s}
\end{figure}

Next, the steady state value of the ionic currents as a function of voltage was calculated by substituting the steady state activation and inactivation functions in the expressions for the ionic currents. The pattern here was that the I-V curves for the excitatory currents $I_{Na}$ and $I_{CaT}$ were shifted to the left, as shown in Figure \ref{fig:excitatory_currents_0s}. The estimated I-V curves tended to match the shape of the true I-V curves less well using the step current protocol than the curves obtained from the complex current protocol, shown in the next section. This is likely due to the fact that the range of parameter sets which assimilate the data from the simpler step current drive is larger than the range of parameter sets assimilating the low frequency complex current protocol, which puts more constraints on the values of the parameters in the assimilation window.

\begin{figure}
\includegraphics[scale=0.5]{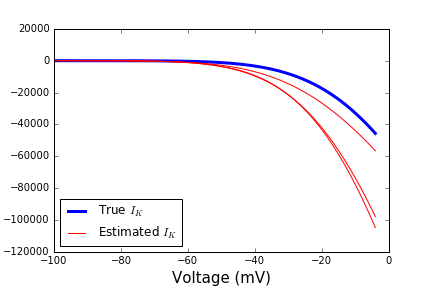}
\includegraphics[scale=0.5]{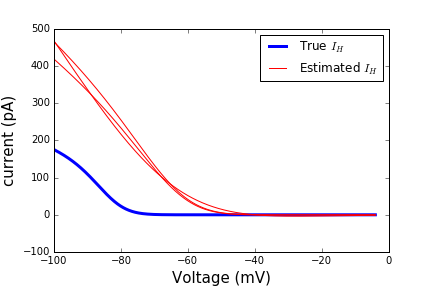}
\includegraphics[scale=0.5]{Dan_twinanalysis_epoch0sparse/current.png}
\caption{Steady state activation functions for $I_{K}$ (top) and $I_{H}$ (middle) for the sparse step current stimulating protocol (bottom). The red curves are sample curves calculated from the estimated parameter sets, and the blue from the parameter sets used to generate the data. $I_K$ seems to be well estimated, though here again $\sigma_n$ was substantially overestimated. To compensate, $g_K$ was often underestimated.}
\label{fig:other_currents_0s}
\end{figure}

The steady state activation curves for $I_K$ and $I_H$ are shown in Figure \ref{fig:other_currents_0s}. The $I_H$ waveform was not estimated as well as other current wave forms, probably due to the fact that it is a current that operates on a much slower timescale and could not be sufficiently sampled in the 1600 ms recording epoch. It also operates at hyperpolarized voltages, while the stimulating protocol used did not explore this regime of the membrane voltage sufficiently.

\subsection{Twin Experiment: Low Frequency Complex Current, Sparse}
\label{sec:twinexperimentscomplexsparse}

The low frequency complex current protocol sampled at 10 kHz produced high quality predictions and did well on the clustering measures. Out of 100 paths examined, 11 (11\%) produced high quality predictions. A subset of these high quality predictions produced the best predictions in the prediction window, but to best explore the distribution of the clustering of the model, all the levels of quality of predictions were retained.

\begin{figure}[h]
\includegraphics[scale=0.5]{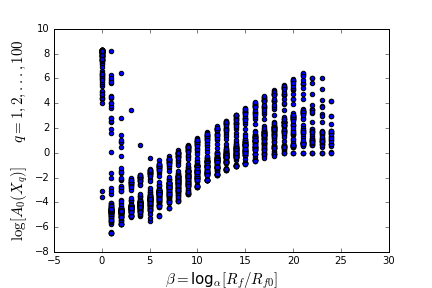}
\caption{Action level (cost function) plot for the sparsely sampled complex current protocol. The natural logarithm of the action is plotted against $\beta$ for $\alpha = 2$. None of the estimated paths produce dominant contributions to the integral of equation \ref{eq:prob_dis} because the corresponding values of the cost function are not appreciatively different. The action flattens out but begins to rise again with increasing $\beta$, in contrast to the situation in Figure \ref{fig:action_levels_0s}. This rising of the action indicates that the model corresponding to the minimum found in the high dimensional landscape is slightly wrong. The paths corresponding to these lowest action levels were still superior to those of Figure \ref{fig:action_levels_0s} as measured by forward prediction and by the shape of the estimated theoretical I-V curves compared to their true shapes, a result of the superior stimulation protocol used.}
\label{fig:action_levels_4s}
\end{figure}

The graph showing clustering of parameters is once again very large, as there were $\sim$40 parameters estimated. Therefore, this graph is omitted. Estimated parameters were clustered in a region close to their true values, but were systematically overestimated or underestimated. In no case were all of the parameters correctly estimated. Although the quality of predictions monotonically increased with the anneal step ($\beta$ value) in the range of $\beta$ completed in 24 hours, the estimated parameter sets were always wrong and therefore the action began to rise again after an initial plateau, as shown in Figure \ref{fig:action_levels_4s}. Despite this inaccuracy, the obtained predictions are still very good, suggesting that the parameter sets obtained are symmetries in the model given the data. Another possible explanation for the inaccurate parameter estimations is that the Gaussian noise added to the data in the estimation window degrades the quality of the estimations.

\begin{figure}
\includegraphics[scale=0.3]{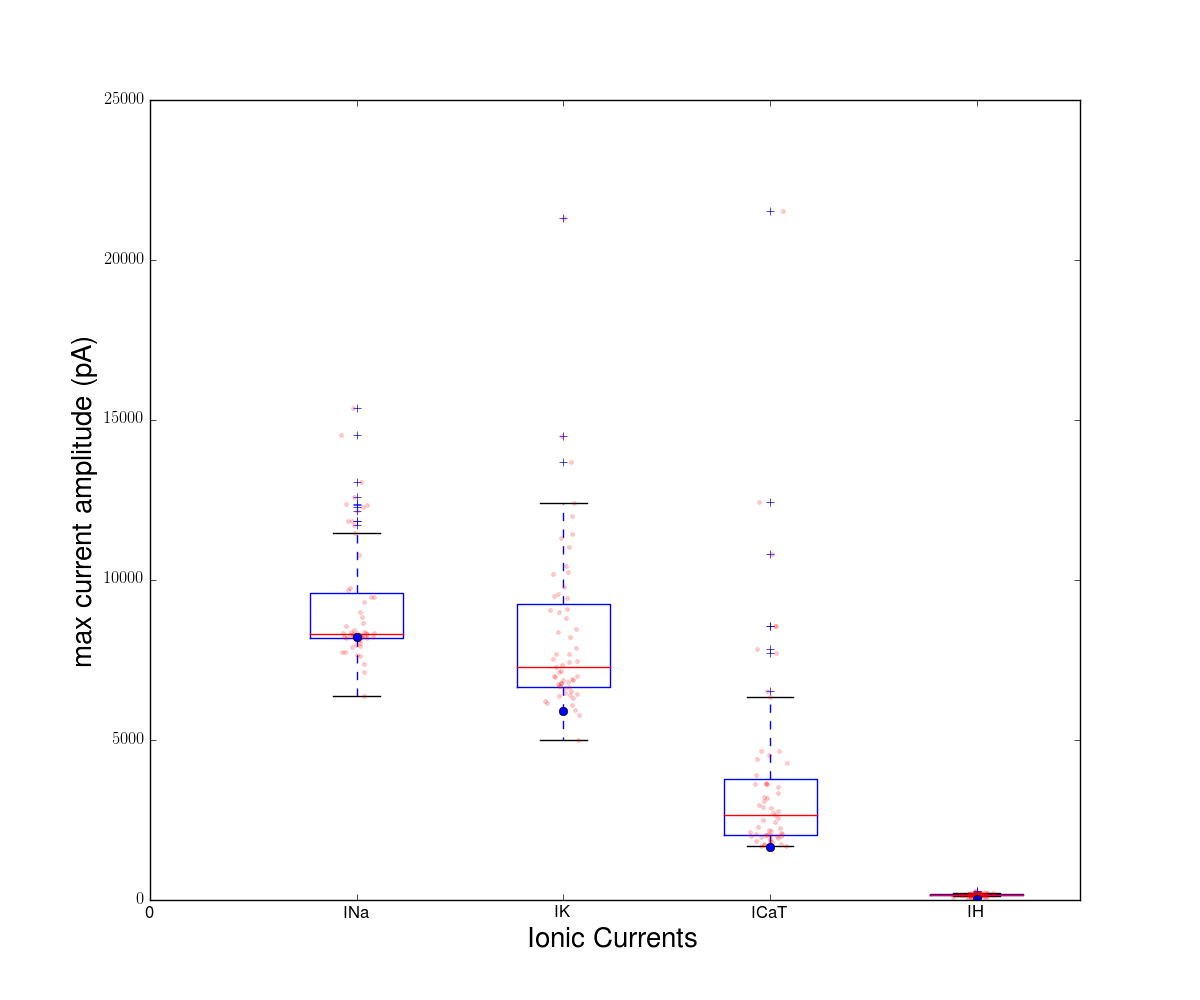}
\includegraphics[scale=0.3]{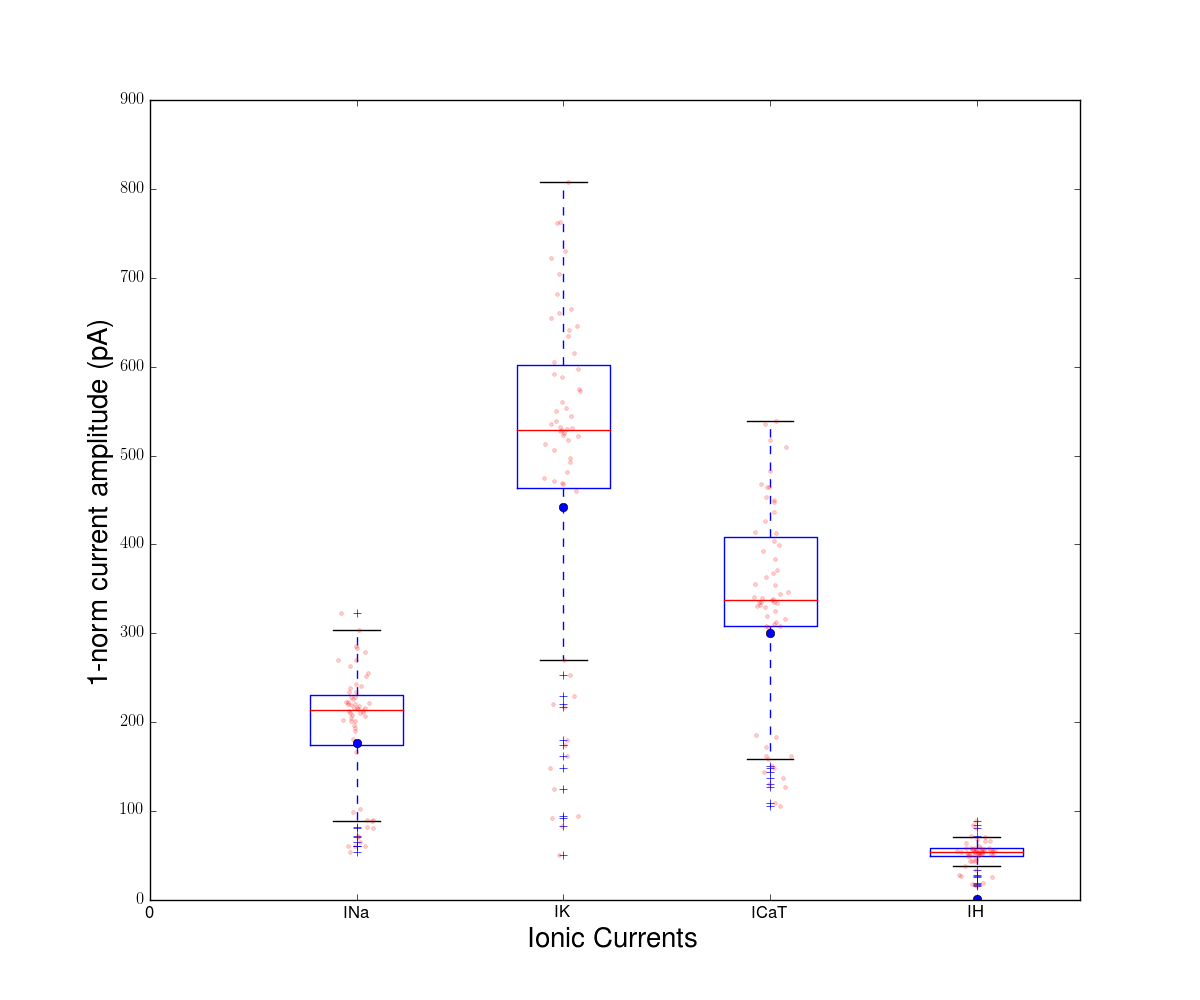}
\caption{Maximum values (top) and time averaged magnitude (bottom) of individual ionic currents over all parameter sets. The red dots are the values calculated from estimated parameter sets. The blue dots are the values calculated from the true parameter set used to generate the data from the sparse complex current stimulating protocol.}
\label{fig:max_currents_4s}
\end{figure}

The maximally attained current amplitude for individual ionic currents over all parameters sets, plotted against their true values is shown in Figure \ref{fig:max_currents_4s}. On average, the estimated values are above the true value. This could be because the upper bounds on the values of the maximal conductances were usually significantly larger than the true values in order to simulate the ignorance likely to be present in a real experiment, and since the model is likely to be degenerate in these parameters given the data, these parameters tended to be estimated in a range of parameter bounds usually above the true value.

The time averaged magnitude of individual ionic currents over all parameter sets is also plotted with the corresponding true value in Figure \ref{fig:max_currents_4s}. As with the max current amplitude measure, the estimated values are somewhat above the true value.

\begin{figure}
\includegraphics[scale=0.5]{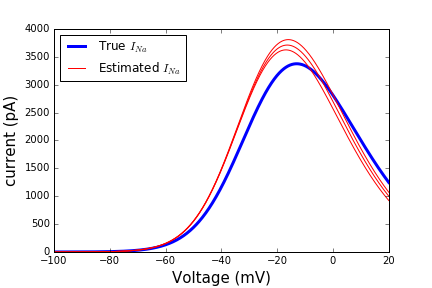}
\includegraphics[scale=0.5]{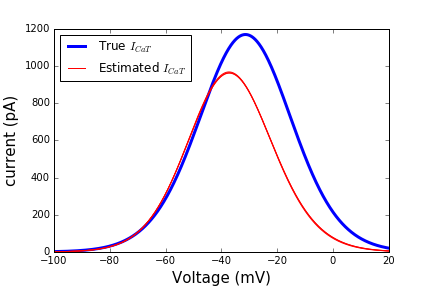}
\includegraphics[scale=0.5]{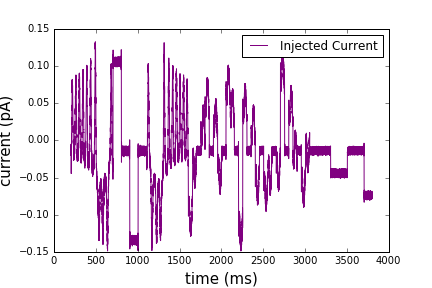}
\caption{Steady state activation functions for $I_{Na}$ (top) and $I_{CaT}$ (middle) for the sparse complex current stimulating protocol (bottom). The red curves are sample curves calculated from the estimated parameter sets, and the blue from the parameter sets used to generate the data. The magnitude of the current was generally overestimated. One respect in which these steady state curves seem to be improved upon the estimations using the step current protocol is that their average shape is much closer to the true shape, if sometimes estimated to be of larger magnitude, and they are much closer to the true value in regions sensitive to turning subthreshold stimulating into an action potential.}
\label{fig:excitatory_currents_4s}
\end{figure}

Next, the steady state value of the ionic currents as a function of voltage was calculated. The pattern here was that the I-V curves for the excitatory currents $I_{Na}$ and $I_{CaT}$ had an estimated shape which was mostly correct, but with a too-large amplitude. The curves for the complex stimulating protocol are shown in Figure \ref{fig:excitatory_currents_4s}. $I_K$ also compares well with its true value, but $I_H$ matches less well, especially at subthreshold voltages, which are not well sampled during the estimation procedure.

\begin{figure}
\includegraphics[scale=0.5]{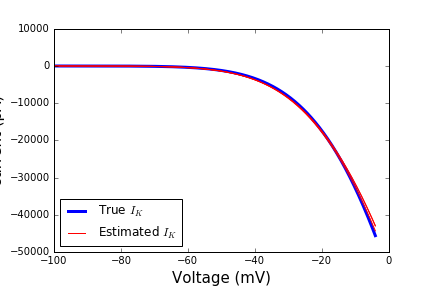}
\includegraphics[scale=0.5]{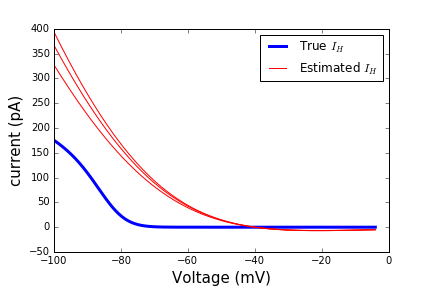}
\includegraphics[scale=0.5]{Dan_twinanalysis_epoch4sparse/current.png}
\caption{Steady state activation functions for $I_{K}$ (top) and $I_{H}$ (middle) for the sparse complex current stimulating protocol (bottom). The red curves are calculated from the estimated parameter sets, and the blue from the parameter sets used to generate the data. $I_K$ seems to be well estimated, though here the parameters $\sigma_n$, $g_K$ and $\theta_n$ were consistently overestimated. $I_H$, however, did not seem to be as well estimated, probably due to the fact that its overall contribution to net behavior is small.}
\label{fig:other_currents_4s}
\end{figure}

The I-V curves for $I_K$ and $I_H$ are shown in Figure \ref{fig:other_currents_4s}. Again, the $I_H$ waveform was not estimated as well as other current wave forms, probably due to the fact that it is a current that operates on a much slower timescale and could not be sufficiently sampled in the 1600 ms recording epoch. It also operates at hyperpolarized voltages, while the stimulating protocol used did not explore this regime of the membrane voltage sufficiently.

\section{Discussion}

\subsection{Conclusions to be drawn from model fit to real data}

Attaining a model fit producing accurate predictions to novel stimuli on voltage recordings from real HVC$_I$ neurons, including estimating \textit{all} parameters entering the model nonlinearly such as the kinetic parameters, is a significant achievement. To our knowledge, our methods of data assimilation are the only methods to date that are capable of this.

The model fit also reproduced qualitatively observed behavior of $\text{HVC}_\text{I}$ neurons, including a sag in the voltage in response to hyperpolarizing current injection and rebound spiking upon release. Additionally, the model was able to reproduce experimentally observed effects of blocking $I_H$ or $I_{CaT}$. This shows that the corresponding degrees of freedom in the model were stimulated during data assimilation with the recorded voltage waveform.

One potential limitation of the current methodology is that all voltage recordings are obtained by injecting a custom current waveform into the neuron which then drives the voltage. Even with a well selected complex current waveform, the regions of phase space of the model neuron are limited to those that can be stimulated at subthreshold membrane voltages and those that are attainable by relatively stereotyped limit cycle trajectories traced out during voltage spikes. A potential methodological improvement could be to control the membrane voltage directly and record the requisite injected current for bringing the membrane voltage to the control values. With cleverly designed control voltage waveforms, the region of phase space traversed and available to the assimilation algorithm could be significantly increased, constraining the number of theoretical I-V and relaxation curves compatible with the data.

\subsection{Conclusions to be drawn from data analysis of real data}
Out of the six different analyses performed using three stimulating current protocols eliciting a voltage measured from $\text{HVC}_\text{I}$ \textit{in vitro}, the low frequency current protocol at 10kHz (1000.1 ms) was the most successful. It appears that the improvement in prediction and estimation from longer estimation windows due to lower frequency sampling surpasses any decline in prediction and estimation quality due to lower sampling rates. It can also be concluded that a step current is less effective at sampling the I-V curves and relaxation properties of a neuron to clearly define minima in the action and prevent the formation of isosurfaces in the cost function which lead to inaccurate models. A chaotic current, with a region of extended negative current to explore subthreshold membrane voltages activating $I_H$ and $I_{CaT}$, is required for higher quality predictions. Both the currents of \autoref{input1} and \autoref{input4} have the characteristics mentioned. The difference between \autoref{input1} and \autoref{input4} is the frequency at which the input current oscillates. The current in \autoref{input1} has a much higher oscillation frequency, changing more rapidly than the membrane voltage can respond. A high sample rate may be required to adequately characterize fast fluctuations in the stimulating current, reducing the number of points in the assimilation window for exploration of slow currents like $I_H$ and $I_{CaT}$. Additionally, the membrane of the neuron is a capacitor, which is a low pass filter. Consequently, effects of high frequency stimulating currents may be indistinguishable from measurement error of the voltage, which is typically on the order of 1 mV. This will cause difficulty in estimating the parameters and characterizing properties of the neuron.
 
 The estimated characterization of electrophysiological properties found in the 10kHz analysis of the stimulating current protocol of \autoref{input4} are most correct based on the our metrics and known properties of $\text{HVC}_\text{I}$ neurons. In general, a chaotic current with slow oscillations relative to the $RC$ time constant of the membrane and an extended hyperpolarizing region within the estimation region are needed for the best predictions. Additionally a longer estimation window is more important than the additional resolution in time of a 50 kHz sampling rate as opposed to a 10 kHz sampling rate.
 
 In the above analyses, the estimated magnitude of $I_H$ is less than or equal in magnitude to the leak current. If a simpler model is desired, this suggests that $I_H$ may be removed from the model. Alternatively our model is missing some other crucial component, or the parameterization of $I_H$ in the model is not realistic.
 
The magnitude of $I_H$ could be as small as it appears, with its main effect in the subthreshold region, being small but contributing significantly to the behavior of the neuron. $I_H$ and $I_{CaT}$ are complementary and crucial mechanisms in the dynamics of some neural circuits \citep{huguenard1996low, mccormick1990properties}. Future work should examine the role of $I_H$ in the dynamics of HVC.

\subsection{Conclusions to be drawn from data analysis of synthetic data}

In our HVC$_I$ neuron model, different combinations of parameters could produce accurate predictions and effectively identical voltage traces. Plotting the theoretical I-V curves from the estimated sets of parameters against their true value shows that max conductances $g$, activation and inactivation thresholds $\theta$, and widths $\sigma$ can trade off with each other to produce I-V curves which are similar in shape, and which produce nearly identical responses in the membrane voltage to different values of the injected current.

There also exist model degeneracies in parameters given the assimilated and validation data. We found that data assimilation on step currents could produce a variety of models producing accurate predictions when current waveforms similar to those found in the estimation window were presented to the neuron. However, the resulting estimated theoretical I-V curves were less similar to the true theoretical I-V curves than in the case where a low frequency complex current waveform was presented to the neuron. If the neuron model estimated from the voltage waveform elicited by the step current stimulating protocol is presented with one of the chaotic current stimulating protocols, it is likely to produce inaccurate predictions.

When the voltage waveform produced by driving the neuron with the high frequency stimulating current was presented to the assimilation algorithm, it did not produce accurate estimated neuron models. This could have been caused by the fact that the high frequency stimulating current did not sample the subthreshold degrees of freedom of the neuron model, producing parameter sets able to predict accurately only in stimulating current regimes producing spiking, but doing relatively poorly when tested in subthreshold regimes. Overall, this stimulating current produced estimated theoretical I-V curves far from their true value.

Another factor explored in our twin experiments on HVC$_I$ neurons was the influence of the tradeoffs in higher or lower sampling rates given limited computational resources. We found that for assimilation on the HVC$_I$ neuron model, 10 kHz time resolution did better than 50 kHz. In this case, this means that the loss in information transfer from data to the model by reducing the time resolution was outweighed by the increase in information transfer by stimulating additional degrees of freedom in the model.


 
\label{sec:discussion}

\section*{Acknowledgments}

Additional information can be given in the template, such as to not include funder information in the acknowledgments section.

\FloatBarrier
\bibliographystyle{spbasic}
\bibliography{mybib}{}

\end{document}